\documentclass[a4paper,11pt]{article}
\pdfoutput=1 
\usepackage{jheppub} 
\usepackage[numbers]{natbib}
\usepackage[T1]{fontenc} 
\usepackage[dvipsnames]{xcolor}\usepackage{multirow}
\usepackage{multicol}
\usepackage{verbatim}
\usepackage{amsmath,mathtools}
\usepackage{relsize}
\usepackage{bm, nicefrac}
\usepackage{hyperref}
\usepackage{cleveref}
\usepackage[normalem]{ulem}
\usepackage{comment}
\usepackage{slashed}

\usepackage{amsmath}
\usepackage{amsfonts}
\usepackage{MnSymbol}
\usepackage{nicematrix}
\usepackage{graphicx}
\usepackage{subcaption}
\usepackage{placeins}
\usepackage{units}

\renewcommand{\to}{\rightarrow}

\newcommand{\THDMa}{2HDM$+a$}

\subheader{\rm IFT-UAM/CSIC-25-167}

\title{Towards precise baryogenesis in the \THDMa}
\author[a]{Tom Gent,}
\author[a]{Stephan Huber,}
\author[b]{Ken Mimasu,}
\author[c,d]{Jose Miguel No}

\affiliation[a]{Department of Physics and Astronomy, University of Sussex, Brighton, BN1 9QH, United Kingdom}
\affiliation[b]{School of Physics and Astronomy, University of Southampton,
Highfield, Southampton S017 1BJ, United Kingdom}
\affiliation[c]{Instituto de Fisica Teorica, IFT-UAM/CSIC,
Cantoblanco, 28049, Madrid, Spain}
\affiliation[d]{Departamento de Fisica Teorica, Universidad Autonoma de Madrid,
Cantoblanco, 28049, Madrid, Spain}

\emailAdd{t.gent@sussex.ac.uk}\emailAdd{s.huber@sussex.ac.uk}\emailAdd{ken.mimasu@soton.ac.uk}
\emailAdd{josemiguel.no@uam.es}

\preprint{
\begin{flushright}
\end{flushright}
}

\abstract{We perform a detailed investigation of the viable baryogenesis parameter space of a non-minimal Higgs sector consisting of two Higgs doublets and a singlet pseudoscalar (\THDMa). In such a model, an early Universe period of transient CP violation may occur, driven by a nonvanishing vacuum expectation value of the CP-odd scalar $a$. This naturally avoids the stringent electric dipole moment experimental constraints on beyond-the-Standard-Model sources of CP violation. We provide a state-of-art computation of the baryon asymmetry, providing several important improvements over existing baryogenesis computations for this model. We show that the required thermal history and successful baryogenesis lead to a predictive scenario, testable in the near future by a combination of LHC searches and low-energy flavour measurements. Our improved predictions for the baryon asymmetry find that it is rather suppressed compared to earlier predictions, requiring larger mixing between the singlet and 2HDM pseudoscalars and hence leading to a more easily testable model at colliders. 
}

\begin{document}
\maketitle
\section{Introduction}
The discovery of a standard model (SM) Higgs-like scalar at the LHC has brought about a fascinating era in high energy physics. Whilst the particle content of the SM appears complete, the consistency of the measured properties of the 125 GeV scalar with SM expectations means that we are no closer to solving some of the big puzzles surrounding this minimal realisation of electroweak (EW) symmetry breaking. One of the major questions surrounds the generation of the observed baryon asymmetry of the Universe (BAU), which can be expressed through the baryon-to-entropy ratio, 
\begin{align}
    \eta_B\equiv\frac{n_{\smash{B}}-n_{\smash{\bar{B}}}}{s}\simeq 8.7\times10^{-11}\,,
\end{align}
where $n_{\smash{B}}$  and $n_{\smash{\bar{B}}}$ are the total baryon and anti-baryon densities, $s$ is the total entropy density of the Universe and the numerical value has been measured by the Planck collaboration~\cite{Planck:2018vyg}.

It is well-known that the requirements for realising the BAU are encapsulated in the Sakharov conditions, namely Baryon number (B), C and CP violation, as well as a departure from thermal equilibrium in the early Universe~\cite{Sakharov:1967dj,Cohen:1993nk,Trodden:1998ym,Morrissey:2012db,Konstandin:2013caa}. The mechanism of EW baryogenesis, where the BAU is generated at temperatures around the EW scale, provides an intriguing connection between the physics of the early Universe and the nature of the Higgs sector. In the SM,
the first Sakharov condition is met through the existence of EW sphaleron processes that are unsuppressed at high temperatures~\cite{Kuzmin:1985mm}. Whilst a departure from thermal equilibrium can be realised through a strongly first-order EW phase transition (EWPT) in the early Universe, this transition is known to be a crossover~\cite{DOnofrio:2015gop} in the SM. Non-minimal Higgs sectors are therefore required in this scenario, and generically predict new states that can be searched for at colliders, as well as potentially observable stochastic gravitational wave backgrounds from the first-order EWPT.

The amount of CP violation present in the mixing of the SM fermion sector is also insufficient to realise EW baryogenesis~\cite{Gavela:1993ts,Gavela:1994ds,Gavela:1994dt}, meaning that candidate models must also include additional sources of CP breaking. However, new sources of CP violation associated to EW-scale particles are strongly constrained by experimental searches for electric dipole moments (EDMs) of the electron~\cite{ACME:2018yjb}, neutron~\cite{nEDM:2020crw, Roussy:2022cmp} 
%\TG{updated electron EDM bound from 2023} 
and atomic elements like mercury~\cite{Griffith:2009zz}, posing a significant problem for models with explicit sources of CP violation. This is exemplified by the case of the two-Higgs doublet model (2HDM), where minimal $\mathbb{Z}_2$-symmetric realisations that naturally suppress flavour-changing neutral currents struggle to reproduce the BAU whilst remaining consistent with EDM measurements~\cite{Fromme:2006cm,Dorsch:2016nrg,Basler:2021kgq}.
%\TG{there are some recents paper on baryogenesis in 2HDM that look slightly more promising \cite{Athron:2025iew, Aiko:2025tbk}}

The tension between EDMs and non-minimal Higgs sectors realising EW baryogenesis arises because the sources of CP violation in the early Universe are also active in the present day. An elegant possibility to decouple EW baryongenesis from EDM measurements is to have CP violation occur only during the time it is needed for generating the BAU, namely during the EWPT in the early Universe. This can be realised if the phase structure of the scalar sector leads to a period of spontaneous CP violation in the early Universe, which then undergoes a strongly first-order phase transition (FOPT) into the EW symmetry breaking vacuum where CP is restored; in other words, the evolution of the Universe features a period of \emph{transient} CP violation.
 The path taken by the scalar fields during the EWPT then involves a varying pseudoscalar degree of freedom which, provided it couples to fermions, can source the CP violating interaction with the bubble wall necessary to generate a net baryon asymmetry in the broken phase. 
 
 In Ref.~\cite{Huber:2022ndk}, we presented a minimal model that realises this scenario, involving extending the 2HDM with an additional $SU(2)_L$-singlet pseudoscalar $a$ (2HDM + $a$).
In our setup, a vacuum expectation value (vev) of the pseudoscalar $a$ in the early Universe triggers the spontaneous breaking of CP. CP is then restored after the EW phase transition, which is first-order from the existence of a %\KM{tree-level [should we keep this?]} 
potential barrier between the CP violating (CPV) and EW minima. The region of parameter space that accommodates this thermal history of the Universe and leads to successful baryogenesis leaves no trace in current EDM experiments, but can be probed via current/future LHC searches and low-energy flavour experiments (rare $B$-meson decays). % This model has also recently been considered as a well-motivated portal to dark matter (DM)~\cite{Ipek:2014gua,No:2015xqa,Goncalves:2016iyg,Bauer:2017ota,Abe:2018bpo,Robens:2021lov} \KM{this sentence is a bit isolated here}.

In this paper,  we continue our investigations into this model, focusing in particular on a more precise computation of the BAU on multiple fronts.
First, we go beyond both the tree-level approximation at zero temperature and the so called Hartree approximation (that only keeps $\propto T^2$ corrections) for the finite-temperature effective potential, including the full one-loop effects. Second, we numerically solve the bounce equation to determine the nucleation temperature, $T_n$, at which the EW phase transition occurs as well as the associated bubble field profiles that interpolate between the CPV and EW minima. Using these initial conditions, we then solve the transport equations to determine the BAU, which we find leads to a drastic reduction compared to the approximate formula used in Ref.~\cite{Huber:2022ndk} and consequently shifts the preferred parameter space region for successful baryogenesis in this model. 
A crucial input to the transport equations is the bubble wall velocity, $v_w$, which is notoriously difficult to compute as it involves solving a complicated system of integro-differential equations involving the plasma fluctuations and the scalar field(s).  Nevertheless, in many cases, it is possible to bound $v_w$ from above and below such that a range of possible BAU predictions can be inferred. Our calculations take this possible range into account, going beyond the typical analyses which are only valid for slow ($v_w\lesssim 0.2$) walls~\cite{Cline:2020jre}, or in which particular $v_w$ are simply put in by hand.

Although we find a reduced phase transition strength when including the full one-loop potential, the expected wall velocities are found to be rather fast ($v_w \gtrsim 0.3$) and the preferred region of parameter space is driven to slightly larger singlet masses and significantly larger mixings than in our original study, which are more experimentally accessible.  
Using this information, we confront the refined region of parameter space with existing experimental searches by considering several benchmark scenarios and discuss potential avenues for further exploring this scenario at colliders.

The paper is organised as follows. In~\cref{2HDM+a model}, we introduce the \THDMa\ scenario, establishing conventions and discuss the zero- and finite-temperature effective potentials that govern the evolution of the scalar fields in the early Universe. In~\cref{ewpt}, we discuss the EWPT in this model, focusing on the two-step phase transition history required to generate the BAU through a period of transient CP violation during the transition to the EW vacuum. \Cref{baryogenesis}
 details our computation of the BAU detailing our treatment of the transport equations and the unknown bubble wall velocity. In~\cref{sec:results}, we discuss the results of our computation, detailing the newly-preferred regions of parameter space with respect to our previous estimates~\cite{Huber:2022ndk}. We elucidate our numerical results with some analytical considerations and establish some benchmark scenarios which we confront with experimental constraints in~\cref{experimental}. Finally, we summarise and conclude in~\cref{sec:conclusions}.

\section{\THDMa: Model and particle content} \label{2HDM+a model}

\subsection{Scalar potential} \label{potential}
The zero-temperature tree-level scalar potential that describes this model is given by $V_0 = V_{\mathrm{2HDM}} + V_a$, where $V_{\mathrm{2HDM}}$ is the 2HDM potential for two electroweak doublets $\Phi_{1, \, 2}$, and $V_a$ is the potential involving the real pseudoscalar singlet $a$. The most general (gauge-invariant and renormalisable) 2HDM scalar potential reads
\begin{align} 
V_{\mathrm{2HDM}} = \, & \mu_{11}^{2} \, |\Phi_1|^2 \,+\, \mu_{22}^{2}\,|\Phi_2|^2 \,-\, \left( \mu_{12}^{2} \Phi_{1}^{\dagger} \Phi_2 + \mathrm{h.c.} \right) \,+\, \frac{\lambda_{1}}{2}\,|\Phi_1|^4 \,+\, \frac{\lambda_{2}}{2}\,|\Phi_2|^4 \,+\, \lambda_{3}\,|\Phi_1|^2 |\Phi_2|^2  \nonumber \\[5pt] 
& + \lambda_{4} \left| \Phi_{1}^{\dagger} \Phi_2 \right|^2 + \frac{1}{2} \left( \lambda_{5} \left(\Phi_{1}^{\dagger} \Phi_2\right)^2 + \mathrm{h.c.} \right) + \bigg\{ \left( \lambda_6 \, |\Phi_1|^2 + \lambda_7 \, |\Phi_2|^2 \right) \left( \Phi_{1}^{\dagger} \Phi_2 \right) + \mathrm{h.c.} \bigg\}   
\label{eq:V_0}
\end{align}
It is customary %common 
to impose a $\mathbb{Z}_2$ symmetry on the EW doublets, $\{ \Phi_1, \, \Phi_2 \} \rightarrow \{ -\Phi_1, \, \Phi_2 \}$, in order to avoid tree-level flavour-changing-neutral currents (FCNCs) mediated by the neutral Higgses~\cite{Paschos:1976ay, Glashow:1976nt}. Imposing this symmetry then leads to the vanishing of the coefficients $\lambda_6, \, \lambda_7$ and $\mu_{12}^2$ in the scalar potential~\eqref{eq:V_0}.\footnote{The basis of scalar doublets $\{ \Phi_1, \, \Phi_2 \}$ in which the $\mathbb{Z}_2$ symmetry is manifest is often called the `$\mathbb{Z}_2$ basis'.} However, it is also common to retain terms which yield a soft-breaking of this $\mathbb{Z}_2$ symmetry,
%break this symmetry but have operator dimension less than four, thus only softly-breaking the symmetry and 
whose coefficients will not contribute to the theory in the ultraviolet (UV). This has the additional benefit of 
preventing %suppressing 
domain wall formation in the early Universe upon EW symmetry breaking (EWSB)~\cite{Battye:2020jeu}. 
%\KM{mention why: domain wall problem if i recall}
%\TG{done and reference added}. 
We thus retain the $\mu_{12}^2$ term in the potential~\eqref{eq:V_0}.  Given this restriction, the potential $V_a$ involving the pseudoscalar singlet is
\begin{align}
\label{eq:Va}
V_a & = \frac{\mu_a^{2}}{2}\,a^2 + \frac{\lambda_a}{4}\,a^4 + \left( i\kappa \, a \, \Phi_{1}^{\dagger} \Phi_2 + \mathrm{h.c.} \right) + \frac{1}{2} \lambda_{a1}\,|\Phi_1|^2 \, a^2  \,+\, \frac{1}{2} \lambda_{a2}\,|\Phi_2|^2 \,a^2  \,,
\end{align}
%
%\KM{Mention that $a$ is also odd-under $Z_2$?} 
%\TG{done}
where the real singlet $a$ is odd under the same $\mathbb{Z}_2$ symmetry imposed on the EW doublets. Due to the $U(1)$ hypercharge re-phasing freedom of the doublets $\Phi_{1, \, 2}$, we can enforce $\lambda_5 \in \mathbb{R}$, so that only one potentially complex parameter in the 2HDM potential remains, $\mu_{12}^2$, which we additionally enforce to be real in order to conserve CP in the 2HDM scalar potential.
%Alternatively, the parameters can be left complex but their phases are directly related to each other. This version of the 2HDM is known in the literature as the CP conserving 2HDM. 
Furthermore, to avoid CP violating contributions at zero temperature from the pseudoscalar potential $V_a$, we require $\kappa \in \mathbb{R}$ and the singlet vev to vanish ($\langle a \rangle = 0$) in the EW minimum. The latter implies (see \cref{minima_App})
\begin{align}
\mu_a^2 + \frac{1}{2} (\lambda_{a1}\, v_1^2 + \lambda_{a2}\, v_2^2) = \mu_a^2 + \frac{1}{2} \lambda_{\beta}\, v^2 \equiv m_a^2 > 0 \label{pseudo mass} \,,\\[5pt]
v = \sqrt{v_1^2 + v_2^2} \,\, , \,\,\, t_{\beta} = \frac{v_2}{v_1} \,\, , \,\,\, \lambda_{\beta} = \frac{\lambda_{a1} + \lambda_{a2} \,t_{\beta}^2}{1 + t_{\beta}^2}\,,
\end{align}
%
%\TG{
%Minimisation condition in the singlet direction in the case of generic vevs reads,
%\begin{align}
%    &\mu_{a}^2 \, v_s + \frac{1}{2}\lambda_{a1}\,v_1\,v_s + \frac{1}{2}\lambda_{a2}\,v_2^2\,v_s + \lambda_a \, v_s^3 = 0 \\[5pt]
%    &\left(\mu_a^2 + \frac{1}{2} \lambda_{\beta}\,v^2 \right) v_s + \lambda_a \, v_s^3 = 0 \\[5pt]
%    &\left(\mu_a^2 + \frac{1}{2} \lambda_{\beta}\,v^2 \right) + \lambda_a \, v_s^2 = 0 \,\,\, , \,\, v_s \neq 0
%\end{align}
%If we impose that $\mu_a^2 + \frac{1}{2} \lambda_{\beta}\,v^2 > 0$ and $\lambda_a > 0$, then the minimisation condition cannot be satisfied.
%}
where $v_{1, \, 2} = \sqrt{2} \, \langle \Phi_{1, \, 2}^0 \rangle$ are the vevs of the neutral components of the doublets after EWSB, $v = 246.22$ GeV is the EW scale and $t_{\varphi} \equiv \tan{\varphi}$, $s_{\varphi} \equiv \sin{\varphi} $, $c_{\varphi} \equiv \cos{\varphi}$. %is a shorthand for $\tan{\varphi}$, likewise, $s_{\varphi}$ and $c_{\varphi}$ are a shorthand for $\sin{\varphi}$ and $\cos{\varphi}$ respectively. 
In this way, %Finally, $\kappa$ is assumed to be real so that 
the whole \THDMa\ scalar potential is CP conserving and there are no beyond-the-Standard-Model (BSM) contributions to EDMs at zero temperature.

\subsection{Particle content} \label{particle content}

The \THDMa\ model contains five additional scalar degrees of freedom (d.o.f) compared to the SM, namely four BSM states from the 2HDM and the CP-odd singlet scalar $a$.
%so what states do these new d.o.f's form and what properties do they possess? First, we examine the 2HDM spectrum, 
Expanding the doublets around the CP conserving EW minimum
\begin{align} \label{doublet expansion}
&\Phi_{1,2} = \begin{bmatrix}
\phi_{1,2}^{+} \\[7pt]
\frac{1}{\sqrt{2}}\left( v_{1,2}+h_{1,2}+i\,\eta_{1,2} \right)
\end{bmatrix} \,,
\end{align}
we obtain %which imposes 
the zero-temperature 2HDM minimisation conditions 
\begin{align}
&\mu_{11}^2 - M^2 \, s_{\beta}^2 +\frac{1}{2}\lambda_1 \, c_{\beta}^2 \, v^2 + \frac{1}{2} \lambda_{345} \, s_{\beta}^2 \, v^2 = 0 \,, \label{h1_min} \\
&\mu_{22}^2 - M^2 \, c_{\beta}^2 +\frac{1}{2}\lambda_2 \, s_{\beta}^2 \, v^2 + \frac{1}{2} \lambda_{345} \, c_{\beta}^2 \, v^2 = 0 \,, \label{h2_min} 
\end{align}
with $\lambda_{345} \equiv \lambda_3 + \lambda_4 + \lambda_5$ and $\mu_{12}^2 \equiv M^2 \, s_{\beta} \, c_{\beta}$. Diagonalising the consequent scalar mass matrices, the 2HDM spectrum contains one massless and one massive CP-odd scalar, $G_0$ and $A_0$, one massless and one massive charged scalar, $G_{\pm}$ and $H_{\pm}$, and two massive CP-even scalars, $h_0$ and $H_0$. The massless scalars correspond to the Goldstone modes absorbed by the massive EW gauge bosons. We stress that the assumption of $\mu_{12}^2$ and $\lambda_5$ being real prevents mixing between $h_{1, \, 2}$ and $\eta_{1, \, 2}$, leading to mass eigenstates with definite CP properties. 

In this work, we also restrict ourselves to the experimentally-favoured 2HDM alignment limit~\cite{Gunion:2002zf, ATLAS:2022vkf} in which the CP-even eigenstate $h_0$ is identified with the physical SM Higgs $h$ (see \cref{align} for more details).
The coupling $\kappa$ in~\cref{eq:Va} induces a mixing between the two CP-odd states $A_0$ and $a$ producing two mass eigenstates $a_1$ and $a_2$ (with $m_{a_1} < m_{a_2}$).\footnote{We note that if $\kappa$ were not real, there would be additional mixing of the pseudoscalar $a$ with the two CP-even scalars $h$ and $H_0$.} %\TG{is this just a convention or does this have consequences?}\KM{We can choose to label them as we like but in our case we are taking $a_2$ to be the mostly 2HDM state, we could also consider the opposite case in principle but have not done so}.
We denote this mixing by $s_{\theta}$, with the limit $s_{\theta} \to 0$ characterizing the absence of pseudoscalar mixing.
% (depending on the alignment limit chosen).}  %Our choice of real couplings allows the eigenstates of the theory to have definite CP properties. 
%\begin{center}
%\begin{tabular}{|p{1cm}|p{0.8cm}|p{0.8cm}|p{0.8cm}|}
%\hline
%State & Mass & CP & d.o.f \\[2pt] \hline
%$h$ & $m_{h}$ & Even & 1 \\[2pt] \hline
%$H_0$ & $m_{H_0}$ & Even & 1 \\[2pt] \hline
%$G_0$ & $0$ & Odd & 1 \\[2pt] \hline
%$a_1$ & $m_{a_1}$ & Odd & 1 \\[2pt] \hline
%$a_2$ & $m_{a_2}$ & Odd & 1 \\[2pt] \hline
%$G_{\pm}$ & $0$ & - & 2 \\[2pt] \hline
%$H_{\pm}$ & $m_{H_{\pm}}$ & - & 2 \\[2pt] \hline
%\end{tabular}
%\captionsetup{width=5.5cm}
%\captionof{table}{Summary of the mass eigenstates from the 2HDM+a sector and whether they are massive, have CP eigenvalues and their number of degrees of freedom\KM{This table may be overly explicit, if we keep it, make the caption textwidth.}}
%\end{center}
%\TG{removed table}
%
%
%
%
%
%Whilst the addition of a second Higgs doublet and a singlet pseudoscalar make the structure of the potential much richer, it also introduces many more parameters that one would like to theoretically and experimentally constrain. 
In the following, we also assume -- for the sake of simplicity -- 
%Therefore, along with assumptions thus far, we assume 
a common mass-scale for some of the new BSM states $M = m_{H_{\pm}} = m_{H_0}\simeq m_{a_2}$ and mass hierarchy $\{ m_{h}, \, m_{a_1} \} < M$. 
Finally, we can trade the singlet quartic coupling $\lambda_a$ for the singlet-field would-be vev $v_s$ in the absence of EWSB, given by $v_s^2 = - \mu_a^2/\lambda_a$.
The parameter set used for our exploration is thus $\{ t_\beta, \, s_\theta, \, v_s, \, m_{a_1}, \, m_{a_2}, \, M, \, \lambda_{a1}, \, \lambda_{a2} \}$. %\KM{missing $\mu_{12}$} \TG{$\mu_{12}^2 = M^2 \, s_{\beta} \,  c_{\beta}$}

\subsection{Higgs basis} \label{higgs basis}
Due to the $U(2)$ invariance of the kinetic terms of the two scalar doublets, the 2HDM potential can be realised in many different bases. An important one is the `Higgs basis' where 
%the two doublets are rotated such that 
only one doublet takes a vev at zero temperature and in this basis, the expanded doublets become
\begin{align}
&H_{1} = \begin{bmatrix}
G_{\pm} \\[7pt]
\frac{1}{\sqrt{2}}\left( v + h + i\,G_0 \right)
\end{bmatrix} \, , \,\,
H_{2} = \begin{bmatrix}
H_{\pm} \\[7pt]
\frac{1}{\sqrt{2}}\left( H_0 + i\,A_0 \right)
\end{bmatrix} \,.
\end{align}
When moving to the Higgs basis, the parameters of the scalar potential (\cref{eq:V_0,eq:Va}) and the minimisation conditions (\cref{h1_min,h2_min}) also transform accordingly (from the $\mathbb{Z}_2$ basis), with the potential reading
\begin{align}
&V_{\mathrm{2HDM}} = \, Y_1 \, |H_1|^2 \,+\, Y_2 \, |H_2|^2 \,-\, Y_3 \left(H_{1}^{\dagger} H_2 + \mathrm{h.c.} \right) \,+\, \frac{Z_1}{2}\,|H_1|^4 \,+\, \frac{Z_2}{2}\,|H_2|^4 \,+\, Z_3 \, |H_1|^2 |H_2|^2  \nonumber \\
& \phantom{V_{\mathrm{2HDM}} =} + Z_4 \left| H_{1}^{\dagger} H_2 \right|^2 + \frac{Z_5}{2} \left( \left(H_{1}^{\dagger} H_2\right)^2 + \mathrm{h.c.} \right) + \left( Z_6 \, |H_1|^2 + Z_7 \, |H_2|^2 \right) \left( H_{1}^{\dagger} H_2 + \mathrm{h.c.} \right)\,, \label{higgs 2hdm pot} \\[5pt]
&V_a =\, \frac{\mu_a^{2}}{2}\,a^2 + \frac{\lambda_a}{4}\,a^4 + i\, \kappa \, a \left( H_{1}^{\dagger} H_2 - \mathrm{h.c.} \right) \nonumber + \frac{1}{2} \lambda_{aH_1}\,|H_1|^2\,a^2  \\ 
&\phantom{V_a = } +\, \frac{1}{2} \lambda_{aH_2}\,|H_2|^2\,a^2 \,+\, \lambda_{aH_3} \left(H_{1}^{\dagger} H_2 + \mathrm{h.c.} \right) \, a^2 \,, \label{higgs pseudo pot}   
\end{align}
with minimisation conditions
\begin{align}
Y_1 = -\frac{1}{2} Z_1 v^2 \,\quad , \,\,\, \quad Y_3 = \frac{1}{2} Z_6 v^2 \, .
\end{align}
In the expressions above, reality of the couplings has been assumed. The masses, mixings and relations between fields and parameters in the two 2HDM bases are gathered for convenience in \cref{mixing,params}. In the Yukawa sector, we assume a Type-I 2HDM \cite{Branco:2011iw} although we stress here that our determination of the BAU is type-independent. Only the experimental constraints depend on 2HDM type, and we display Type-I constraints for concreteness. More details of the 2HDM Yukawa sector are given in \cref{yukawa}.
% \JMN{\textbf{JMN: Mention Yukawa sector (very briefly, we can cite a couple of papers, as all this is not necessary to be repeated in detail). We focus on Type I, do we? And refer to \Cref{yukawa}.}} 
%\TG{BAU analysis is type-independent but for collider side we assume and display type-I. We have flavour bounds for type-II also but I only describe these in words.}

\subsection{Effective potential} \label{effective potential}
We now discuss the effect of accounting for loop corrections to the scalar potential -- through the effective potential formalism --, which
can be split into zero- and finite-temperature parts. This already constitutes an extension of our previous analysis of the \THDMa\ model~\cite{Huber:2022ndk}, where only a certain part of the finite-temperature potential was retained and the zero-temperature potential was considered at tree-level.
%We now extend the analysis of to include one-loop effects which 
At zero temperature, these quantum corrections at one-loop are encapsulated by the well-known Coleman-Weinberg potential, $V_\mathrm{CW}$ \cite{Weinberg:1973am}. This, along with the tree-level potential $V_\mathrm{Tree}$ evaluated on classical field 
%\KM{what do you meant by constant here?} 
%\TG{changed constant to field} 
configurations (denoted by the `$c$' superscript), form the zero-temperature one-loop effective potential. Working in the Higgs basis, our classical field configurations are taken as
\begin{align}
    &H_{1}^{c} =
\begin{bmatrix}
0 \\[7pt]
\left(H_1^0\right)^c \end{bmatrix} \,\, , \,\,\,
H_{2}^{c} =
\begin{bmatrix}
0 \\[7pt]
\left(H_2^0\right)^c \end{bmatrix} \,\, , \,\,\,
a^c \label{classical config}\,,
\end{align}
\emph{i.e.} we consider CP conserving and CP violating field configurations at finite temperature but not charge breaking ones. 
The tree and one-loop contributions to the zero-temperature effective potential are
\begin{align} 
&V_{\mathrm{Tree}} = V_0 \left( H_1^c, \, H_2^c, \, a^c \right)=V_\mathrm{2HDM}+V_a \,,\\
&V_{\mathrm{CW}} = \frac{1}{64\pi^2} \sum_{i} (-1)^{2s_i} \, n_i \, m_i^4 \left( \log{\left( \frac{m_i^2}{\mu^2} \right)} - c_i \right) \,,\label{cw formula} 
\end{align}
where $i$ sums over each particle species and $s_i$, $n_i$ and $m_i^2$  are the corresponding spin, number of d.o.f, and field-dependent mass-squared eigenvalues. We set the renormalisation scale $\mu = v = 246.22$ GeV and $c_i$ are 
renormalisation-scheme dependent constants that depend on the polarisation of the particle. For all particles except transverse gauge bosons, $c_i = 3/2$, otherwise, $c_i = 1/2$ (see e.g.~\cite{Bernon:2017jgv}). From here onwards, we will drop the superscript `$c$' for brevity. 
At finite temperature $T$, there is an extra one-loop contribution 
to the scalar potential~\cite{Dolan:1973qd},
\begin{align}
\label{therm formula}
&V_{\mathrm{T}} = \frac{T^4}{2\pi^2} \left( \, \sum_{i \in \mathrm{B}} n_i J_{\mathrm{B}} \left(\beta^2 m_i^2\right) + \sum_{i \in \mathrm{F}} n_i J_{\mathrm{F}} \left(\beta^2 m_i^2\right) \right) \,,
\\
\label{therm int}
& J_{\mathrm{B}/\mathrm{F}} = \pm \int_{0}^{\infty} \, dx \, x^2 \log{\left( 1 \mp e^{-\sqrt{x^2 + \beta^2 m_i^2}} \right)} \,,
\end{align}
where $\beta = 1/T$  
and B and F denote bosons and fermions, 
respectively.

A further ingredient needed to fully characterise the one-loop 
effective potential is the field-dependent mass eigenvalues of the particles that will be included the sums of \cref{cw formula,therm formula}. In this work 
we include 
the top quark, massive
EW
gauge bosons and all \THDMa\  scalars. The top quark and EW gauge boson mass eigenvalues can be calculated from the SM Lagrangian, and \THDMa\ mass eigenvalues are found by diagonalising the Hessian of the scalar potential $V_0$ evaluated on the classical configurations in \cref{classical config},
\begin{align}
& m_t^2 = \frac{y_t^2}{2} \, \left| H_1^0 + t_{\beta}^{-1}\,H_2^0 \right|^{\,2} \,, \\[5pt]
& m_{W_{\pm}}^2 = \frac{g^2}{4} \, \left( \left| H_1^0 \right|^2 + \left| H_2^0 \right|^2 \right) \,, \\[5pt]
& m_{Z_0}^2 = \frac{g^2 + {g'}^2}{4} \, \left( \left| H_1^0 \right|^2 + \left| H_2^0 \right|^2 \right) \,, \\[5pt]
& m_{\mathrm{2HDM+a}}^2 = \mathrm{Eigenvalues} \left( \frac{\partial^2 V_0}{\partial \phi_i \partial \phi_j} \right) \Bigg|_{H_i = H_i^c, \, a = a^c} \,\,,
\end{align}
where $\phi = \{ G_{\pm, \, 1}, G_{\pm, \, 2}, H_{\pm, \, 1}, H_{\pm, \, 2}, h, H, G^0, A^0, a \}$ and the charged states $G_{\pm}$ and $H_{\pm}$ have been further decomposed into their real and imaginary parts, $G_{\pm, \, 1}$, $H_{\pm, \, 1}$ and $G_{\pm, \, 2}$, $H_{\pm, \, 2}$ respectively. $y_t$ is the top Yukawa coupling and $g$ and $g'$ are the $SU(2)_L$ and $U(1)_Y$ gauge couplings.
We do not include any other SM quarks (besides the top quark) or leptons due to their very small Yukawa couplings, which yield negligible contributions to the one-loop effective potential.    

Our Coleman-Weinberg potential must include a counterterm piece, $V_{\mathrm{CT}}$, that we fix by requiring that the configuration of the zero-temperature EW minimum and zero-temperature tree-level masses are preserved. 
We note that in the parametrisation of \cref{classical config} we have the freedom to remove the phase of either complex configuration of the vevs of the neutral fields, %neutral $H$ vevs 
since the potential depends only on the difference between their phases. Taking advantage of this freedom, we can enforce the first doublet in either basis to be real and rotate the phase into the second doublet. More explicitly, in the $\mathbb{Z}_2$ and Higgs bases,
%\begin{align}
%&\Phi_1^0 = h_1 + i \eta_1 = \chi_1 \, e^{i\varphi_1} \, , \,\, \Phi_2^0 = h_2 + i \, \eta_2 %= \chi_2 \, e^{i\varphi_2} \quad \longrightarrow \quad \Phi_1^0 = \chi_1 \, , \,\, \Phi_2^0 = %\chi_2 \, e^{i\varphi} , \,\, \varphi = \varphi_2 - \varphi_1 \label{z2 phases} \\[5pt]
%&H_1^0 = h + i G_0 = \rho_1 \, e^{i\delta_1} \, , \,\, H_2^0 = H_0 + i A_0= \rho_2 \, %e^{i\delta_2} \quad \longrightarrow \quad H_1^0 = \rho_1 \, , \,\, H_2^0 = \rho_2 \, %e^{i\delta} , \,\, \delta = \delta_2 - \delta_1 \label{hg phases}  
%\end{align}
%
\begin{align}
&\mathbb{Z}_2 \, \mathrm{basis}: \quad \,\,\,\ \Phi_1^0 = \chi_1 \, , \,\, \Phi_2^0 = \chi_2 \, e^{i\varphi} , \,\, \varphi = \varphi_2 - \varphi_1 \label{z2 phases} \,, \\[10pt]
&\mathrm{Higgs\,basis}: \,\, H_1^0 = \rho_1 \, , \,\, H_2^0 = \rho_2 \, e^{i\delta} , \,\, \delta = \delta_2 - \delta_1 \,.\label{hg phases}  
\end{align}
%
%\KM{maybe we can keep one equation and just say it is the same for the other basis with different labels for the moduli and phases.} \TG{is this any better?}
In particular, using the linear parametrisation $\{\rho_1$, $\rho_2 \, e^{i\delta} = \rho_2^{\mathrm{R}} + i\rho_2^{\mathrm{I}}, \, a\}$ we preserve the location of the zero-temperature EW minimum $\left( \langle \rho_1 \rangle, \langle \rho_2^{\mathrm{R}} \rangle, \langle \rho_2^{\mathrm{I}} \rangle, \langle a \rangle \right) = \left( v, 0, 0, 0 \right)$ and the physical tree-level masses. 
%\KM{not clear here, you mention linear parametrisation but the refer to symbols of the non-linear parametrisation ($\rho_i$). Also $rho$'s should be real according ot the above equation, no? } 
%\TG{yes, incorrect notation, have fixed this now} 
The counterterm potential reads,
\begin{align}
V_{\mathrm{CT}} = \sum_j \frac{\partial V_{\mathrm{Tree}}}{\partial p_j} \delta p_j \,\,,
\end{align}
where $j$ labels the number of parameters in the tree-level potential, $p_j$ are the parameters and $\delta p_j$ are the counterterm parameters. Up to now, the Landau gauge has been assumed as doing so decouples ghosts from the rest of the SM, simplifying calculations. However, as a consequence, Goldstone bosons are now massless when evaluated at the zero-temperature vev and this introduces an IR divergence when calculating the aforementioned counterterms, which we determine numerically. Their expressions and how the IR divergences are dealt with can be found in \cref{counterterms}.

\vspace{1mm}

In addition, we include in our analysis -- as it will be relevant for the wall velocity determination, see~\cref{sec:Vwall} -- the potential $\tilde{V}_0$ stemming from the approximately massless degrees of freedom in the plasma that are not captured in the sums of~\cref{cw formula} and~\cref{therm formula}, whose number is given by $g_{*}$:  
\begin{align}
\tilde{V}_0 = -\frac{\pi^2 g_{*} T^4}{90} \,.%\, , \,\, g_{*} = 83.25 
\end{align} 
%KM{$g^\ast=83.25$ is presumably for a particular temperature range, should we mention it?} \TG{valid until b quarks hadronise, $\simeq \mathcal{O}(100 \, \mathrm{MeV})$}
%\KM{Explain what goes into $g^\ast$}

\vspace{1mm}

Finally, it is well known that in finite-temperature quantum field theory (QFT) the perturbative expansion breaks down in the infrared (IR) limit due to the existence of Matsubara zero modes for scalars and gauge bosons \cite{Linde:1978px, Gross:1980br}. The leading IR divergences are caused by a class of multi-loop contributions called `Daisy' diagrams and resumming these diagrams shifts the field-dependent masses by factors proportional to $T^2$. We use the `Arnold-Espinosa' resummation approach~\cite{Arnold:1992rz} to take this effect into account  and it ultimately leads to a new `Daisy' %`ring' 
term being included in the effective potential,
\begin{align}
V_{\mathrm{Daisy}} = - \frac{T}{12\pi} \sum_{i \in B} \left( \left(\overline{m}_i^2 \right)^{3/2} - \left(m_i^2 \right)^{3/2} \right) \label{daisy formula} \,,
\end{align}
where $\overline{m_i}^2$ are the thermally corrected masses, whose expressions are collected in~\cref{debye}.
%For brevity, the expressions for the thermally corrected masses . 
Many other %resummation 
methods that resum additional sub-leading multi-loop diagrams %divergences 
do exist for single and multiple fields \cite{Espinosa:1992gq, Parwani:1991gq, Dine:1992wr,Boyd:1993tz, Andersen:2004fp, Curtin:2016urg, Curtin:2022ovx, Bahl:2024ykv, Bittar:2025lcr}. Yet, going beyond the widely-used `Arnold-Espinosa' or `Parwani'~\cite{Parwani:1991gq} resummation schemes is not only difficult but numerically intensive, and beyond the scope of this work. 
%\KM{justify this a bit?} 
%\TG{justify using Ar-Esp scheme vs Parwani or justify usage of this resummation scheme compared to the fancier ones referenced?}
%\TG{I think we decided not to comment further}
%
Altogether, our resummed one-loop effective potential reads
\begin{align}
\label{Veff_total}
&V_{\mathrm{eff}}\left({\rho_1, \rho_2^{\mathrm{R}}, \rho_2^{\mathrm{I}}, a}\right) = \tilde{V}_0 + V_{\mathrm{Tree}} + V_{\mathrm{CW}} + V_{\mathrm{CT}} + V_{\mathrm{T}} + V_{\mathrm{Daisy}} \,.
\end{align}

Before moving to the next section, we briefly discuss several issues related to %Of course, one can improve on 
the effective potential $V_{\mathrm{eff}}$ that is presented here, which is ubiquitously used in the literature for similar models. %For example, 
For certain regions in field space, some field-dependent masses can become tachyonic and one must then decide how to interpret \cref{cw formula} and \cref{therm formula}. It has been argued that the effective potential at zero temperature becoming complex and its consequent imaginary part can be interpreted as the decay rate of some well-defined state \cite{Weinberg:1987vp}. Therefore, we take the real part of \cref{cw formula} when $m^2 < 0$. However, a tachyonic mass is harder to interpret in \cref{therm formula} as the bosonic integral would no longer be convergent due to a divergence inside of the limits of integration, as well as both integrals becoming complex. Therefore, we extend the bosonic and fermionic integrals \cref{therm int} to $m^2 < 0$ by first defining %creating 
an interpolating function for $m^2 > 0$ whose second derivative is continuous. The function then passes through $m^2=0$ such that the second derivative is also continuous at that point. Lastly, the interpolating function decays to zero as $m^2 \rightarrow -\infty$ with the usual Boltzmann suppression.

Another issue concerns the
uncertainty introduced by the renormalisation scale, $\mu$, in \cref{cw formula}. It has been shown in \cite{Gould:2021oba} that for a scalar toy model, the one-loop effective potential at zero temperature is invariant under variations of this scale. However, this is not the case when finite-temperature effects are included, and the potential would only be invariant (up to some order in the perturbative expansion) if two-loop effects were included. Therefore, the impact of a variation of the renormalisation scale 
can be regarded as an estimate of neglected two-loop effects. Two-loop perturbative and non-perturbative studies of EWPTs have been carried out in the literature for simpler models such as singlet and triplet extensions of the SM, or the 2HDM~\cite{Laine:2000rm, Laine:2017hdk, Kainulainen:2019kyp, Niemi:2020hto, Niemi:2021qvp, Ramsey-Musolf:2024ykk, Niemi:2024axp, Niemi:2024vzw} but this is yet to be performed for the \THDMa, and lies beyond the scope of this work. 

\section{The electroweak phase transition} \label{ewpt}

As outlined in the introduction (and previously discussed in~\cite{Huber:2022ndk}), the \THDMa\ allows for
two-step phase transitions: CP and EW symmetries are initially restored at some high temperature and as the temperature decreases CP is eventually spontaneously broken; decreasing the temperature further, EW symmetry is then broken and a transition from the CP breaking vacuum to the EW vacuum %two vacua 
may occur. In terms of fields, the latter transition -- the EW phase transition -- reads
%It is the second transition that is of interest and in terms of fields,
%
\begin{align}
\left( \langle \rho_1 \rangle, \langle \rho_2^{\mathrm{R}} \rangle, \langle \rho_2^{\mathrm{I}} \rangle, \langle a \rangle \right) = \left( 0, 0, 0, v_s(T) \right) \rightarrow \left( v_1(T), v_2(T), 0, 0 \right) \nonumber \,.
\end{align}
The Universe then sits in the (finite-temperature) EW vacuum until it evolves to the present (zero-temperature) EW vacuum that we live in. 

Asserting the above symmetry breaking pattern in the early Universe constrains the mass of the pseudoscalar $a_1$. The corresponding bounds on $m_{a_1}$
were obtained analytically in \cite{Huber:2022ndk}, for the tree-level zero-temperature potential and assuming the Hartree approximation that only keeps $\propto T^2$ corrections to the finite-temperature potential. These bounds read
\begin{align}
\label{eq:ma1_min}
&m_{a_1, \, \mathrm{min}}^2 = \frac{1}{c_{\theta}^2 - s_{\theta}^2} \left( \frac{1}{2} c_{\theta}^2 \left( \lambda_{\beta} - \frac{m_{h}^2}{v_s^2} \right) - s_{\theta}^2 \, m_{A_0}^2 \right) \,,\\[6pt]
&m_{a_1, \, \mathrm{max}}^2 = \frac{1}{c_{\theta}^2 - s_{\theta}^2} \left( \frac{1}{2} c_{\theta}^2 \left( \lambda_{\beta} - 2F \right) - s_{\theta}^2 \, m_{A_0}^2 \right)\,,
\label{eq:ma1_max}
\end{align}
with
\begin{align}
&F = \frac{\left( 2\lambda_{a1} + 2\lambda_{a2} + 3\lambda_a \right) m_h^2}{\Delta + 10 m_h^2 + 12 m_t^2 + 12 m_{W_{\pm}}^2 + 6 m_{Z_0}^2 + \lambda_{\beta} v^2} \,,\\[6pt]
&\Delta = 4(m_{H_{\pm}}^2 - M^2) + 2(m_{H_0}^2 - M^2) + 2(m_{A_0}^2 - M^2)\,,
\end{align}
and $m_{A_0}$ being the would-be mass of the 2HDM pseudoscalar $A_0$ in the absence of mixing (i.e. for $s_{\theta} = 0$).
The upper and lower bounds on $m_{a_1}$ stem, respectively, from requiring that spontaneous CP breaking occurs before EWSB, and from demanding the EW minimum to be the global minimum at $T=0$. The temperatures at which the CP and EW breaking phase transitions occur are approximated by the destabilisation of the origin of field-space in the pseudoscalar and Higgs-direction, respectively.
Since in this work we use the full one-loop effective potential,~\cref{eq:ma1_min,eq:ma1_max} become only approximate, and analogous analytic bounds at one-loop are not possible to obtain. Yet, they serve as a useful guide for exploring the \THDMa\ parameter space.

\begin{figure}[h] 
\centering
\includegraphics[scale=0.55]{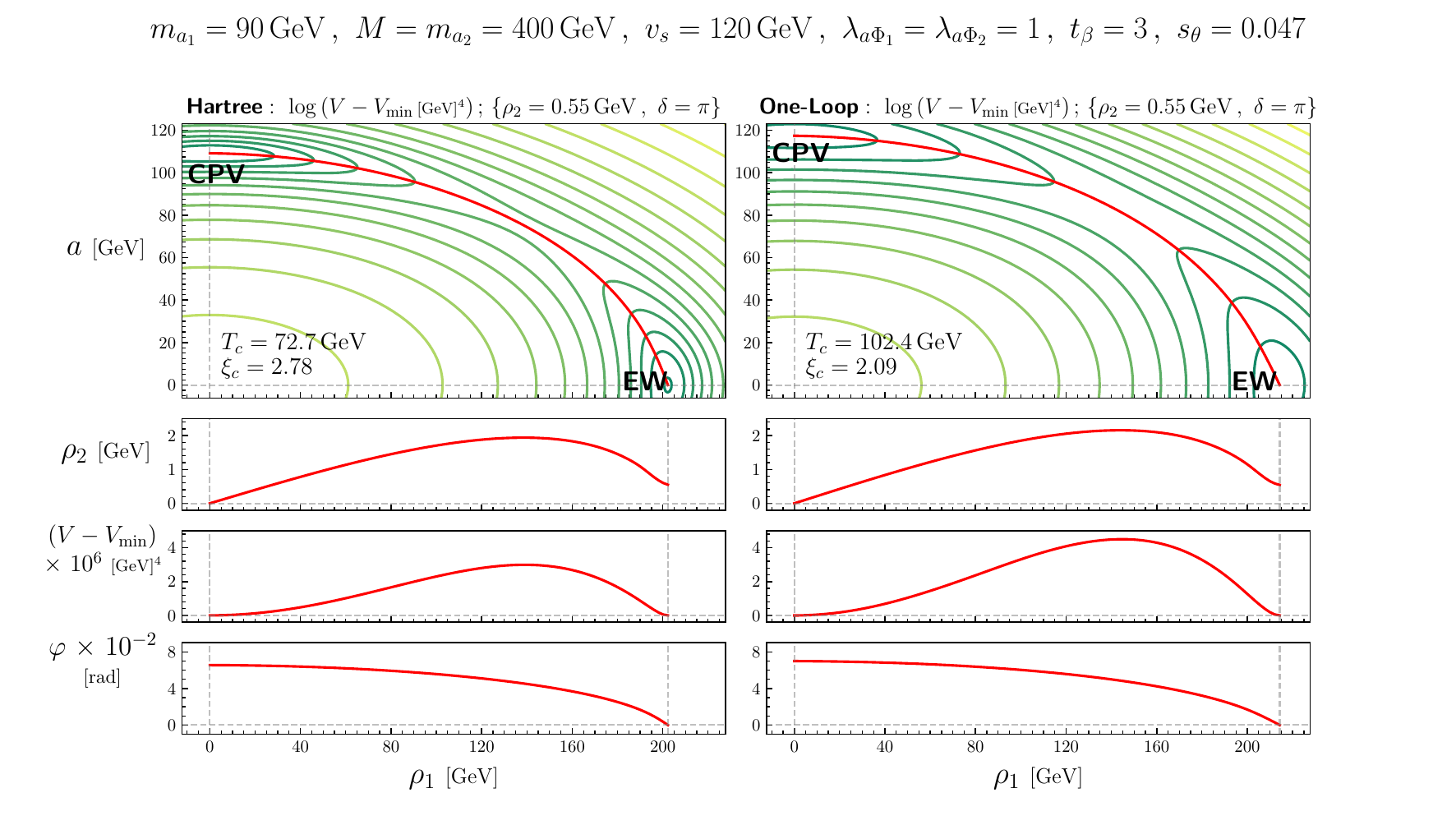}

\vspace{-3mm}

\caption{\small Iso-contours of the effective potential at the critical temperature in the $\rho_1$-$a$ plane for the Hartree (left) and full one-loop effective potential (right), with $\rho_2$ and $\delta$ fixed at their value in the EW minimum. The red curve traces along the valley (the MEP, see text for details) connecting the CPV and EW minima. The figures below the potential landscape show how $\rho_2$, $\varphi$ and $V - V_{\mathrm{min}}$ vary along this valley.}
\label{fig:crit_temp_figure}
\end{figure}

%\begin{align}
%&m_{a_1, \, \mathrm{min}}^2 = \frac{1}{c_{\theta}^2 - s_{\theta}^2} \left( \frac{1}{2} c_{\theta}^2 \left( \lambda_{\beta} - \frac{m_{h}^2}{v_s^2} \right) - s_{\theta}^2 \, m_{A_0}^2 \right) \\[6pt]
%&m_{a_1, \, \mathrm{max}}^2 = \frac{1}{c_{\theta}^2 - s_{\theta}^2} \left( \frac{1}{2} c_{\theta}^2 \left( \lambda_{\beta} - 2F \right) - s_{\theta}^2 \, m_{A_0}^2 \right) \\[6pt]
%&F = \frac{\left( 2\lambda_{a, \Phi_1} + 2\lambda_{a, \Phi_2} + 3\lambda_a \right) m_h^2}{\Delta + 10 m_h^2 + 12 m_t^2 + 12 m_{W_{\pm}}^2 + 6 m_{Z_0}^2 + \lambda_{\beta} v^2} \\[6pt]
%&\Delta = 4(m_{H_{\pm}}^2 - M^2) + 2(m_{H_0}^2 - M^2) + 2(m_{A_0}^2 - M^2)
%\end{align}
%\KM{explain where the lower and upper bounds come from as only one includes finite temperature effects.} 
%\TG{done} 
We also note that for spontaneous CP breaking in the early Universe to be possible, the potential $V_a$ in \cref{higgs pseudo pot} must have a negative squared-mass term at zero temperature ($\mu_a^2 < 0$), which yields a weaker upper bound on $m_{a_1}$~\cite{Huber:2022ndk}.
Within the range $m_{a_1} \in [m_{a_1, \, \mathrm{min}},\,m_{a_1, \, \mathrm{max}}]$ there exists a critical temperature $T_c$ for which the CP and EW breaking extrema are degenerate, and the presence of a barrier between the two vacua is indicative of a FOPT.

\Cref{fig:crit_temp_figure} provides a visual example of a parameter point that features such a transition for the Hartree (left) and resummed one-loop (right) effective potentials,
where iso-contours of the potential are shown in the $\rho_1 - a$ plane (top plot) -- with $\rho_2$ and $\delta$ fixed at their values in EW minimum. The minimum energy path (MEP) which tracks the valley connecting the CPV and EW minima (evaluated using the same method in \cite{Huber:2022ndk}) is shown as a red curve, and the variation of $\rho_2$, $\varphi$ and $V - V_{\mathrm{min}}$ along the MEP is depicted in the bottom plots. The variation of $V - V_{\mathrm{min}}$ depicts the potential barrier between the two minima, whilst that of $\varphi$ indicates the transient CP violation during the transition between them, where a pseudoscalar degree of freedom has a varying expectation value.

In the following, we always require an appropriate thermal history numerically -- yielding a transient period of CP violation as discussed above -- for our resummed one-loop effective potential. In addition, we require boundedness from below of the zero-temperature effective potential -- which can again be checked numerically -- as well as perturbative unitarity of the tree-level scalar potential couplings.\footnote{These perturbative unitarity constraints are taken to hold approximately at the one-loop level.} We note that in the case of the Hartree approximation used in~\cite{Huber:2022ndk}, these bounds can be cast in an analytic form, and we summarise the corresponding results in \cref{bfb,per uni}.

\begin{figure}[h]
\centering
\includegraphics[scale=0.55]{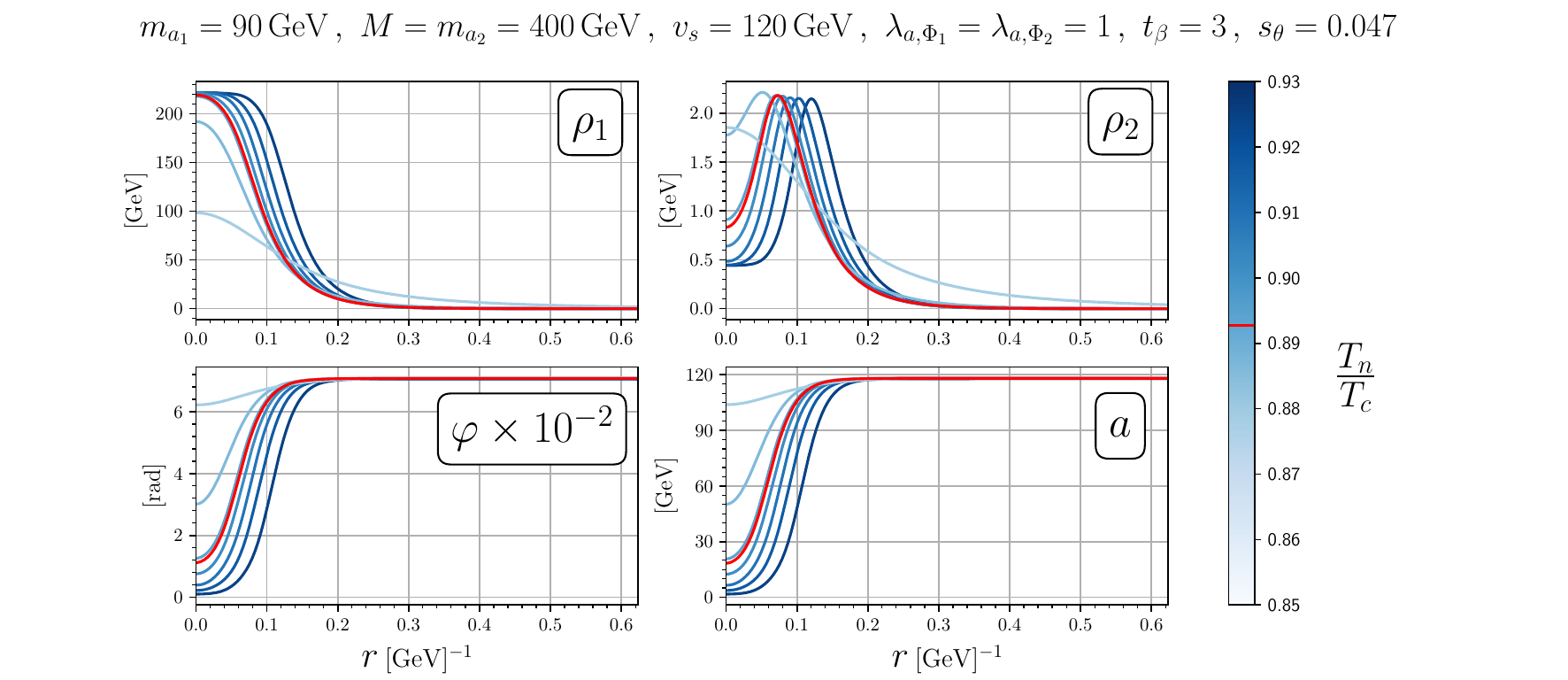}
\caption{\small One-loop bounce profiles at differing values of supercooling $T_n/T_c$ for the same benchmark point in~\cref{fig:crit_temp_figure}. The solid red curves are the configurations which satisfy the nucleation condition, $S_3/T \approx 140$.}
\label{fig:nucl_profiles}
\end{figure}

If a potential barrier exists between the CP breaking and EW vacua
at $T_c$,
the phase transition may proceed through bubble nucleation at a temperature $T_n < T_c$~\cite{Coleman:1977py, Callan:1977pt, Linde:1980tt}. Such a FOPT 
involves bubbles of the true (EW) vacuum nucleating in a sea of the false (CPV) vacuum. These bubbles (when larger than a critical size) expand into the false vacuum, converting it to the true vacuum as they grow.
%and, if the bubbles are of critical size, they expand into the false vacuum until it is converted into the true vacuum. 
The rate of nucleation of such true vacuum bubbles is controlled by the three-dimensional Euclidean action of the theory, $S_3$, given by
\begin{align}
& S_3 = 4\pi \int_0^{\infty} dr \, r^2 \left( \sum_i \frac{1}{2} \left( \frac{d \phi_i}{dr} \right)^2 + V_{\mathrm{eff}}(\{\phi\}, T) \right) \,, \label{action}
\end{align}
under the assumption of spherically symmetric bubbles. The collection of scalar fields $\phi_i$ in the effective potential is denoted as $\{ \phi \}$, and $r = |\vec{x}|$ where $x^\mu = (t,\vec{x})$.
The action $S_3$
%~\cref{action} 
is then evaluated on the field configurations which interpolate between the true and false vacua and are determined by solving the equations of motion (EOMs) subject to the appropriate boundary conditions,
\begin{align}
&\left( \frac{d^2}{d r^2} + \frac{2}{r} \frac{d}{dr} \right) \, \phi_i = \frac{\partial V_{\mathrm{eff}}}{\partial \phi_i} \,\, , \,\,\, \{\phi\} = \{\rho_1, \, \rho_2^{\mathrm{R}}, \, \rho_2^{\mathrm{I}}, \, a\}  \, ,\\[5pt]
& \lim_{r \, \rightarrow \, 0^{+}} \frac{d \phi_{i}}{dr} = 0 \, , \,\, \lim_{r\, \rightarrow \, +\infty} \phi_{i} = \phi_i^{\mathrm{f}} \, ,
\end{align}  
where $\phi_i^{f}$ is the false vacuum. This field configuration is known as the \textit{bounce} solution, and we showcase in~\cref{fig:nucl_profiles} an example of such a solution for the same \THDMa\ benchmark point as~\cref{fig:crit_temp_figure} which satisfies the criteria discussed above. 

The temperature $T_n$ at which bubbles start nucleating is determined by the condition~$S_3(T_n) / T_n \approx 140$ \cite{Anderson:1991zb} -- corresponding to the average nucleation of one true vacuum bubble per Hubble volume. The bounce profile that satisfies this condition is shown in red in \cref{fig:nucl_profiles}.
%Whether or not the bubbles nucleate is determined by satisfying the condition .
Note that if the potential barrier between the two vacua persists down to $T=0$, this nucleation condition may not be satisfied for any $T < T_c$, and as a consequence, the Universe will stay trapped in the false vacuum. This phenomenon, referred to as `vacuum-trapping', has been studied in various BSM extensions \cite{Cline:1999wi, Baum:2020vfl, Biekotter:2021ysx, Goncalves:2021egx, Biekotter:2022kgf} and is naturally accounted for in our analysis.
%\TG{since we attempt to solve for the bounce}. 
In the following we use $T_n$ as the phase transition temperature.\footnote{It is also possible for the bubbles to expand at a velocity and temperature such that the Hubble expansion prevails and the bubbles may not ever reach each other. In such a case the phase transition is very slow (or might not even complete) and the temperature at which it completes is significantly lower than $T_n$,
see~\cite{Athron:2022mmm} for a detailed discussion. However, this is often a feature of phase transitions with large supercooling ($T_n/T_c \ll 1$) which will not be the case in this work.}

In general, solving for the bounce solution is numerically challenging as the EOMs form a coupled system of non-linear boundary value problems. However, a multitude of numerical tools to determine the bubble profiles and whether or not nucleation occurs are publicly available \cite{Wainwright:2011kj, Athron:2019nbd, Guada:2020xnz, Basler:2024aaf, Athron:2024xrh, Hua:2025fap}. We calculate the bounce solution using an in-house code that is based on the algorithm presented in~\cite{Athron:2019nbd}, and we have further cross-checked a sample of our parameter points with the \texttt{PhaseTracer2}~\cite{Athron:2024xrh} public tool. 

Assuming a FOPT occurs with a nucleation temperature $T_n$, the strength of the transition can be characterised by the quantity $\xi_n = v_n / T_n$, with $v_n = \sqrt{v_1(T_n)^2 + v_2(T_n)^2}$, recalling that $v_1$ and $v_2$ are the vevs of the two scalar doublets, here evaluated at the temperature $T_n$. To ensure that the baryon asymmetry generated during the EWPT (see~\cref{baryogenesis}) is not washed-out in the true vacuum, one requires 
$\xi_n \gtrsim 1.0$ (see e.g.~\cite{Quiros:1999jp}).
%
% We note here that within our two-step symmetry breaking regime, it is possible at one-loop for the spontaneous breaking of CP to be a FOPT. However, given that the barrier is potentially generated by radiative corrections, we expect the transition to be relatively weak and therefore unlikely to be detectable by gravitational wave experiments.  
 %

\begin{figure}[h] 
\centering
\includegraphics[scale=0.5]{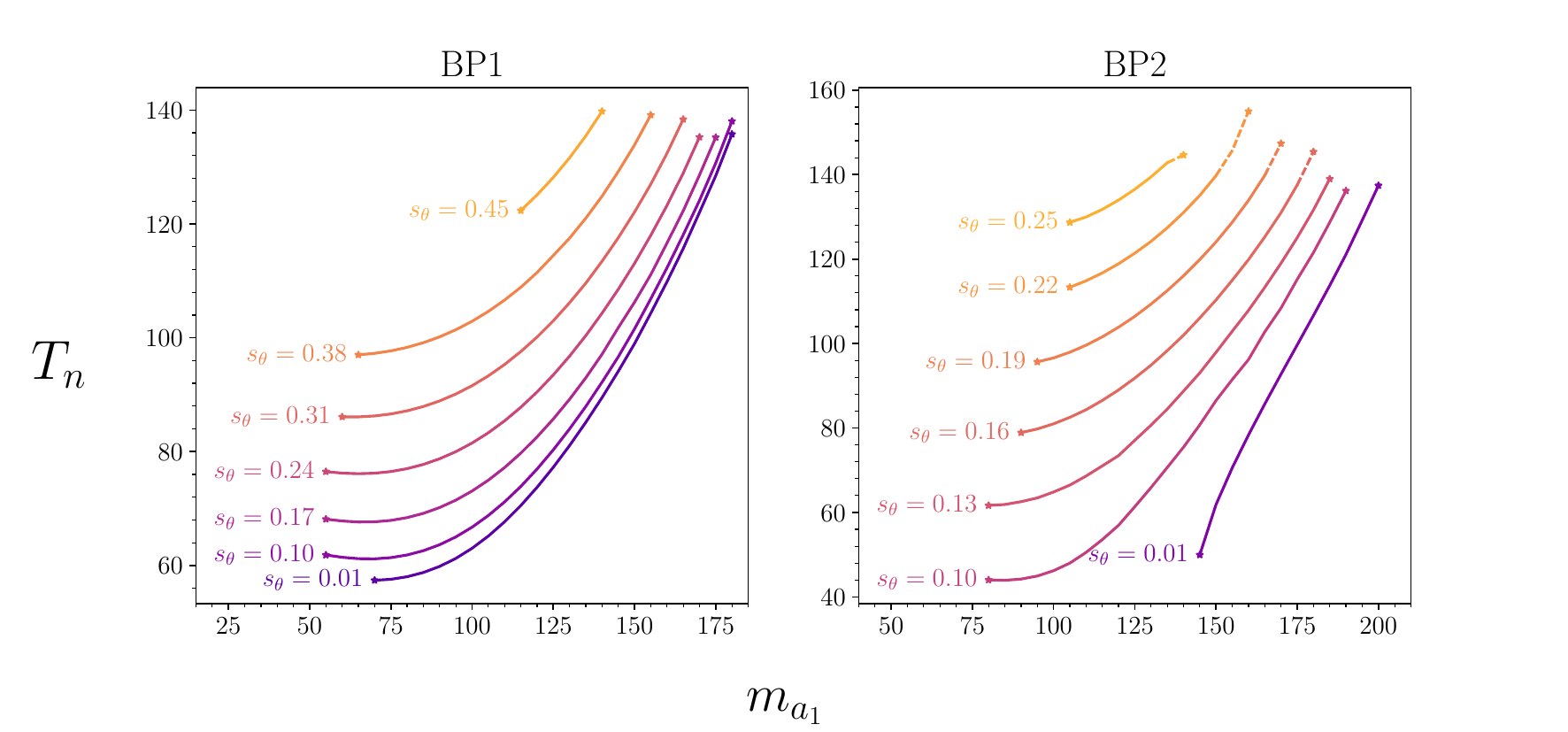}
\caption{\small Sample phase diagrams for the \THDMa, showing the nucleation temperature, $T_n$ - at fixed mixing angles $s_{\theta}$ - as a function of the singlet mass $m_{a_1}$ for BP1 (left) and BP2 (right). Lines are dashed where $\xi_n < 1$.
}
\label{fig:phase_diagram}
\end{figure}

Finally, we show in~\cref{fig:phase_diagram} the dependence of the nucleation temperature with the mass $m_{a_1}$, for various values of the pseudoscalar mixing $s_{\theta}$ and two choices of the remaining \THDMa\ parameters, respectively labeled BP1 (left) and BP2 (right) and given explicitly in~\cref{tab:benchmarks ma1 stheta}.
We stress that each line in~\cref{fig:phase_diagram} has two mass endpoints, beyond which a first-order EWPT is not possible for the benchmark chosen. For each line, the region depicted as dashed corresponds to $\xi_n < 1$, indicating that in such case successful baryogenesis would not be possible. Additionally, for such weak transitions, perturbative methods may not be a reliable assessor of the order of the transition.

\section{Baryogenesis} 
\label{baryogenesis}
In this section, we describe how the interplay between CP violation and a strongly first-order EWPT can lead to successful baryogenesis. 
As discussed in~\cref{ewpt}, during the EW phase transition, EW vacuum bubbles nucleate into a Universe that is filled with the CP violating vacuum. Therefore, outside of each bubble, quarks of one chirality and anti-quarks of the opposite
chirality interact with the bubble wall differently to one another. This results in a difference between their respective transmission and reflection coefficients, which produces
a net baryon asymmetry inside the bubble. In the outer vicinity of 
the bubble, the opposite baryon asymmetry also exists but it can partly be erased by EW sphaleron processes, which are unsuppressed in regions where there is no 
EWSB~\cite{Khlebnikov:1988sr, Rubakov:1996vz}. What fraction of this outer asymmetry 
is erased then depends on the EW sphaleron interaction rate and the velocity of the bubble wall $v_w$. Additionally, the assumption of a strongly FOPT ($\xi_n \gtrsim 1$) is required to ensure that the baryon asymmetry inside of the bubbles is preserved; otherwise, EW sphalerons will also erase it. One should however keep in mind that this $\xi_n \gtrsim 1$ criterion is only approximate, as discussed in \cite{Patel:2011th}, yet still serves as a useful guide for exploring parameter space. 

\subsection{Transport equations} \label{transport equations}

The diffusion of CP-sources through the bubble wall and their subsequent conversion into a baryon asymmetry can be described by a set of transport equations, whose derivation is based on a  Wentzel-Kramers-Brillouin (WKB) expansion in gradients of the bubble profiles. Boltzmann equations can then be written that include a semi-classical force term originating from the CP-sources. By taking moments of these equations, it is possible to convert the system of Boltzmann equations into a system of coupled diffusion equations for the chemical potentials, $\mu$, and plasma velocities, $u$, of the relevant species. Again, due to the hierarchy in Yukawa couplings, the most relevant quark in this scenario will be the top quark. The diffusive equation for a single particle species takes the form (see e.g.~\cite{Cline:2020jre}) 
\begin{align}
A \, \frac{d w}{dz} + \frac{d m^2}{dz} B \, w = S + \delta C \label{transport ode} \, ,
\end{align}
where $w = (\mu , \, u)^{T}$, $z$ is the co-ordinate perpendicular to the bubble wall, $m(z)$ is the mass profile for that species, $S(z)$ is the corresponding CP violating source and $\{A, \, B \}$ are functions which depend on the wall velocity $v_w$ and the combination $m/T$. 
The different  species then couple through $\delta C$, which encodes number-density changing interactions, e.g.~$W$-boson interactions and helicity flips (whose rates also scale with temperature). In this paper, the relevant species are left- and right-handed top quarks, left-handed bottom quarks and the scalar fields of the \THDMa. 

In addition to the above,
%Furthermore, 
it is assumed that the bubble has sufficiently expanded such that the phase transition boundary is locally a planar wall to a very good approximation.
We also assume that the process happens at the nucleation temperature $T_n$. More details on the structure of~\cref{transport ode} are given in~\cref{transport detail} (following the discussion in \cite{Cline:2020jre}). 

Given that the EW phase transition is singlet-assisted in our setup, we expect relatively strong transitions and hence potentially large bubble wall velocities. Therefore, the transport formalism
of \cite{Joyce:1994zt, Cline:2000nw, Kainulainen:2002th, Prokopec:2004ic, Fromme:2006wx} is unsuitable as the equations therein are expanded to linear order in $v_w$. 
On account of this, we use the transport equations of~\cite{Cline:2020jre} which are valid for all values of $v_w$. For the field profiles that enter the transport equations, we solve for the bounce configuration in the Higgs basis, construct the $\chi_1$, $\chi_2$ and $\varphi$ profiles from the $\mathbb{Z}_2$ basis (\cref{z2 phases}) and then fit these profiles to a tanh-ansatz centred around $z=0$. Schematically these profiles look like 
\begin{align}
\phi = \langle \phi \rangle_{\mathrm{CPV}} + \frac{\langle \phi \rangle_{\mathrm{EW}} - \langle \phi \rangle_{\mathrm{CPV}}}{2} \left( 1 - \tanh{\bigg( \frac{z}{L_w^{\phi}} \bigg)} \right) \,\, , \,\,\, \phi \in \{ \chi_1, \, \chi_2, \, \varphi \} \,, \label{prof param}
\end{align}
where $\langle \phi \rangle_{\mathrm{EW}}$ is evaluated in the minimum at $T_n$ and not at the endpoint of the tunnelling path. This is due to our assumption that the wall sufficiently expands (i.e. the thin-wall approximation is valid) before baryogenesis occurs. The $L_w^{\phi}$ parameter is the bubble wall thickness (for each field), and 
the validity of the WKB expansion used to derive the transport equations requires
that the bubble walls are sufficiently thick, $L_w^{\phi} \, T_n \gtrsim 2$. We find this to be generally satisfied across the regions we investigate in this work.
More advanced parametrisations for the profiles can be performed\footnote{e.g. the above procedure is inaccurate for very thick bubble walls. However, the BAU is approximately proportional to $(L_w^{\phi})^{-1}$ 
and hence these parameter points are not of interest for successful baryogenesis.} but we find that~\cref{prof param} is sufficiently accurate for parameter points relevant for producing the BAU. 

\subsection{CP violating source} \label{cp-source}
As outlined in \cite{Huber:2022ndk}, CP violation is generated transiently by the pseudoscalar singlet field $a$ developing a vev at some finite temperature, $\langle a \rangle = v_s(T)$. Working in the $\mathbb{Z}_2$ basis of the 2HDM, this generates a physical (non-removable) and temperature-dependent phase in the mixed doublet mass term
\begin{align}
\left( \mu_{12}^{2} \, \Phi_{1}^{\dagger} \Phi_2 + \mathrm{h.c.} \right) \nonumber \rightarrow \left( \mu_{12}^{2}(T) \, \Phi_{1}^{\dagger} \Phi_2 + \mathrm{h.c.} \right) \nonumber \,,
\end{align}
where $\mu_{12}^2(T) = \mu_{12}^2 - i\, \kappa \, v_s(T)$. The physical CP violating phase is $\delta_s \equiv \arg{\left( \mu_{12}^2(T)^{*} \, \mu_{12}^2 \right)}$. The tree-level potential for the two Higgs doublets essentially becomes 
a version of the complex-2HDM (C2HDM) with temperature-dependent $\mu_{11}^2, \, \mu_{22}^2$ and $\mu_{12}^2$ coefficients. The CP violating phase $\delta_s$ is precisely equal to the phase $\varphi$ from \cref{z2 phases} evaluated in the CP violating minimum.\footnote{Strictly speaking, $\varphi$ is ill-defined at the CP violating minimum, since both Higgs vevs vanish. However, it is well-defined by its limiting value, as the Higgs field trajectories approach the CP violating minimum.} The (space-time dependent) value of $\varphi$ across the phase transition boundary can be expressed in terms of the fields $\rho_1$, $\rho_2^{\mathrm{R}}$ and $\rho_2^{\mathrm{I}}$,
\begin{align}
t_{\varphi} = \frac{\rho_1 \rho_2^{\mathrm{I}}}{\frac{1}{2} s_{2\beta} \left(\rho_1^2 - \left(\rho_2^{\mathrm{R}}\right)^2 - \left(\rho_2^{\mathrm{I}}\right)^2 \right) + c_{2\beta} \, \rho_1 \rho_2^{\mathrm{R}}} \label{cp vio angle} \,.
\end{align} 

The transient CP violation between the CPV and EW minima, generates a complex-valued mass term for the top quark, $m_t(z) = |m_t(z)| \, e^{i \gamma_5 \, \theta_t(z)} $, and it is possible to relate the CP violating phase $\varphi$ to the phase of the top quark $\theta_t$. Again, working in the $\mathbb{Z}_2$ basis and the gauge where $\varphi_1 = 0$~\cite{Cline:2011mm} (recall \cref{z2 phases}),
\begin{align}
&m_t = \frac{y_t}{\sqrt{2}\, s_{\beta}} \, \chi_2  \quad , \,\,\,\,
\theta_t^{\prime} = \frac{\chi_1^2}{\chi_1^2 + \left|\chi_2\right|^2} \, \varphi^{\prime}  \,\,\, \Rightarrow \,\,\,\Delta \theta_t \approx \frac{1}{1 + t_{\beta}^2} \, \Delta \varphi \label{top quark phase} \,,
\end{align}
where the prime represents the derivative in the $z$-direction. Interestingly, we see that the change in phase of the top quark is strongly suppressed with $t_{\beta}$ and therefore, lower values of $t_{\beta}$ will benefit baryogenesis. Only derivatives of the angular profiles enter the transport equations and thus~\cref{top quark phase} is sufficient.

\subsection{Bubble wall velocity}
\label{sec:Vwall}
The remaining required input to the transport equations is the bubble wall velocity $v_w$. In order to compute it, one in principle needs to solve a coupled set of non-linear EOMs for the order-parameters of the phase transition (the space-time dependent vevs of the various scalar fields) and Boltzmann equations for particle species that are coupled to these order-parameters, including both equilibrium and out-of-equilibrium contributions. 
There exist different methods (of varying accuracy) to solve this unwieldy system of integro-differential equations   
\cite{Espinosa:2010hh, Konstandin:2014zta, Dorsch:2018pat, Laurent:2020gpg, Friedlander:2020tnq, Dorsch:2021ubz, DeCurtis:2022hlx, Dorsch:2021nje, Laurent:2022jrs, DeCurtis:2023hil, Dorsch:2023tss, DeCurtis:2024hvh, Ai:2025bjw, Branchina:2025jou, Ramsey-Musolf:2025jyk}.  In this work we however, we bypass this by placing lower and upper bounds on the wall velocity using the `ballistic' and `Local Thermal Equilibrium' (LTE) approximations respectively \cite{Ai:2021kak, Ai:2023see, Ai:2024btx}. 

In the following we give a brief account of how the two bounds are calculated for a given BSM model. For a bubble in steady-state expansion, the hydrodynamics of the system is described by the plasma velocity and temperature distributions. The thermodynamics is governed by the pressure density which in our case is the negative of the effective potential. Additionally, due to the self-similar growth of these bubbles, the distributions depend on the dimensionless variable $\xi = r/t$~\cite{Steinhardt:1981ct} (see also~\cite{Espinosa:2010hh} for a detailed discussion) where $r$ is the distance from the centre of the bubble to the bubble wall and $t$ is the time since nucleation. Using conservation of energy-momentum yields two equations,
\begin{align}
v_{+} \, v_{-} = \frac{p_{+} - p_{-}}{e_{+} - e_{-}} \quad,\quad \frac{v_{+}}{v_{-}} = \frac{e_{-} + p_{+}}{e_{+} + p_{-}} \label{hydro} \,,
\end{align}
where $\pm$ denotes in-front of (CPV phase) and behind (EW phase) the bubble wall respectively, $p_{\pm} = p(T_{\pm})$ and $e_{\pm} = e(T_{\pm})$ are the pressure and energy densities derived from the effective potential evaluated in their respective minima, and $v_{\pm}$ are the plasma fluid velocities in the reference frame of the bubble wall. Therefore, we want to solve for the five quantities $\{ v_{\pm}, T_{\pm}, v_w \}$ and \cref{hydro} provides two equations, resulting in three unknowns. The specific hydrodynamic mode of expansion -- deflagration, detonation or hybrid~\cite{Espinosa:2010hh} -- of the bubbles also provides two more initial conditions (e.g. for a detonation $v_{+} = v_w$ and $T_{+} = T_n$). Hence, one is left with one unknown from the five quantities and an additional matching condition is needed.

In the LTE approximation, interactions with the bubble wall are perfectly efficient and as a result, entropy is locally conserved across the bubble wall. This results in an extra matching condition for the wall velocity
\begin{align}
\gamma_{+} T_{+} = \gamma_{-} T_{-} \label{match ub} \,,
\end{align}
where $\gamma = 1/\sqrt{1-v^2}$ is the Lorentz factor. 
The LTE value corresponds to an upper bound on $v_w$ when a stationary state for the expanding bubbles is assumed.\footnote{It might still be possible for a bubble not to reach its stationary state before the would-be LTE solution for $v_w$ has been surpassed, therefore violating the LTE criterion. See~\cite{Krajewski:2024gma} for a discussion on this possibility.} 
It may be surprising that even in LTE, there still exists friction (which seems to be an inherently out-of-equilibrium effect) on the bubble wall: this effect, due to the different temperatures at both sides of the wall, is known as `hydrodynamic obstruction' \cite{Konstandin:2010dm}. We also note that including fluctuations around classical field configurations may prevent the LTE bound from being saturated, as explored in \cite{Eriksson:2025owh}. 
For the phase transition parameters of interest regarding baryogenesis, we have observed the departure from local thermal equilibrium to be small, and hence the model-independent code snippet from~\cite{Ai:2023see} can reliably be used to derive the wall velocity upper bound, as we do in this work. In comparison to using the full effective potential, the deviations observed were at the percent level. 

In the ballistic approximation, interaction rates with the bubble wall approach zero and particles either reflect, transmit into or out of the bubble wall, and this approximation yields a lower bound on the wall velocity $v_w$. Summing up all contributions from particles that couple strongly to the bubble wall, the final matching condition for the wall velocity
in this case is
\begin{align}
\Delta \overline{V} \equiv \overline{V}_{+} - \overline{V}_{-} = \sum_i \mathcal{P}_i{(v_{\pm}, T_{\pm})} \label{match lb} \,,
\end{align}
where $\overline{V}_{\pm} = \left( V_{\mathrm{Tree}} + V_{\mathrm{CW}} + V_{\mathrm{CT}} \right)_{\pm}$ is the zero-temperature one-loop effective potential evaluated in the symmetric ($+$) and broken ($-$) phases, and $\mathcal{P}_i$ is the pressure from each relevant species in the ballistic approximation, whose expressions can be found in \cite{Ai:2025bjw}. In this work, the sum above is performed over the top quark, $W$-boson and $Z$-boson d.o.f, as they provide the dominant contributions to the friction. 
Numerically we have observed that the additional contributions from the \THDMa\ scalar sector have a negligible effect on the wall velocity lower bound. 

Finally, it should be noted that generically, steady-state solutions are not always realised when taking into account the early-time formation of cosmological bubbles \cite{Krajewski:2024gma, Krajewski:2024xuz}. However, this is beyond the scope of this analysis and we assume in the remainder of this work that the upper and lower wall velocity bounds are physically realised. 

\section{Results}
\label{sec:results}
\begin{figure}[h]
\centering
\includegraphics[scale=0.53]{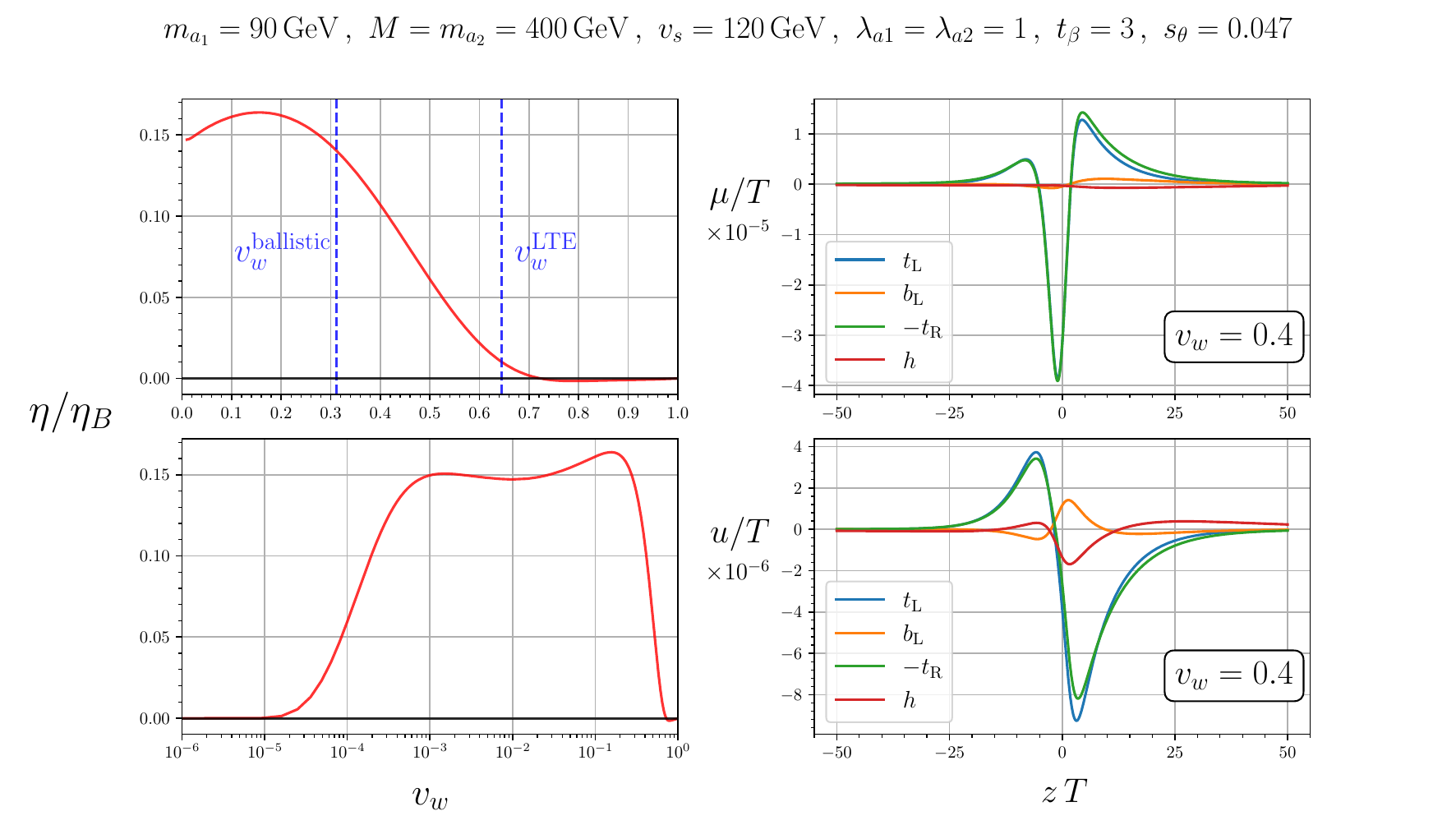}
\caption{\small Left panels: The normalised BAU as function of wall velocity for the same benchmark point in \cref{fig:crit_temp_figure}. The lower (ballistic) and upper (LTE) bounds for the wall velocity are indicated by dashed blue lines. Right panels: The variation of the chemical potential $\mu$ and plasma velocity $u$ of the relevant species with respect to the distance from the planar bubble wall ($z=0$) at a wall velocity $v_w = 0.4$. In our convention, $z<0$ is the EW phase and $z>0$ is the CPV phase.}
\label{fig:baryo curve}
\end{figure}

Using the analysis strategy described in the previous sections, we are ready to generate bounds on the potentially achievable baryogenesis across the \THDMa\ parameter space. The outline of our procedure is listed below:
\begin{enumerate}
\item Choosing a \THDMa\ parameter point, we first check tree-level perturbative unitarity, as well as boundedness from below and absolute stability
of the EW minimum of the zero-temperature one-loop effective potential. 
\item For the points which satisfy the previous constraints, we use the finite-temperature effective potential $V_{\mathrm{eff}}$, as defined in~\cref{Veff_total}, to determine the corresponding thermal history. 
We demand symmetry restoration at high-temperatures, and the existence of a (two-step) first-order EWPT as the temperature decreases, for which we compute the critical temperature.
\item We check whether the phase transition completes via bubble nucleation, and in such case evaluate the runaway criterion following~\cite{Bodeker:2009qy} (to determine if the expanding bubbles reach a stationary state).\footnote{We note that even if the bubble walls runaway (keep accelerating) according to the criterium from~\cite{Bodeker:2009qy}, it was later found by the same authors~\cite{Bodeker:2017cim} that higher-order contributions to the bubble wall friction in the ultrarelativistic limit will drive the bubbles to reach a stationary state, albeit with $\gamma_w = 1/\sqrt{1-v_w^2} \gg 1$. Being ultrarelativistic, the expanding bubbles cannot give rise to baryogenesis in such a scenario.}

\item If the bubble wall does not runaway, we determine the upper and lower wall velocity bounds as discussed in~\cref{sec:Vwall}. Specifically, we use the model-independent approach from~\cite{Ai:2023see} to obtain the upper bound $v_w^{\mathrm{LTE}}$ and the full \THDMa\ effective-potential $V_{\mathrm{eff}}$ to obtain the lower bound $v_w^{\mathrm{ballistic}}$.
\item We evaluate the transport equations as a function of $v_w$ in the range $[v_w^{\mathrm{ballistic}} ,\, v_w^{\mathrm{LTE}} ]$ to build an interpolation of the resulting BAU. 
\end{enumerate}
As an example, we follow the above procedure for the same \THDMa\ parameter point used in~\cref{fig:crit_temp_figure,fig:nucl_profiles}
%given by $m_{a_1} = 90$ GeV, $M = m_{a_2} = 400$ GeV, $v_s = 120$ GeV, $\lambda_{a1} = \lambda_{a2} = 1$, $t_{\beta} = 3$, $s_{\theta} = -0.047$ 
, that fulfills all of the required criteria.
We show in~\cref{fig:baryo curve} the dependence of the BAU on $v_w$ (left panels), as well as the variation across the bubble wall of the chemical potential (top-right panel) and plasma velocity (bottom-right panel) for various relevant particle species.
The top-left panel of~\cref{fig:baryo curve} 
additionally shows the wall velocity upper and lower bounds in this case. The lower bound is sizeable, $v_w^{\mathrm{ballistic}} \approx 0.3$, and the upper bound, $v_w^{\mathrm{LTE}} \approx 0.64$, is slightly larger than the plasma speed of sound.

As compared to the results in~\cite{Huber:2022ndk}, which found a very similar \THDMa\ benchmark point to yield the observed BAU $\eta_{B}$,~\cref{fig:baryo curve} shows that 
the maximum BAU realisable for this benchmark is $\eta_{\mathrm{max}} \approx 0.14 \,\eta_{\smash{B}}$. 
This showcases the impact of our improved computation regarding the use of $V_{\mathrm{eff}}$ and the transport equations with respect to the analysis of \cite{Huber:2022ndk}, an issue we will discuss in more detail in the following sections.  
Furthermore, the minimum realisable BAU in this case is $\eta_{\mathrm{min}} \approx 0.01 \, \eta_{\smash{B}}$, which yields a very weak lower bound on the BAU. We will see that these conclusions persist in the next section.

\subsection{Analytical considerations and baryogenesis estimates}
\label{sec:analyticalBAU}
The semi-analytical estimate for the baryon asymmetry in the \THDMa\ used in \cite{Huber:2022ndk} is
\begin{align}
\overline{\eta}_{\mathrm{est}} \equiv \eta_{\mathrm{est}} / \eta_{\smash{B}} = \frac{C}{8.7} \times \frac{\sin{({\Delta \theta_{\mathrm{t}}})} \,\, \xi_c^2}{L_w T_c} \label{bau estimate}  \,,
\end{align}
with $C \approx 600$. This value of $C$ was fitted using BAU data from various %FOPT scenarios in the 
2HDM scenarios following the analysis in~\cite{Fromme:2006wx}, and was assumed to remain relatively constant across parameter space. Besides, this estimate relies on  %but, unlike this paper, the 
an old version of the transport equations~\cite{Fromme:2006wx} to calculate %were used in calculating 
the BAU after deriving bubble nucleation profiles within the 2HDM.
%\KM{a few words on the assumptions used in deriving this approx so its clear what we are improving.} 
%\TG{done but double-check with Stephan}
Despite these shortcomings (which we will address in detail in this work), %Nevertheless, 
\cref{bau estimate} allows to have an idea of where one might achieve successful %expect to see suitable 
baryogenesis in the parameter space of the model. In this section we will tease out the dependence of the key parameters, $\{ \Delta \theta_\mathrm{t} \, , \, \xi_c \, , \, L_w T_c\}$, with respect to the pseudoscalar mass $m_{a_1}$ and mixing angle $s_{\theta}$. According to \cref{bau estimate}, successful baryogenesis is favoured for %we would like to find 
strong phase transitions with thin bubble walls, accompanied by a large amount of transient CP violation, encoded in ${\Delta \theta_{\mathrm{t}}}$. We also use the values of these quantities at the critical temperature %critical quantities 
as a proxy for how the corresponding quantities evaluated at the nucleation temperature
%nucleation quantities 
change in parameter space. This is reasonable as we do not find %nor expect 
strong supercooling in the \THDMa. 
%\sout{Furthermore, we will investigate the efficacy of this estimate when compared to a more complete treatment via the transport equations.}

As already shown in~\cref{fig:crit_temp_figure}, the primary fields for determining the \THDMa\ vacuum structure in the Higgs basis are $\rho_1$ and $a$. Therefore, the potential can be well-approximated by a reduced potential (with $\rho_2 = 0$) within the Hartree approximation,
\begin{align}
V^{\mathrm{red}}(\rho_1, %\, \rho_2=0, 
\, \, a) = \frac{1}{2} \left( Y_1 + \Pi_{11} \, T^2 \right) \rho_1^2 + \frac{Z_1}{8} \rho_1^4 + \frac{1}{2} \left( \mu_a^2 + \Pi_{33}\,  T^2 \right) a^2 + \frac{\lambda_a}{4} a^4 + \frac{\lambda_{a H_1}}{4} \rho_1^2 \, a^2
\,,\label{reduced pot}
\end{align}
where the coefficients $\Pi$ are given in~\cref{debye} and we note here that $\lambda_{a H_1} = \lambda_{\beta}$. The simplicity of this reduced potential allows us to determine the location of the non-trivial temperature-dependent extrema and the critical temperature $T_c$ analytically,     
\begin{align}
&{\rho_1^2}(T) \, \big|_{\mathrm{EW}} = \frac{2}{Z_1} \left( \frac{1}{2} Z_1 v^2 - \Pi_{11} T^2 \right) = \left(1 + \frac{\lambda_{\beta}}{\sqrt{2Z_1 \lambda_{a}}} \right) {\rho_1^2}(T) \, \big|_{*} \label{extrema approx EW} \,\,,\\[5pt]
&{a^2}(T) \, \big|_{\mathrm{CP}} = \frac{1}{\lambda_a} \left( \lambda_a v_s^2 - \Pi_{33} T^2 \right) = \left(1 + \frac{\lambda_{\beta}}{\sqrt{2 Z_1 \lambda_{a}}} \right) {a^2}(T) \, \big|_{*} \label{extrema approx CP} \,\,,\\[5pt]
&{\rho_1^2}(T) \, \big|_{\mathrm{EW}} = \sqrt{\frac{2 \lambda_a}{Z_1}} \, {a^2}(T) \, \big|_{\mathrm{CP}} \quad , \quad {\rho_1^2}(T) \, \big|_{*} = \sqrt{\frac{2 \lambda_a}{Z_1}} \, {a^2}(T) \, \big|_{*} \label{extrema relation} \,\,,\\[8pt]  
&T_c^2 = \frac{\sqrt{2 Z_1} \, \lambda_a v_s^2 - \sqrt{\lambda_a} \, Z_1 v^2}{\Pi_{33} \, \sqrt{2 Z_1} - 2 \, \Pi_{11} \, \sqrt{\lambda_a}} \,,\label{crit temp approx}
\end{align} 
where ${\rho_1}(T) \big|_{\mathrm{EW}}$ and ${a}(T) \big|_{\mathrm{CP}}$ are the local EW and CPV minima respectively and ${\rho_1}(T) \big|_{*}$ and ${a}(T) \big|_{*}$ are the local maxima, i.e., the location of the top of the barrier that separates the two minima. Curiously, it is seen from \cref{extrema relation} that all four extrema are determined by each other. We then analytically determine the value of the potential at the critical temperature, $V_{\mathrm{c}}^{\mathrm{red}}$, and consequently the barrier height, $V_b^{\mathrm{red}}$, between the EW and CPV minima,
%
%Now, we can evaluate the potential at the critical temperature and analytically determine the barrier height $V_b^{\mathrm{red}}$ between the EW and CPV minima, 
\begin{align}
&V_{\mathrm{c}}^{\mathrm{red}} = - \frac{Z_1}{8} {\rho_1^4}(T_c) \, \big|_{\mathrm{EW}} \, = \, -\frac{\lambda_a}{4} a^4(T_c) \big|_{\mathrm{CP}} \,\,,\\[6pt]
&V_{\mathrm{b}}^{\mathrm{red}} = V_{*}^{\mathrm{red}} - V_{\mathrm{c}}^{\mathrm{red}} = - \frac{\lambda_{\beta} - \sqrt{2 Z_1 \lambda_a}}{\lambda_{\beta} + \sqrt{2 Z_1 \lambda_a}} \, V_{\mathrm{c}}^{\mathrm{red}} \label{barrier approx} \,\,.
\end{align}
It should be noted that even if all fields (not only $\rho_1$ and $a$) were to be included in the computation, \cref{barrier approx} is only an approximation to the true barrier height as the MEP or bounce configuration may not pass through the local maxima. This aspect of multi-field phase transitions makes it difficult to obtain precise %predictive 
analytic estimates, %for the barrier height
as we will discuss later in this section.

Furthermore, we can find an analytic approximation for the change in the CP violating phase, $\Delta \varphi$, across the transition. For this, the reduced potential \cref{reduced pot} is however insufficient, as it is seen from \cref{cp vio angle} that this phase is strongly dependent on all field directions. We can expand the full Hartree potential around the CPV minimum, drop higher-order terms in $\chi_{1,2}$, since these tend to zero in the CPV phase, and then find the phase $\varphi_{\scriptscriptstyle\mathrm{CP}}$ that minimises the resulting potential. Working in the $\mathbb{Z}_2$ basis and non-linear parametrisation of \cref{z2 phases}, the Hartree potential simplifies to
\begin{align}
V_{\mathrm{Hartree}} &\approx - \mu_{12}^{2} \, \chi_1 \chi_2 \, c_{\varphi} - \kappa \, \chi_1 \chi_2 \, a(T) \big|_{\mathrm{CP}} \, s_{\varphi} + \{ \varphi-\mathrm{independent} \, \mathrm{terms} \} \nonumber \,, \
\\[5pt]
\Rightarrow \,\,\, t_{\varphi_\mathrm{CP}} &=\frac{\kappa}{\mu_{12}^2}\,a(T) \big|_{\mathrm{CP}} = \frac{m_{a_2}^2 - m_{a_1}^2}{M^2} \, \frac{a(T) \big|_{\mathrm{CP}}}{v} \, \frac{s_{2\theta}}{s_{2\beta}}\,.
\label{phase approx}
\end{align} 
Since in the EW minimum $a(T)|_{\mathrm{EW}}=0$ and the potential is minimised at $\varphi_{\scriptscriptstyle \mathrm{EW}}=0$\footnote{Technically $\varphi_{\scriptscriptstyle{\mathrm{EW}}} = n\pi \, , \,\, n \in \mathbb{Z}$ but we can choose $n=0$ if we restrict $\varphi_{\scriptscriptstyle{\mathrm{CP}}} \in (-\pi/2, \,\pi/2)$. Besides, any choice of $n$ makes no difference to $\Delta \varphi$.}, $\Delta\varphi\equiv\varphi_{\scriptscriptstyle \mathrm{EW}}-\varphi_{\scriptscriptstyle \mathrm{CP}}=-\varphi_{\scriptscriptstyle \mathrm{CP}}\approx -t_{\varphi_{\scriptscriptstyle \mathrm{CP}}}$,
where the last approximation  roughly holds due to boundedness from below preventing overly large magnitudes of the singlet mixing angle 
across the parameter space. Using \cref{top quark phase}, we can approximate the total change in the phase of the top quark across the transition at $T_c$, 
\begin{align}
\Delta \theta_{\mathrm{t}} \approx \frac{\Delta \varphi_c}{1+t_{\beta}^2} \approx \frac{m_{a_2}^2 - m_{a_1}^2}{2M^2} \, \frac{a(T_c) \big|_{\mathrm{CP}}}{v} \, \frac{s_{2\theta}}{t_{\beta}} \label{top phase approx} \,.
\end{align}
To demonstrate the reasonable accuracy of the above approximations, we have evaluated our analytic solutions at $T_c$ using the parameter inputs in~\cref{fig:crit_temp_figure} and compared the results with the full numerical quantities involving all fields for the Hartree and full one-loop scalar potentials. For the  Hartree and one-loop potential, the numerical barrier height, $V_b$, is taken to be the maximum value along the MEP. The results of this comparison are shown in \cref{tab:analytic_approx}.

\begin{center}
\begin{tabular}{|p{1.55cm}|p{1.6cm}|p{1.5cm}|p{1.3cm}|p{1.2cm}|p{1.0cm}|p{1.3cm}|p{1.8cm}|}
\hline
%@ $T_c$ 
& ${\rho_1}(T_c) \big|_{\mathrm{EW}}$ \newline $\scriptstyle [\mathrm{GeV}]$ & ${a}(T_c) \big|_{\mathrm{CP}}$ \newline $\scriptstyle [\mathrm{GeV}]$ & ${\rho_1}(T_c) \big|_{*}$ \newline $\scriptstyle [\mathrm{GeV}]$ & ${a}(T_c) \big|_{*}$ \newline $\scriptstyle [\mathrm{GeV}]$ & $T_c$ \newline $\scriptstyle [\mathrm{GeV}]$ & $ V_b \times 10^{6}$ \newline $\scriptstyle [\mathrm{GeV}]^4$ & $\Delta\varphi_c \times 10^{-2}$ \newline $\scriptstyle [\mathrm{rad}]$ \\[2.5pt] \hline
Analytic & $202.3 $ & $109.2 $ & $138.6 $ & $74.8 $ & $72.6 $ & $3.27 $ & $-6.58$ \\[3pt] \hline
Hartree & $202.2 $ & $109.2 $ & $139.0 $ & $75.0 $ & $72.7 $ & $2.99 $ & $-6.55 $ \\[3pt] \hline
One-loop & $214.3 $ & $117.4 $ & $144.7 $ & $81.2 $ & $102.4 $ & $4.50 $ & $-6.98 $ \\[3pt]
\hline

\end{tabular}
\captionof{table}{\small Comparison between the analytic formulas of \cref{extrema approx EW} - \cref{phase approx} and the numerical values for the Hartree and one-loop potential (including all fields) evaluated at the critical temperature with parameters displayed in~\cref{fig:crit_temp_figure}.}
%\KM{since we have a benchmark here, perhaps we could add a row for the one-loop finite temp result to help the discussion below.}
\label{tab:analytic_approx}
\end{center}
% \KM{\cref{tab:analytic_approx} shows that our analytic estimates reproduce the true properties of the finite temperature potential at the critical temperature rather well, with the CP violating phase in particular being reproduced to within 10\%.}

Using these analytic estimates, we can investigate where the largest observed BAU across the \THDMa\ parameter space in $(m_{a_1},\,s_{\theta})$ occurs, based on the key quantities $\{ \Delta \theta_\mathrm{t} \, , \, \xi_c \, , \, L_w T_c\}$. Details are left for~\cref{BAU dependence}, and we summarise the conclusions below: 
\begin{itemize}
\item For the vast majority of parameter space allowed by theoretical considerations, the strongest transitions -- those with largest $\xi_c$ -- are found for low $m_{a_1}$ and low $s_{\theta}$. 
\item The largest changes in the top quark phase across the phase transition $\Delta \theta_{\mathrm{t}}$ are found at low $m_{a_1}$ and high $s_{\theta}$. Interestingly, once the details are worked out, $\Delta \theta_{\mathrm{t}}$ scales as $t_{\beta}^{-1}$ and not $(1 + t_{\beta}^2)^{-1}$ as \cref{top quark phase} may initially suggest. Furthermore, for all other parameters fixed, $\Delta \theta_{\mathrm{t}}$ is larger for lower values of the 2HDM mass scale $M$. 
%
%\item Arguments are inconclusive with respect to the estimated wall thickness $L_w T_c$. The dependencies on $m_{a_1}$ and $s_{\theta}$ are rather complicated, and, the issue of the local maxima not always being a good approximation of the barrier height %of the true tunnelling path for multi-field transitions further complicates the issue.   
\end{itemize}
In contrast, the dependencies on $m_{a_1}$ and $s_{\theta}$ for the normalised bubble wall thickness $L_w T_c$ are rather complicated, and do not lead to a clear preference for a specific region of \THDMa\ parameter space. The aforementioned issue of the local maxima of the potential not always being an accurate approximation to the true barrier height %of the true tunnelling path for multi-field transitions
further complicates the analysis in this case.

\vspace{1mm}

\begin{figure}[h] 
\centering
\includegraphics[scale=0.53]{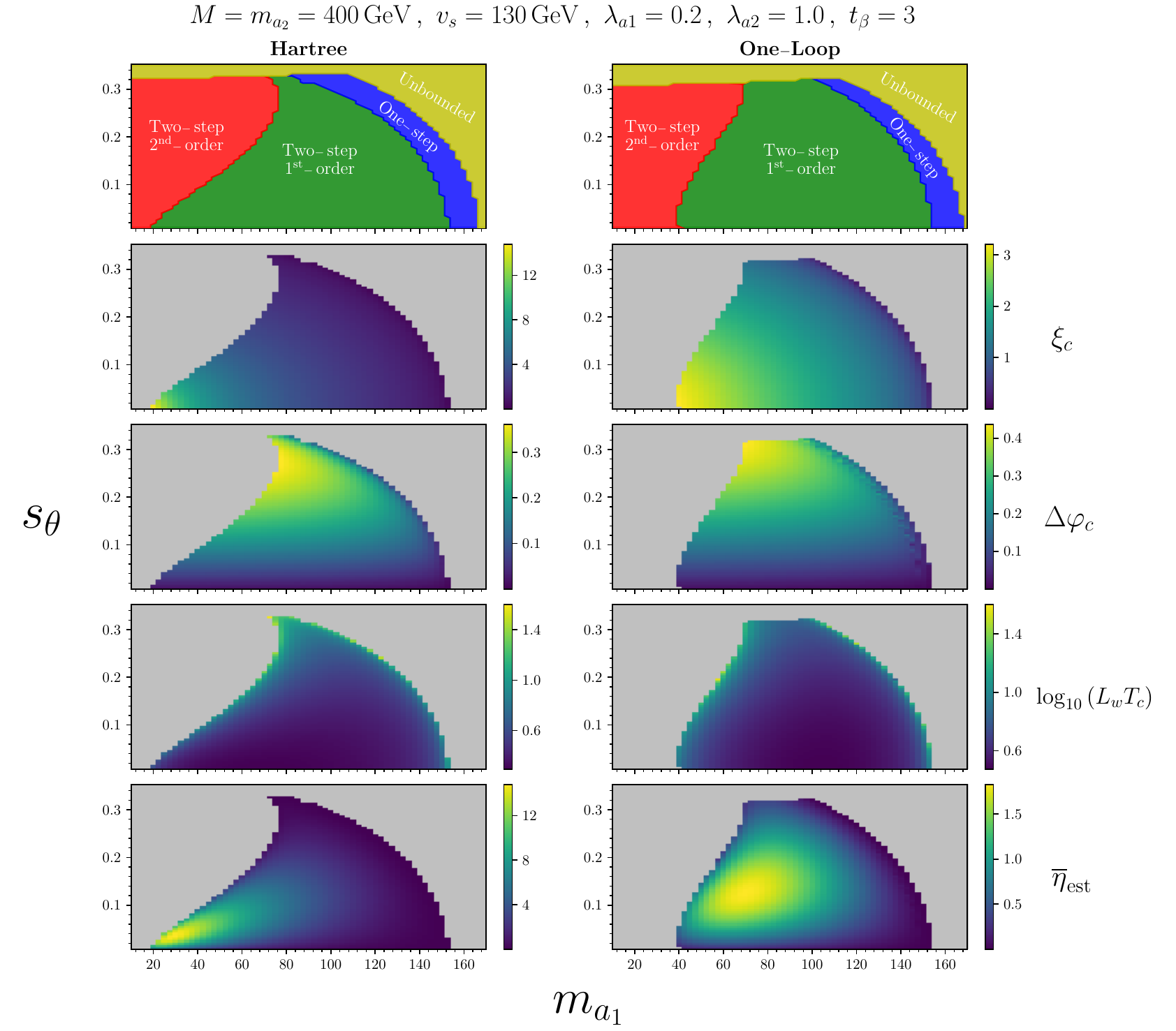}
\caption{\small A comparison of the Hartree and one-loop potential with respect to the singlet pseudoscalar mass $m_{a_1}$ and mixing angle $s_{\theta}$ for the first benchmark point in Figure 3 from \cite{Huber:2022ndk}. The plots on the first row display the phase diagram and whether the transition is one-step (blue) or two-step, and, if the latter is true, whether the second transition is first (green) or second order (red). The yellow region features a potential that is unbounded from below. The second to fourth row demonstrates how key quantities in estimating the BAU vary and the final row shows how the normalised BAU estimate \cref{bau estimate} varies, with respect to $m_{a_1}$ and $s_{\theta}$ for the desired FOPTs.}
\label{fig:crit key quants}
\end{figure}

We re-stress that the above analytical treatment stems from the scalar potential in the Hartree approximation, and we would like to further ascertain how these arguments are affected when using the full one-loop potential. In~\cref{fig:crit key quants} we show such a comparison between the Hartree and full one-loop potentials in the ($m_{a_1},\, s_{\theta}$) plane, with the remaining parameters fixed to those of the benchmark shown in~\cref{fig:crit_temp_figure,fig:nucl_profiles,fig:baryo curve}.
%with the remaining parameters fixed to $M = m_{a_2} = 400$ GeV, $v_s = 130$ GeV, $\lambda_{a1} = 0.2$, $\lambda_{a2} = 1$, $t_{\beta} = 3$ (corresponding to a specific benchmark used in~\cite{Huber:2022ndk}).
The top panels identify the parameter region of interest (green region), satisfying the various theoretical constraints and predicting the desired thermal history, which features a FOPT from the CP violating to EW minimum. %: show the viable points of parameter space with respect to theoretical constraints and  in green. 
The red region yields a second-order CPV-to-EW %EW to CPV minima 
transition at the corresponding level of approximation considered. The yellow region yields a scalar potential unbounded from below, and in the blue region the EW minimum remains the global minimum until symmetry restoration happens at high temperatures, such that the desired transitional CP violation does not occur.
The remaining panels of~\cref{fig:crit key quants} show the value of the transition strength $\xi_c$ (second panel), change in CP violating phase $\Delta \varphi_c$ (third panel) and normalised bubble wall thickness (fourth panel), for the parameter region of interest. Finally, the normalised BAU from \cref{bau estimate} is shown in the bottom panels.
As already argued, %argued previously, 
the transition strength $\xi_c$ increases for both 
small $m_{a_1}$ and $s_{\theta}$. Yet, the full one-loop effective potential leads to higher values of $T_c$ whilst the vevs remain comparatively unchanged, resulting in weaker transitions, as the comparison between left and right panels in~\cref{fig:crit key quants} shows (note the difference in scale between the two panels). 
Larger values of CP violating phase favour small $m_{a_1}$ and large $s_{\theta}$, and this trend is maintained once the full one-loop scalar potential is accounted for, as seen by the similarity between the corresponding panels.
The same conclusions are also observed in \cref{tab:analytic_approx}. 

Due to the fact that the BAU estimate from~\cref{bau estimate} more heavily weights the transition strength $\xi_c$ compared to the change in the top quark phase across the transition $\Delta \theta_{\mathrm{t}}\propto\Delta\varphi_c$, one expects larger values of the BAU to still occur for lower $s_{\theta}$ (and small masses $m_{a_1}$), a trend which is seen explicitly in the bottom panels of~\cref{fig:crit key quants}. These also show the anticipated reduction in the BAU from Hartree to one-loop, and we find that this dramatic reduction by about an order of magnitude is not just particular to this choice of parameters, but rather is a generic effect observed across parameter space (more details can be found in~\cref{BAU dependence}).

%The effect of the wall-thickness is to push the maximum BAU to higher values of $m_{a_1}$ as the walls begin to get thicker the closer one gets to the FOPT boundary (fourth panel). It is difficult to comment on the wall-thickness, however, we discover numerically that the estimated wall thickness $L_w T_c$ remains similar. More relevant to the computation of the BAU, the wall-thicknesses $L_w T_n$ obtained at nucleation tend to be larger than the critical estimate $L_w T_c$ for the $\chi_1$ and $\chi_2$ profiles and smaller for the $\varphi$ profile. 

%We summarise the conclusions below with detail provide in~\cref{BAU dependence}.
%\begin{itemize}
%\item The one-loop effective potential leads to higher values of $T_c$ whilst the vevs remain comparatively unchanged, resulting in weaker transitions across parameter space. 
%\item From \cref{cp vio angle}, we observe that the change in CP violating angle, and hence the change in top-quark phase, is heavily determined by input parameters. The only non-input parameter is the pseudoscalar vev at finite temperature, and, as already stated, the changes in vev are comparatively small. Therefore, $\Delta \theta_{\mathrm{t}}$ remains similar from tree to one-loop and this agrees with what we observe numerically. The same conclusions are also observed in \cref{tab:analytic_approx}. 
%\end{itemize}

\subsection{Impact of transport equations and bubble wall velocity}

The analytical estimates of the previous section do not take into account several key aspects of the baryogenesis process, most notably the impact of the  bubble wall velocity on the generated BAU and the details of the transport network of particle asymmetries.
%bubble nucleation and the wall velocity of the expanding bubble. 
The faster the bubble walls, the less time the EW sphalerons have to wash-out the asymmetry in front of the bubble wall, ultimately leading to a reduction in the generated BAU. The analytical estimate from~\cref{bau estimate} (and previously used in \cite{Huber:2022ndk}) encodes the effect of the bubble wall velocity by introducing the efficiency factor $C \approx 600$, which is assumed to remain (roughly) constant for phase transitions of differing strengths and wall thicknesses. 
To test this last assumption in a simple setup %where the analytical estimate is and isn't a suitable approximation, 
we use a %the same 
toy model (see~\cite{Cline:2020jre} for a related analysis) consisting of the SM enlarged by a BSM singlet scalar $s$, with the presence of 
a dimension-five CP violating operator (suppressed by a cutoff scale, $\Lambda$) which yields a varying, complex mass-term for the top quark,
\begin{align}
& \mathcal{L}_{\mathrm{Yukawa}} \supset y_t \, h(z) \, \overline{t}_{\mathrm{L}} \left( 1 + i \, \frac{s(z)}{\Lambda} \right) t_{\mathrm{R}} \,,
\end{align}
with the varying Higgs and singlet field profiles across the phase transition boundary given by 
\begin{align}
& \,\, h(z) = \frac{v_h}{2} \left( 1 - \tanh{\left( \frac{z}{L_h} \right)} \right) \,\, , \,\,\, s(z) = \frac{v_s}{2} \left( 1 + \tanh{\left( \frac{z}{L_s} \right)} \right) \,.
\end{align}
These result in a varying top mass profile $m_t (z)$ and its complex phase $\theta (z)$ during the phase transition
\begin{align}
& \,\, m_t(z) = y_t \, h(z) \, \sqrt{1 + s(z)^2 / \Lambda^2} \,\, , \,\,\, \theta(z) = \arctan{\left( \frac{s(z)}{\Lambda} \right)} \,,
\end{align}
and in the following we take $L_h = L_s = L_w$,  $T_n = 100$ GeV, $\Lambda = 1$ TeV and $v_s$ such that the change in CP violating phase is $\Delta \theta = 0.1$.  Then, fixing the %For a fixed 
bubble wall velocity, we vary the wall thickness $L_w$ and transition strength $\xi_n =  v_h/T_n$, generating contours in the produced BAU which can be compared to the analytic estimate of \cref{bau estimate}. An effective constant $C_\mathrm{eff}$ is defined by taking the ratio of the numerically computed BAU and that of the analytical estimate, such that the estimate is valid when $C_\mathrm{eff} \approx C \approx 600$. Whilst the value of $C$ was derived from 2HDM EWPTs, we expect the value to remain similar in this toy model. In both models, the phase $\theta(z)$ and Higgs profile $h(z)$ are tanh-like, and there is only a minor difference in the structure of the transport equations between the two.
\begin{figure}[h] 
\centering
\includegraphics[scale=0.53]{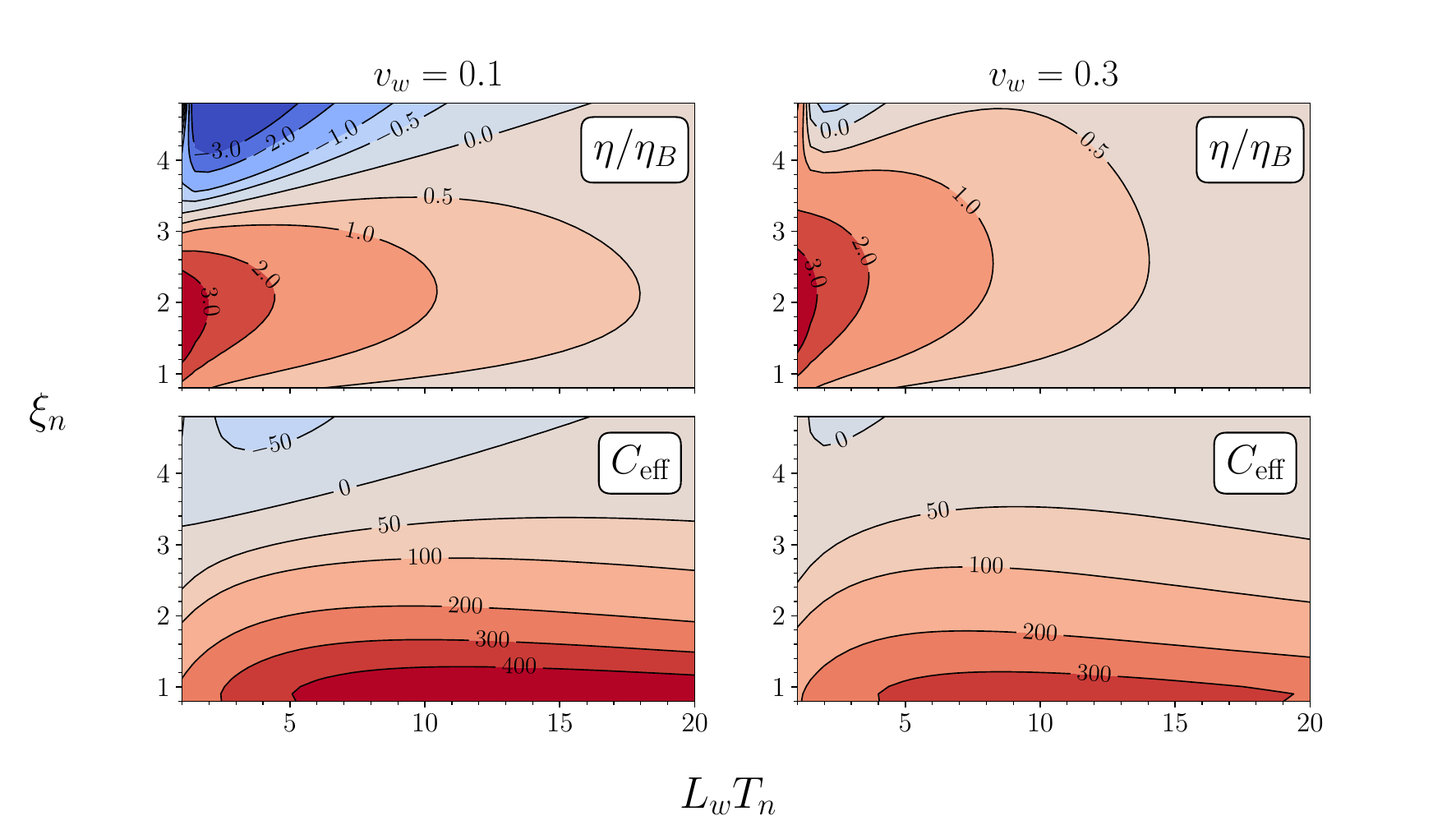}
\caption{\small Contour plots of the numerically computed normalised BAU (top panels) and $C_{\mathrm{eff}}$ (bottom panels) with respect to transition strength $\xi_n$ and wall thickness $L_w T_n$ for the toy model discussed in the text. We again note here that \cref{bau estimate} uses $C_\mathrm{eff} \approx C \approx 600$.%$C \approx 600$.
}
\label{fig:baryo efficiency}
\end{figure}

Fixing first $v_w = 0.1$ (a value commonly used in the baryogenesis literature), %Taking a common estimate for the wall velocity in the literature of , 
we observe from \cref{fig:baryo efficiency} - left that the analytic result %efficiency factor in the best case scenario 
overestimates the BAU by at least %approximately 
$25 \%$ (corresponding to $C_{\mathrm{eff}} \approx 450$), but that it is  reasonable for strengths and thicknesses $\xi \lesssim 1.75$ and $L_w T \gtrsim 2.5$ respectively. The performance of the analytic estimate then quickly degrades as one moves towards thinner walls and stronger transitions. The same behaviour is observed for different bubble wall velocities. In particular, moving to wall velocities faster than $v_w = 0.1$ further degrades the analytic estimate, e.g. for a wall velocity of $v_w = 0.3$ (\cref{fig:baryo efficiency} - right), the efficiency factor in the best-case scenario overestimates the BAU by approximately $40 \%$ (corresponding to $C_{\mathrm{eff}} \approx 350$). In addition, the analytic estimate does not capture the observed dip in the BAU at transition strengths of $\xi \approx 4$. This dip can be explained by the increase in the helicity flipping, Higgs damping and $W$-boson scattering interaction rates in the transport equations (see~\cref{transport detail} for more details), since these rates are directly proportional to the Higgs vev and hence the transition strength. Additionally,~\cref{fig:baryo efficiency} shows %there is 
a sign change in the BAU for very strong transitions.
%that may potentially suppress the BAU stemming from said transitions in the parameter spaces investigated. 
Therefore, in comparison to the analytic estimate from~\cref{bau estimate}, we expect to see a sizeable 
decrease in the BAU across the parameter space of the model, with transitions stronger than $\xi \gtrsim 2.5$ and velocities $v_w \gtrsim 0.3$ being particularly affected.

Coming back to the \THDMa\, it is also useful to determine how the bubble wall velocity upper and lower bounds vary across the  model parameter space, e.g. with respect to $m_{a_1}$ and $s_{\theta}$. Whilst this is difficult to know precisely without numerical exploration, we can deduce from~\cref{match lb} that $v_w^{\mathrm{ballistic}}$ should be proportional to the amount of supercooling, encoded in the ratio $T_n/T_c$. The larger the supercooling, the larger the zero-temperature potential difference $\Delta \overline{V}$, which results in a larger ballistic plasma pressure (since it balances the potential energy difference). For deflagrations and hybrids, the pressure increases monotonically with the wall velocity, and so larger supercooling yields a larger wall velocity lower bound. Likewise, from \cref{match ub}, the larger the supercooling, the larger the ratio $T_{+} / T_{-} = \gamma_{-} / \gamma_{+}$ becomes as more latent heat is released into the fluid in front of the bubble wall. This implies that the difference between $v_{-}$ and $v_{+}$ also increases. Since this difference grows with the bubble wall velocity for deflagrations and hybrids, $v_w^{\mathrm{LTE}}$ also increases for larger supercooling, in analogy to $v_w^{\mathrm{ballistic}}$. 

Finally, we stress that,
despite the numerous improvements made in our computation of the BAU for the \THDMa,
there remain significant sources of uncertainty in deriving this quantity, arising from different approximations made throughout our analysis. 
Most notably, we do not precisely compute the bubble wall velocity -- quite a daunting task -- but instead bound it from above and below. Besides, our computation of the BAU is based on a WKB approximation that requires $L_w\,T_n \gtrsim 2$ to not break down. Yet, this criterion is barely satisfied for some regions of our \THDMa\ parameter space, which might put into question the accuracy of the derivative expansion leading to this requirement.
Beyond these, the system of transport equations is built from finite moments of the full Boltzmann equations where various factorization schemes and \emph{ans\"atze} for the distribution functions must be made \cite{Kainulainen:2024qpm, Barni:2025ifb}. Furthermore, even solving the set of full Boltzmann equations presents its own problems due to the enhancement of particular collision integrals~\cite{vandeVis:2025plm} and poor treatment of infrared gauge boson  modes~\cite{DeCurtis:2024hvh}. Finally, it is generally assumed that EW baryogenesis occurs at a fixed wall velocity and temperature, which is not exactly what would happen in a concrete physical scenario. For all these reasons, we anticipate that our refined BAU prediction still comes with $\mathcal{O}(1)$ uncertainties. We therefore find it prudent not to be overly strict in definitively ruling out parameter points that do not appear to account for the entire BAU. Going forward, we will use the conservative criterion $\overline{\eta}_{\mathrm{max}} = \eta_{\mathrm{max}} / \eta_{\smash{B}} \geq 0.5$ to determine the \THDMa\ parameter region featuring potentially successful EW baryogenesis.\footnote{Incidentally, our refined analysis determines that EW baryogenesis is unsuccessful (\textit{i.e.}~$\overline{\eta}_{\mathrm{max}} < 0.5$) for the two benchmark points used in~\cite{Huber:2022ndk}.}
\subsection{\THDMa\ benchmarks for EW baryogenesis\label{subsec:benchmarks}}
From this point forward, we define and analyse \THDMa\ benchmark scenarios for viable EW baryogenesis, according to our previous discussion. We first perform a scan in $m_{a_1}$ and $s_\theta$ whilst leaving the remaining parameters fixed, which leads to our benchmark scenarios BP1 and BP2, defined explicitly in~\cref{tab:benchmarks ma1 stheta}.
\begin{center}
\begin{tabular}{c|c|c|c|c|c|c|c|c|c|c}
& $t_{\beta}$ & $s_{\theta}$ & $v_s$ & $M$ & $m_{a_1}$ & $m_{a_2}$ & $\lambda_{a1}$ & $\lambda_{a2}$ \\ \hline
$\mathrm{BP1}$ & $2.5$ & $[0.1, 0.555]$ & $110$ & $300$ & $[10, 210]$ & $M$ & $1.0$ & $1.5$ \\ \hline 
$\mathrm{BP2}$ & $2.0$ & $[0.1, 0.355]$ & $110$ & $600$ & $[10, 210]$ & $M+80$ & $1.4$ & $1.6$
\end{tabular}
\captionof{table}{\small Two benchmark points chosen in this paper for a dedicated baryogenesis scan. All masses are in units of $\mathrm{GeV}$ and the input parameters scanned over are the light pseudoscalar mass $m_{a_1}$ and CP-odd mixing angle $s_{\theta}$, remembering that $M = m_{H_0} = m_{H^{\pm}}$. \label{tab:benchmarks ma1 stheta}}
\end{center}
BP1 takes a relatively low (common) mass scale of the 2HDM states, $M=300$ GeV, whereas BP2 has $M=600$ GeV and introduces a splitting between the heavy pseudoscalar ($m_{a_2}=680$ GeV) and the other 2HDM heavy states. 
In~\cref{fig:bp1_velocity} we show the dependence of various physical quantities with ($m_{a_1}$, $s_\theta$) for BP1, to showcase some of the conclusions from the previous sections:
\begin{figure}[h] 
\centering
\includegraphics[scale=0.53]{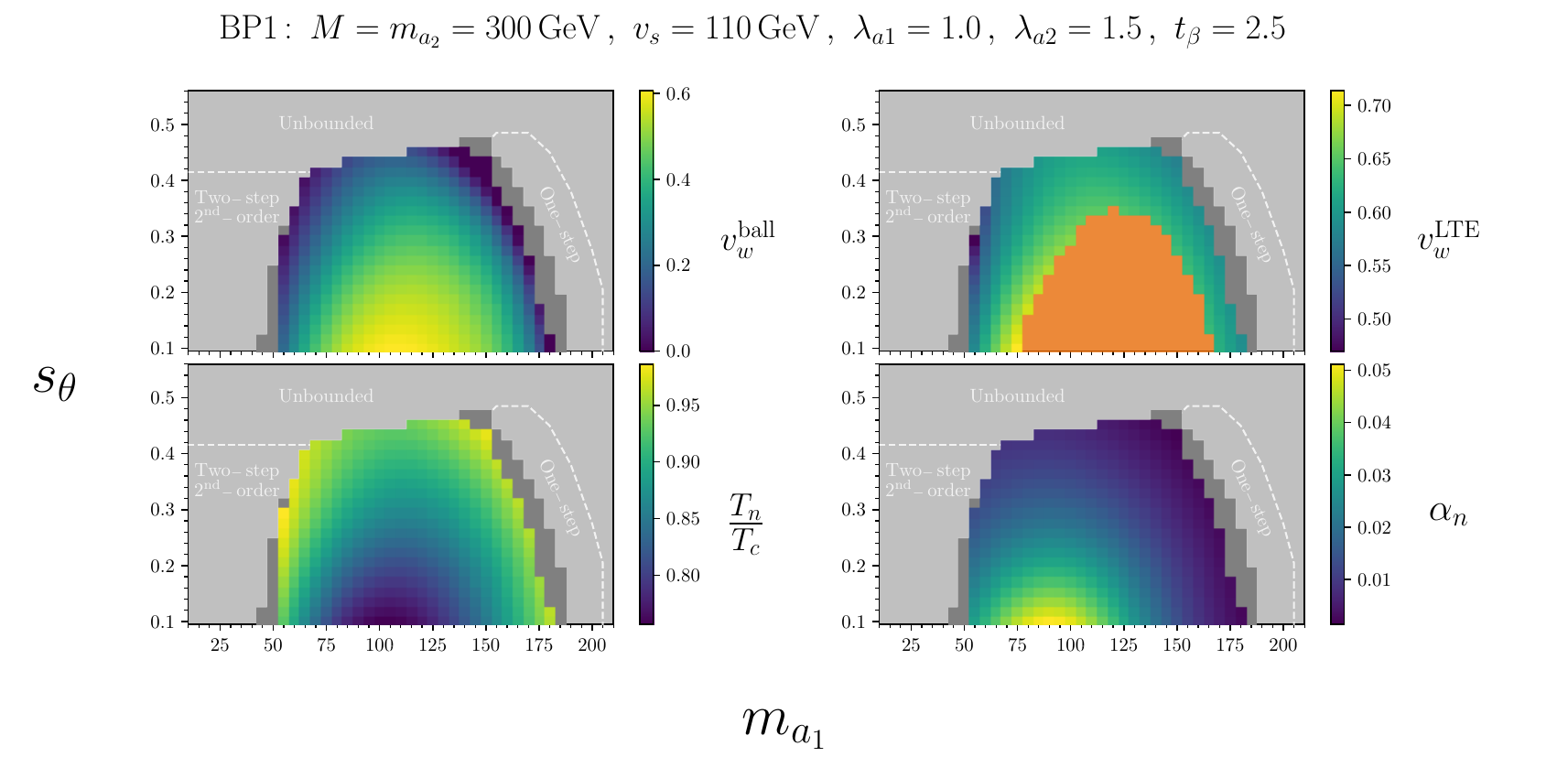}
\caption{\small Variation of the wall velocity lower bound $v_w^{\mathrm{ball}}$ and upper bound $v_w^{\mathrm{LTE}}$ (top row), supercooling $T_n/T_c$ (bottom left) and latent heat released $\alpha_n$ (bottom right) with respect to $m_{a_1}$ and $s_{\theta}$. Points are coloured light grey if the desired FOPT does not occur, dark grey if the bounce solver failed to find $T_n$, and orange if no wall velocity upper bound exists within the LTE approximation, i.e. $v_w^{\mathrm{LTE}} = 1$.}
\label{fig:bp1_velocity}
\end{figure}
%One can see the conclusions of above playing out in~\cref{fig:bp1_velocity} for a scan over $m_{a_1}$ and $s_\theta$ in BP1. 
Light-grey regions are theoretically disallowed or do not yield the desired thermal history.
The specific shortcoming which disallows these points is labelled in white and sub-regions are delineated with dashed white lines.
Dark-grey points yield the desired thermal history (with a two-step FOPT), but our bounce-solving algorithm fails to find the nucleation temperature due to numerical instabilities. For the allowed (coloured) region, the lower and upper bounds on the bubble wall velocity from the respective ballistic and LTE approximations are shown in the top row, along with the supercooling measure $T_n/T_c$ and the latent heat released in the transition  $\alpha_n$ (using the definition from \cite{Ai:2021kak}), respectively shown in the bottom-left and bottom-right panels. 

As expected, the wall velocity bounds $v_w^{\mathrm{ballistic}}$ and $v_w^{\mathrm{LTE}}$, as well as the latent heat $\alpha_n$, all increase as $T_n/T_c$ decreases (larger supercooling).
Orange points in the upper-right panel indicate that no upper bound on the bubble wall velocity was found from the LTE approximation, \textit{i.e.} $v_w^{\mathrm{LTE}} = 1$, which implies a vanishing lower bound on the BAU for such points. Whilst such points do not possess a sub-luminal wall velocity upper bound, we know that these points do not runaway as the runaway criterion has been evaluated~\cite{Moore:1997im}. This aligns with the fact that the EW phase transition is not particularly strong in this region, $\alpha_n = \mathcal{O}(10^{-2})$. Recent results from~\cite{vandeVis:2025plm} (see e.g. their Figure 2) indicate that when all relevant out-of-equilibrium species are taken into account, the bubble wall velocity for models with $\mathcal{O}(1)$ scalar couplings can be significantly lower than the predicted LTE upper bound. 
We therefore choose to retain the orange region in the final BAU analysis as baryogenesis may still be achievable there given a precise computation of the bubble wall velocity. 

Next, we use BP1 and BP2 to highlight 
how the parameter region for which the BAU is maximised gets 
shifted 
when incorporating bubble
nucleation, the wall velocity bounds and the transport equations into our analysis. 
As~\cref{fig:bp1_velocity} shows (for BP1), the majority of the small $s_{\theta}$-region features a large lower bound $v_{w}^{\mathrm{ballistic}}$ on the wall velocity, which heavily penalises successful baryogenesis. In this parameter region, increasing or decreasing $m_{a_1}$ generally does not help because either the bubble walls become too thick or the phase transition too weak for baryogenesis. Therefore, the only option left to increase the BAU is to move to higher values of $s_{\theta}$. 
Another reason that increases the mixing angle is the non-trivial dependence of the transport equations on the transition strength: as discussed in the previous section, the transport equations predict a dip in the BAU for relatively strong transitions, and the primary way to overcome this is to then increase the change in CP violating phase by decreasing $m_{a_1}$ and increasing $s_{\theta}$ (recall~\cref{fig:crit key quants}). Yet, once again we cannot go too low in $m_{a_1}$ due to the bubble walls becoming very thick. In conclusion, we expect the region of maximum BAU to be significantly shifted to larger mixing angles. 

\begin{figure}[h!] 
\centering
\includegraphics[scale=0.53]{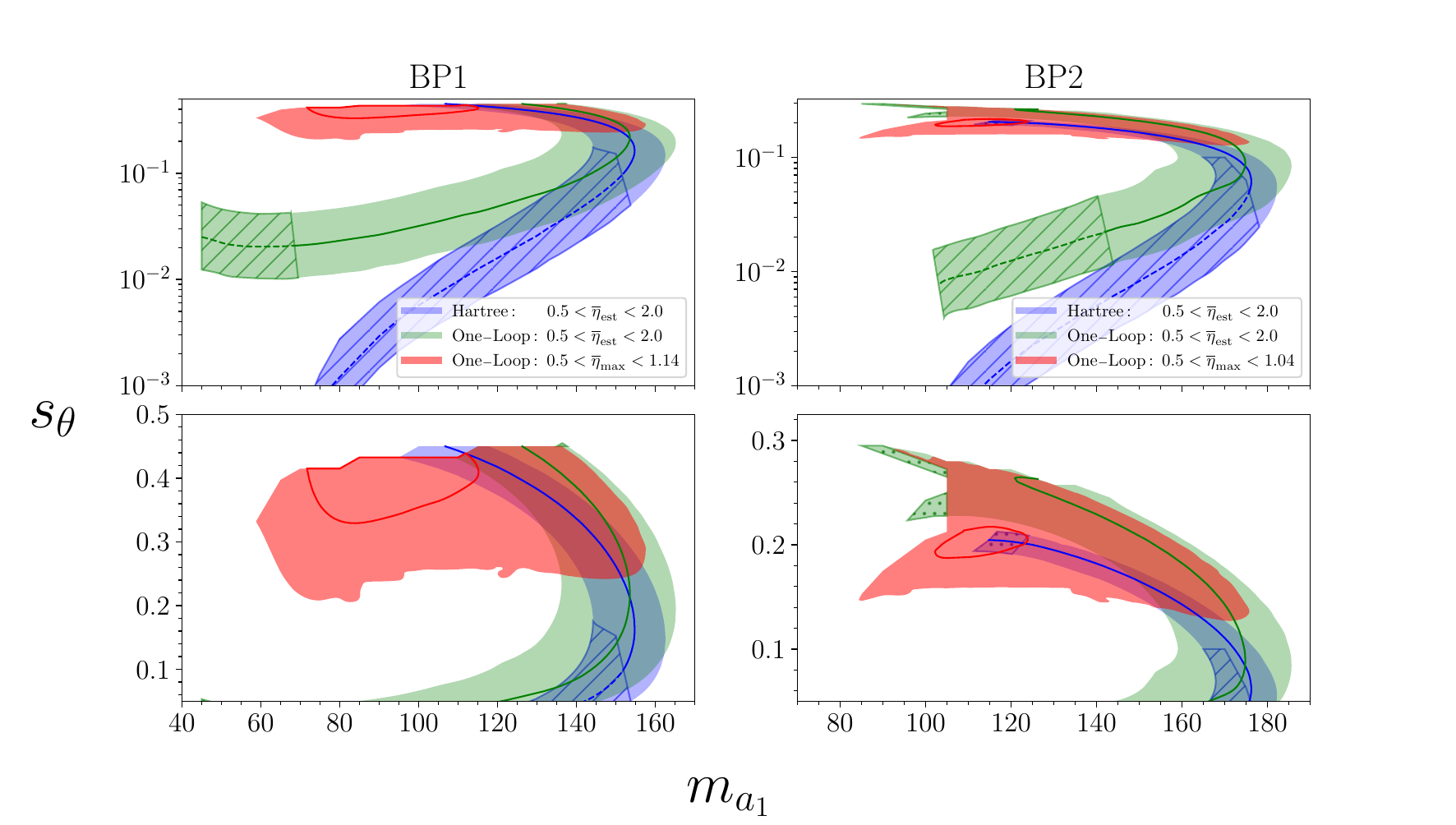}
\caption{\small Viable baryogenesis regions within three different approximations - analysis of \cite{Huber:2022ndk} (blue), analysis of \cite{Huber:2022ndk} with one-loop potential (green), and this analysis (red) - for BP1 (left) and BP2 (right) with respect to $m_{a_1}$ and $s_{\theta}$ on a log (top) and linear (bottom) plot. Solid lines denote $\overline{\eta}_{\mathrm{est/max}} = 1$. Regions are bar-hatched if they are vacuum-trapped and dot-hatched if our bounce solver failed.}
\label{fig:Approx Comp}
\end{figure}

The conclusions drawn above are confirmed in
\cref{fig:Approx Comp}, which depicts for BP1 (left) and BP2 (right) the viable BAU regions found with our full treatment including transport equations (red regions) compared to those found using the BAU estimate in~\cref{bau estimate}, with either the Hartree approximation (blue regions) or the full one-loop potential (green regions). 
The value of $s_{\theta}$
required for successful baryogenesis with our full treatment is significantly increased with respect to the regions in blue (corresponding to the previous estimate in~\cite{Huber:2022ndk}) and in green, and small mixing angles $|s_{\theta}| < 0.1$ are no longer viable for either benchmark. In contrast, we see in~\cref{fig:Approx Comp} that using the BAU estimate in~\cref{bau estimate} yields a viable region with a characteristic shape that features an upper and a lower branch of $s_{\theta}$. This is due to the existence of a region in between the two branches of significant baryon overabundance ($\overline{\eta}_{\mathrm{est}} > 2$). Interestingly, in the Hartree approximation (blue regions), mixing angles $|s_{\theta}| < 0.1$ are highly constrained by vacuum-trapping for both benchmarks BP1 and BP2. This is ameliorated when using instead the full one-loop potential (green regions), which would permit small mixing angles $|s_{\theta}| = \mathcal{O}(10^{-2})$ when using the BAU estimate from~\cref{bau estimate}.
Instead, the effect of the transport equations and bubble wall velocity dependence is to cluster the viable BAU region around
$m_{a_1} \sim 100$ GeV and $|s_{\theta}| \gtrsim 0.1$. Moreover, our observation that the maximum possible BAU in the \THDMa\ parameter space is strongly reduced in our full baryogenesis treatment is apparent in~\cref{fig:Approx Comp}, with BP1 and BP2 yielding a maximum \,$\overline{\eta}_{\mathrm{max}} = \{1.14, 1.04\}$. 

\begin{figure}[h] 
%\centering
\hspace{-5mm}\includegraphics[scale=0.58]{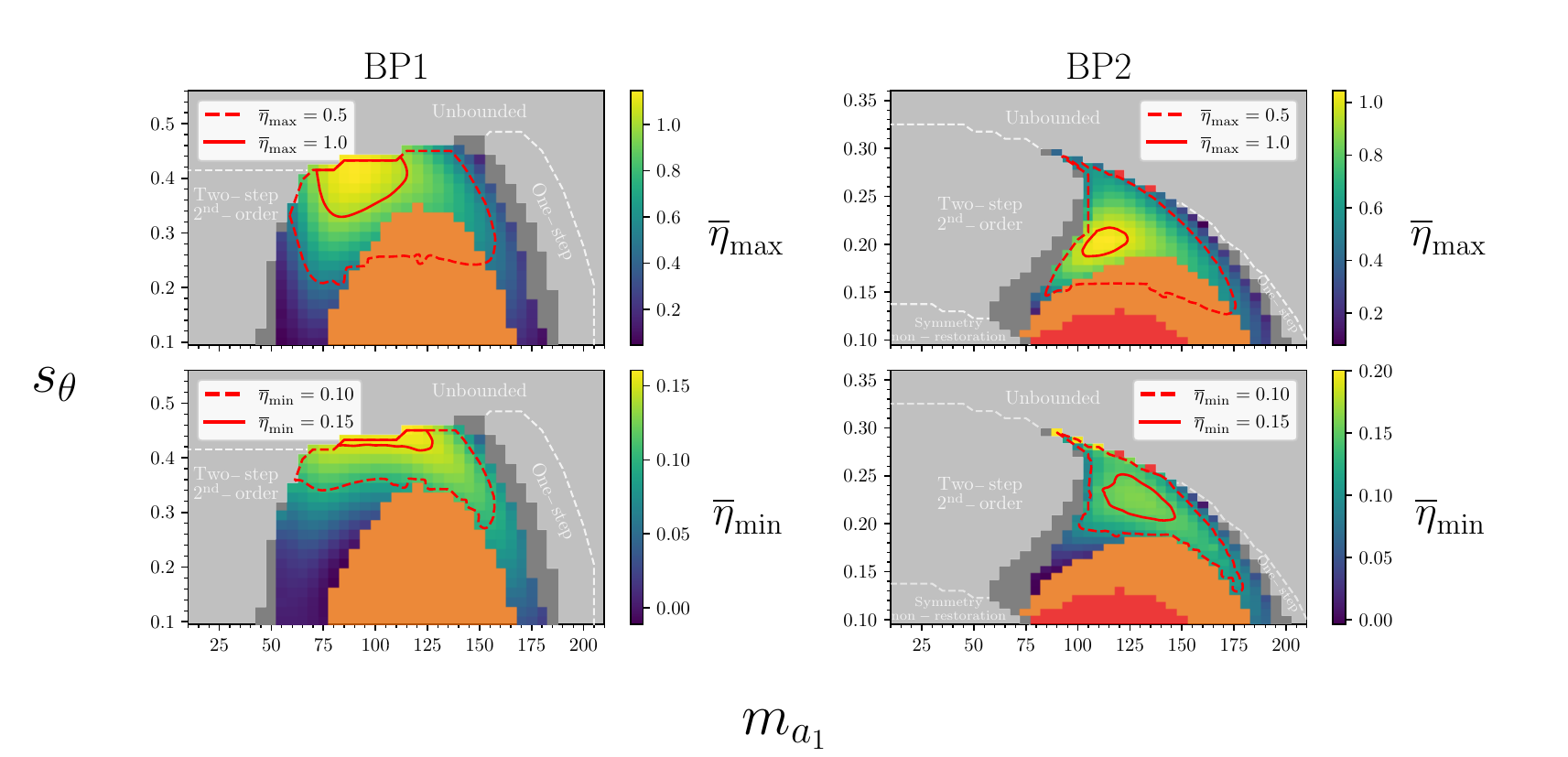}
\caption{\small Variation of $\overline{\eta}_{\mathrm{max}}$ (top) and $\overline{\eta}_{\mathrm{min}}$ (bottom) with respect to $m_{a_1}$ and $s_{\theta}$ for benchmarks BP1 (left) and BP2 (right). Points are coloured light grey if the desired FOPT does not occur, dark grey if the bounce solver failed to find $T_n$, orange if no wall velocity upper bound exists within the LTE approximation, and, red if the bubble wall runs away. Particular lines of constant $\overline{\eta}_{\mathrm{max/min}}$ are overlaid.}
\label{fig:bp1_bp2_bau}
\end{figure}

The heatmaps in \cref{fig:bp1_bp2_bau} display in greater detail how the maximum BAU \,$\overline{\eta}_{\mathrm{max}}$ and minimum BAU \,$\overline{\eta}_{\mathrm{min}}$ vary across the ($m_{a_1}$,\,$s_{\theta}$) plane for BP1 (left panels) and BP2 (right panels). 
As previously alluded to when discussing~\cref{fig:baryo curve}, it is in general difficult to obtain a sizable lower bound on the BAU, $\overline{\eta}_{\mathrm{min}}$: a ubiquitous feature of the transport equations is that the BAU crosses zero at some supersonic wall velocity $v_w^0$ before vanishing again as $v_w \rightarrow 1$. However, our upper bound $v_w^{\mathrm{LTE}}$ is often close to $v_w^0$ and hence the lower bound on the BAU is weak. Again, both BP1 and BP2 feature points where $v_w^{\mathrm{LTE}} = 1$ whilst BP2 additionally displays runaway behaviour for lower values of $s_{\theta}$. 

Besides $m_{a_1}$ and $s_{\theta}$, it is important to explore how the viable BAU regions depend on other \THDMa\ parameters. For this purpose, 
we choose a single  value of ($m_{a_1}$,\,$s_{\theta}$) for BP1 and BP2, denoted respectively as $\overline{\mathrm{BP1}}$ and $\overline{\mathrm{BP2}}$ -- compatible with relevant experimental constraints (see the discussion in \cref{experimental}) -- and scan over the 2HDM parameter $t_{\beta}$ and heavy mass scale $M$ to produce two new benchmarks BP3 and BP4, shown in \cref{tab:benchmarks m tbeta}. Scanning over these two parameters allows to analyse the interplay between the viable BAU regions and the experimental bounds from flavour physics, discussed in the next section.

\begin{center}
\begin{tabular}{c|c|c|c|c|c|c|c|c|c|c}
& $t_{\beta}$ & $s_{\theta}$ & $v_s$ & $M$ & $m_{a_1}$ & $m_{a_2}$ & $\lambda_{a1}$ & $\lambda_{a2}$ \\ \hline
$\mathrm{BP3}$ & $[1.0, 4.0]$ & $0.35$ & $110$ & $[125, 475]$ & $100$ & $M$ & $1.0$ & $1.5$ \\ \hline 
$\mathrm{BP4}$ & $[0.75, 4.0]$ & $0.16$ & $110$ & $[400, 1000]$ & $90$ & $M+80$ & $1.4$ & $1.6$
\end{tabular}
\captionof{table}{\small Two further benchmark points chosen in this paper for a dedicated baryogenesis scan. All masses are in units of $\mathrm{GeV}$ and the input parameters scanned over are the 2HDM mixing angle $t_{\beta}$ and common mass scale $M$, remembering that $M = m_{H_0} = m_{H^{\pm}}$. \label{tab:benchmarks m tbeta}}
\end{center}

\begin{figure}[h] 
%\centering
\hspace{-5mm}\includegraphics[scale=0.58]{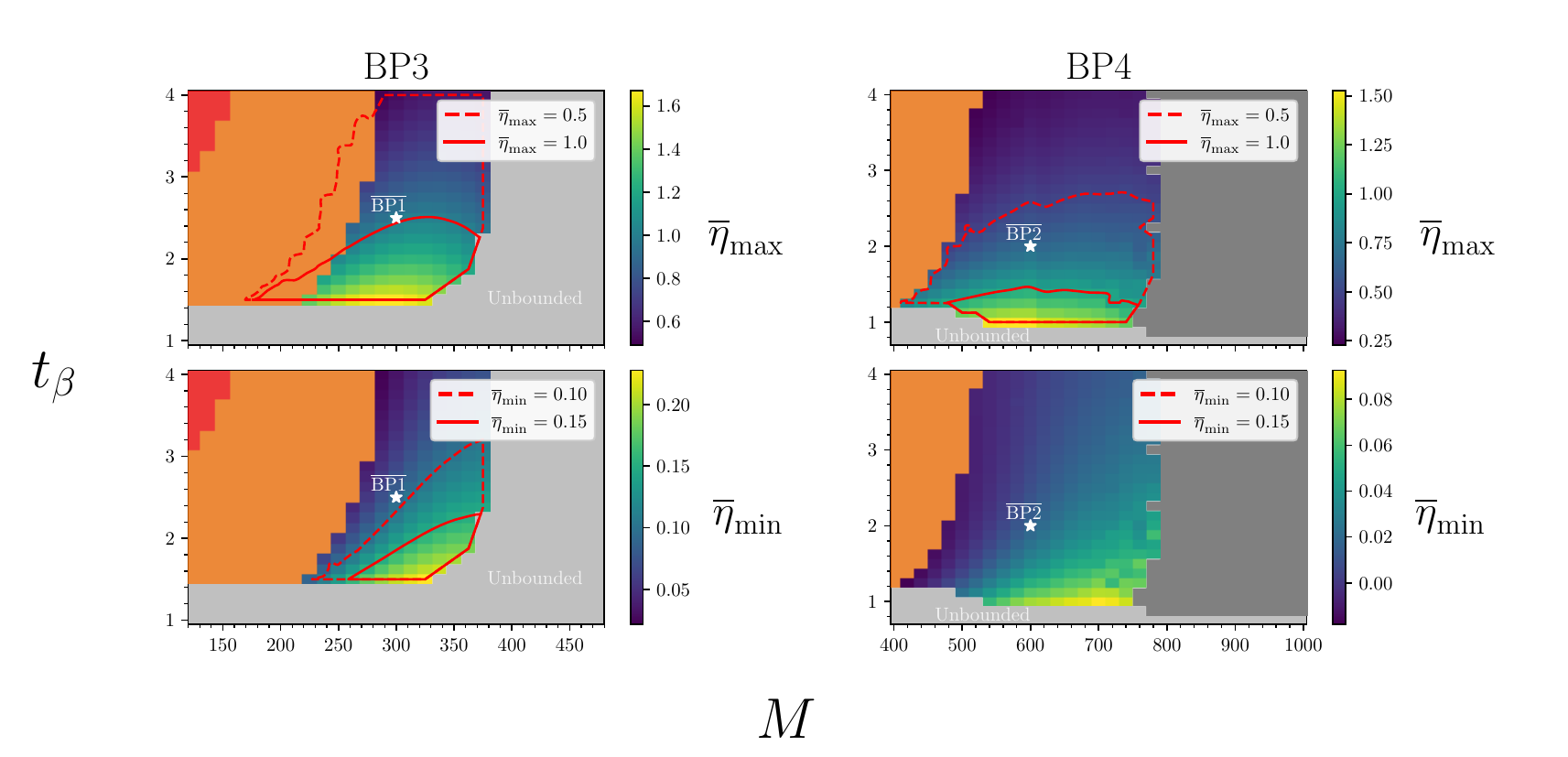}
\caption{\small Variation of $\overline{\eta}_{\mathrm{max}}$ (top) and $\overline{\eta}_{\mathrm{min}}$ (bottom) with respect to $M$ and $t_{\beta}$ for benchmarks BP3 (left) and BP4 (right). Points are coloured light grey if the desired FOPT does not occur, dark grey if the bounce solver failed to find $T_n$, orange if no wall velocity upper bound exists within the LTE approximation, and, red if the bubble wall runs away. Particular lines of constant $\overline{\eta}_{\mathrm{max/min}}$ are overlaid.}
\label{fig:bp3_bp4_bau}
\end{figure}

The viable BAU regions 
%(encoded in $\overline{\eta}_{\mathrm{max}}$ and $\overline{\eta}_{\mathrm{min}}$) 
for BP3 and BP4 are respectively shown in the left and right panels of~\cref{fig:bp3_bp4_bau} , with the $\overline{\mathrm{BP1}}$ and $\overline{\mathrm{BP2}}$ points depicted by stars in their corresponding panel. As is clear from~\cref{fig:bp3_bp4_bau}, increasing the BAU is possible by lowering $t_{\beta}$, eventually being limited by the boundedness of the scalar potential. We note that, as outlined in~\cref{sec:analyticalBAU}, the BAU is seen to feature an approximate $t_\beta^{-1}$ dependence (rather than the naive $t_\beta^{-2}$ which could be inferred from~\cref{top quark phase} for $t_{\beta} \gg 1$). The benchmarks BP3 and BP4 yield an improved maximum possible BAU, for the lowest allowed values of $t_{\beta}$, of \,$\overline{\eta}_{\mathrm{max}} = \{1.67, 1.52\}$ respectively. One striking contrast between BP3 and BP4 benchmarks is the region for which FOPTs are theoretically possible. For BP3, values of $M$ beyond 350 GeV are quickly disallowed as the potential becomes unbounded from below (light-grey region in~\cref{fig:bp3_bp4_bau}). In comparison, BP4 does not suffer from stringent boundedness-from-below constraints up to masses $M\sim 1$ TeV, thanks to the mild mass splitting $m_{a_2} > M$.  Yet, the allowed parameter region above $M\sim 800$ GeV 
is difficult to access numerically: physically, $T_n$ lies very close to the temperature at which the barrier disappears and the CP violating vacuum converts into a saddle point. This leads to poor convergence in the iterative procedure of our bounce solver, despite being sure that a FOPT will occur for these points.\footnote{This is because there exists a non-zero temperature at which the potential barrier disappears, and the EW minimum is still the global one. However, the bubble nucleation field profiles and other quantities needed as inputs for the transport equations cannot be reliably determined by our numerical method.} Having identified the preferred regions of parameter space for baryogenesis within our 2HDM$+a$ benchmark scenarios, in the next section we will confront these regions with existing experimental searches from the LHC, as well as bounds from flavour physics.

\subsection{$\mathbb{Z}_2$ symmetry and the sign of the BAU}
Before moving on we stress an important subtlety regarding successful baryogenesis in this model, already discussed in~\cite{Huber:2022ndk}. 
According to our analysis thus far, the 2HDM$+a$ scalar potential, given by \cref{eq:V_0,eq:Va} enables the generation of a BAU with an \emph{absolute value} that is consistent with observations. The sign of the generated BAU, however, is controlled by the sign of the vev of the $a$ field, $v_s$, since this is what determines the sign of the phase change across the EW phase transition, $\Delta\varphi$. 
The manifest CP symmetry of the 2HDM$+a$ scalar potential is tied to the $\mathbb{Z}_2$ symmetry under which the field $a$ is odd. When this symmetry is spontaneously broken in the early Universe to yield the desired \emph{transient} CP-violation, there is no preference for either sign of $v_s$. 
As a consequence, prior to the EWPT the Universe would comprise an equal volume of spatial patches with differing signs of $v_s$ separated by domain walls. The subsequent EWPT would then yield opposite signs of the BAU locally in these patches, with the net BAU averaging to zero over a Hubble volume. 

This seemingly daunting problem can be resolved via the introduction of a small $\mathbb{Z}_2$-breaking term for $a$ in the scalar potential, 
\emph{e.g.} a cubic term, $\mu_3\,a^3$, that leads to a lower vacuum energy for a particular sign of $v_s$. 
The vacuum energy bias $\Delta V$ introduced by such a term, of order $\sim \mu_3\, v_s^3$~\cite{Huber:2022ndk}, would cause the regions of deeper $v_s$ minima to rapidly engulf those with opposite $v_s$ sign provided that $\Delta V/T^4 \gg 10^{-16}$~\cite{McDonald:1995hp,Espinosa:2011eu}, leaving a Universe filled with only a single sign of $v_s$ and therefore a non-zero net BAU. 
Yet, since the $\mathbb{Z}_2$ and CP symmetries are linked, the presence of this biasing term would simultaneously signal explicit CP violation in the model, as discussed in detail in the Appendix of~\cite{Huber:2022ndk} (see also~\cite{Chao:2017oux}). One then may wonder whether such an explicit CP violation source could be constrained by EDM experiments. Fortunately, it has been shown in~\cite{Huber:2022ndk} that the required amount of bias to solve this problem and ensure successful BAU generation in the 2HDM$+a$ is unobservably small. Assuming that the explicit CP violation is sourced by a complex $\mu_{12}^2$ parameter (with phase $\delta_{12}$) within~\cref{eq:V_0}, which then generates a $\mu_3\,a^3$ biasing term at the one-loop level, the approximate required size of the CP violating phase to bias the Universe towards one sign of $v_s$ is $\delta_{12} >10^{-15}$~\cite{Huber:2022ndk}. This is to be compared with the current bounds on this phase from the electron EDM experimental searches~\cite{Roussy:2022cmp}, of order $\delta_{12} < 10^{-2} - 10^{-4}$ depending on 2HDM parameter choices (see e.g.~\cite{Altmannshofer:2020shb}).

\section{Experimental constraints: LHC searches \& flavour physics}
\label{experimental}
%\subsection{Impact of Experimental Constraints}
 As discussed in~\cite{Huber:2022ndk}, broadly speaking the \THDMa~has very similar phenomenological features to the usual 2HDM, with the additional pseudoscalar $a_1$ state that mixes with the usual pseudoscalar of the 2HDM inheriting all of its couplings with a suppression factor of $s_{\theta}$. Since we work in the alignment limit, the new 2HDM scalars do not couple at tree-level to a pair of SM gauge bosons, and their only gauge interactions involve couplings between a pair of new 2HDM states and a $W$ or $Z$ boson. In this limit, they can only be accessed experimentally through their couplings to SM fermions and/or cascade decays involving the aforementioned gauge interaction or their mutual self-interactions. Our investigation of the parameter space has led us to identify benchmark regions with a relatively heavy common mass scale for the (mostly-)2HDM states, $M$, and a lighter $a_1$ with a mass on the order of 100 GeV. In this region, $a_1$  predominantly decays to $b\bar{b}$, and the presence of a mass hierarchy leads to the possibility of cascade decay signatures where a heavier scalar is produced and subsequently decays to one or more $a_1$. 
 
As already outlined in~\cite{Huber:2022ndk}, the parameter space for successful baryogenesis in the \THDMa\ can be probed by a number of existing LHC searches, such as those targeting the cascade decay $H_0 \to Z a_1$ in the $a_1 \to b \bar{b}$ final state by CMS~\cite{CMS:2016xnc,CMS:2019wml} and ATLAS~\cite{ATLAS:2018oht}, as well as direct searches for scalar resonances decaying into a $\tau^+\tau^-$ pair~\cite{CMS:2018rmh,ATLAS:2020zms}.\footnote{In~\cite{Huber:2022ndk}, a potential probe of the baryogenesis parameter space (for $\lambda_{aH_1} \neq \lambda_{aH_2}$) via a
$p p \to H_0 \to a_1 a_1 \to b\bar{b} b\bar{b}$ search was considered (by using the LHC sensitivity study done in~\cite{Barducci:2019xkq} through a recast of the latest di-Higgs CMS search in the $4b$ final state~\cite{CMS:2018qmt}). We do not consider this signature here (as there is no public ATLAS or CMS search of it), yet we stress that it would provide a strong probe of the baryogenesis region.} Moreover, the region where $m_{a_1}< m_h/2$ is completely ruled out by the introduction of a unobserved decay channel for the observed, 125 GeV Higgs boson that would impact global signal strength measurements obtained by the CMS and ATLAS collaborations~\cite{ATLAS:2022vkf,CMS:2022dwd}. In the absence of any other modifications of Higgs properties these forbid an unobserved branching fraction below about 5\%. In addition, new in this work, we also find that a recent CMS search for the production of a heavy BSM scalar decaying into another BSM scalar and the 125 GeV Higgs boson yielding a 4 $b$-quark final state~\cite{CMS:2024bds} constitutes a very strong probe of the baryogenesis-favoured parameter space of the \THDMa, that predicts significant signals via the $p p \to a_2 \to h \, a_1 \to b \bar{b} b \bar{b}$ channel.

One of the main effects of our more accurate computation of the BAU in this model is the fact that this quantity is more suppressed than originally anticipated. This leads to a need for larger mixing angles between the singlet and the heavy 2HDM pseudoscalars.
One then expects the viable parameter region of this scenario to be more experimentally accessible. Indeed, our present study shows that these searches are more sensitive to the baryogenesis-favoured region of the \THDMa\ than originally discussed in~\cite{Huber:2022ndk}, since the BAU region shifts to larger values of $s_{\theta}$ when one includes the effects discussed in this work, with respect to the naive Hartree approximation, recall \cref{fig:Approx Comp}. Crucially, the region with very small $s_\theta$ identified in our previous work which has the most experimentally challenging to rule out is now deemed incapable of reproducing the BAU.

%In Ref.~\cite{Huber:2022ndk}, we identified several collider searches that probe the interesting regions of this model. 

\begin{figure}[h] 
\centering
\includegraphics[scale=0.57]{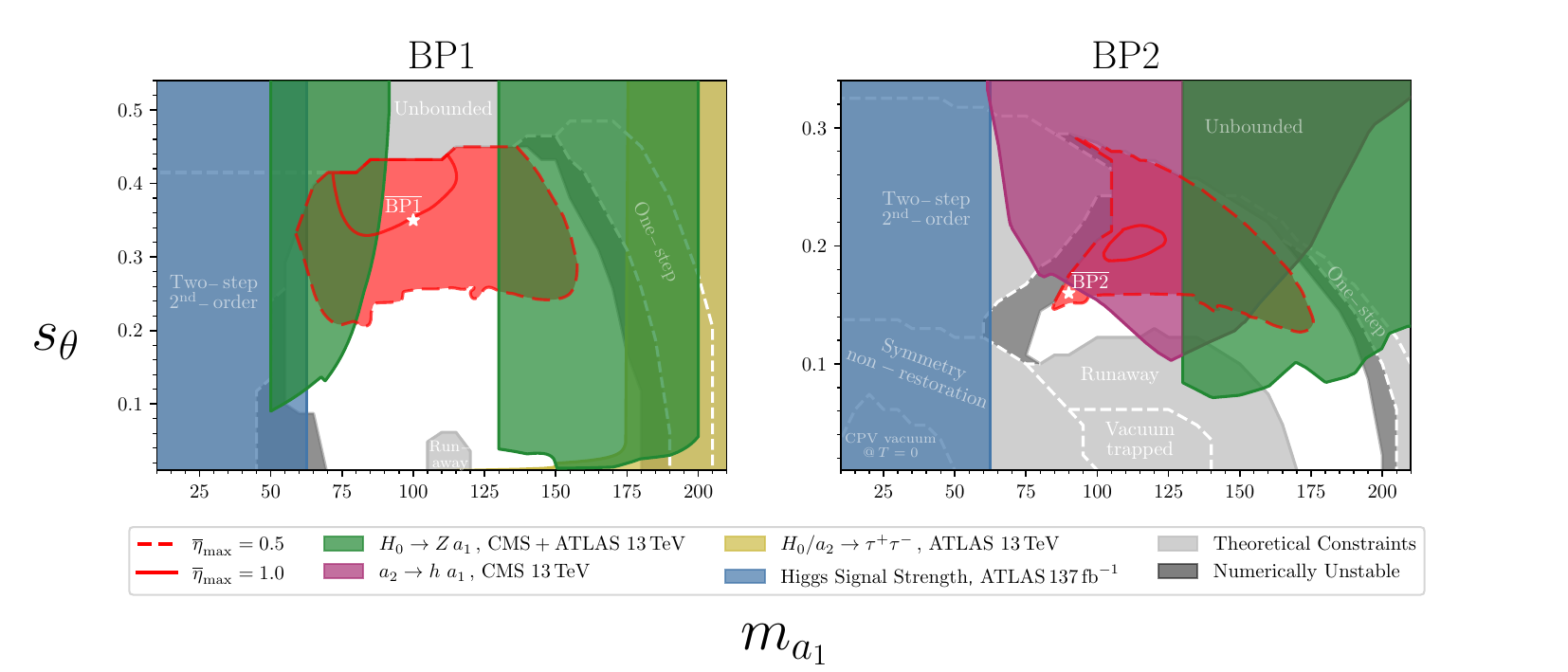}\\
\includegraphics[scale=0.57]{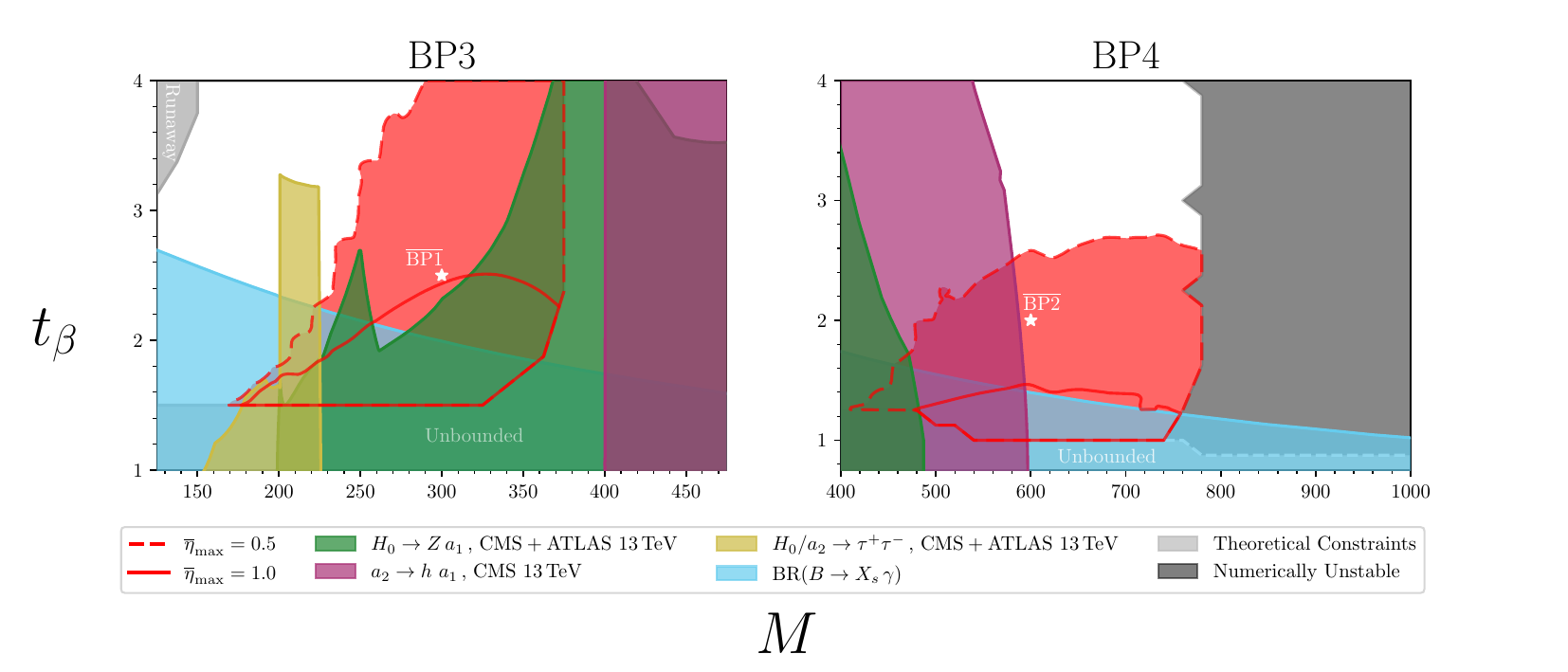}
\caption{\small In red, the region of viable baryogenesis ($\overline{\eta}_{\mathrm{max}} \geq 0.5$) for BP1 (top left) and BP2 (top right) with respect to $m_{a_1}$ and $s_{\theta}$, and for BP3 (bottom left) and BP4 (bottom right) with respect to $M$ and $t_{\beta}$. Overlaid in blue, green, orange and purple are experimentally excluded regions at the LHC. Grey regions are theoretically excluded and are delineated by dashed white lines into labelled subregions.}
\label{fig:bp1_bp2_theory_experiment}
\end{figure}

% \begin{figure}[h] 
% \centering
% \includegraphics[scale=0.5]{bp3_bp4_theory_experiment_v2.pdf}
% \caption{In red, the region of viable baryogenesis ($\overline{\eta}_{\mathrm{max}} \geq 0.5$) for BP3 (left) and BP4 (right) with respect to $m_{a_1}$ and $s_{\theta}$. Overlaid in green and purple are experimentally excluded regions at the LHC. Grey regions are theoretically excluded and are delineated by dashed white lines into labelled subregions.}
% \label{fig:bp3_bp4_theory_experiment}
% \end{figure}

At the same time, the viable baryogenesis parameter space is also constrained by flavour observables, particularly from rare $B$-meson decays: we take into account existing constraints from $b \to s \gamma$ transitions (the world average of the branching ratio $\mathrm{BR}(B\to X_s\gamma)$~\cite{HFLAV:2022esi}), that set a strong lower bound on the charged Higgs mass above 500-600 GeV
for the
Type-II 2HDM (the specific value of this limit varies as new theoretical and experimental results appear~\cite{Hermann:2012fc,HFLAV:2019otj,Atkinson:2021eox}), and also bound $t_\beta$ from below in Type-I. 
The existence of a light pseudoscalar state $a_1$ coupling to SM fermions could also be probed by its contributions to the decay $B_s\to\mu^+\mu^-$~\cite{Skiba:1992mg,Logan:2000iv}, but only for Type-II 2HDM and $t_\beta \gg 1$, which would however suppress the generated BAU. For the $a_1$ masses considered in our analyses we have verified, by interpreting the results of~\cite{Dolan:2014ska}, that the $a_1/a_2$ contributions to these decays are negligible compared to that of the charged Higgs, such that the constraints from flavour physics in this model reduce to those of the usual 2HDM. 

\Cref{fig:bp1_bp2_theory_experiment} displays the interplay between our improved BAU predictions and existing experimental constraints for our benchmarks BP1 and BP2 in the $(m_{a_1}, s_\theta)$ plane (upper row) and the related benchmarks BP3 and BP4 in the $(M,t_\beta)$ plane. The viable regions reproduced from \cref{fig:bp1_bp2_bau,fig:bp3_bp4_bau} are shaded in red with the dashed and solid contours denoting $\overline{\eta}_{\mathrm{max}}=0.5$ and 1, respectively. Grey regions indicate theoretically disallowed regions, or those with numerical instabilities, as discussed in the previous section. The remaining shaded areas indicate various experimental bounds, with those from $H_0\to Za_1\to \ell^+\ell^-b\bar{b}$ searches shaded in green, direct searches for 2HDM scalars decaying to $\tau^+\tau^-$ in yellow, $b\to s\gamma$ transitions for the Type-I \THDMa\ in pale blue, 125 GeV Higgs signal strengths in dark blue and the new $a_2\to ha_1\to4b$ search in purple.

It is clear that the experimental bounds very strongly constrain the possibility of our benchmark scenarios generating the observed BAU. We begin by highlighting the key phenomenological features of our benchmark selections -- which assume a hierarchy between the common mass scale $M$ and $m_{a_1}\!\sim m_h$ -- before moving to a discussion of the specific benchmark points. Firstly, any viable region with $m_{a_1}<m_h/2$ is ruled out by Higgs signal strength data, as the coupling to the Higgs boson, $\lambda_\beta$, plays a key role in yielding the correct thermal history and cannot be dialled down to avoid the Higgs boson decaying to a pair of $a_1$. Secondly, as previously mentioned, the experimental sensitivity of a given parameter point in our benchmark analysis is driven by two key 2HDM couplings: the $H_0$-$Z$-$a_1$ gauge interaction and the $a_2$-$h$-$a_1$ scalar self-interaction. The former originates from the gauge covariant derivative of the Higgs doublets which is subsequently passed down through $a_1$-$a_2$ mixing and is crucially not suppressed in the alignment limit. Given our choice of mass hierarchy, the $H_0\to Z a_1$ decay is the predominant kinematically accessible channel mediated by this coupling in the absence of a mass splitting between $a_2$ and $H_0$ greater than $m_Z$. In this case the $a_2\to H_0Z$ or $H_0\to a_2 Z$ channels would open, depending on the mass ordering, and these could also be constrained using the same LHC searches. 
The latter coupling comes from the extended scalar potential and is proportional to the following combination of input parameters:
\begin{align}
    \lambda_{a_2\,h\,a_1}\propto s_\theta c_\theta(
2M^2-m_{a_2}^2-m_{a_1}^2-m_h^2+2\lambda_\beta v^2
    ).
\end{align}
Once again, the presence of $\lambda_\beta$ (which must be positive) means that this coupling is expected to be present in our scenario. Moreover, the large common mass scale also ensures that this coupling is large in our benchmarks. Consequently the $a_2\to h a_1$ channel quickly comes to dominate the $a_2$ width once an appreciable $s_\theta$ is introduced, which explains the strong sensitivity of the $a_2\to h a_1 \to 4b$ search, when applicable.

In BP1, where the common mass scale for the heavy states is set to $M=300$ GeV, we see that a large part of the interesting parameter space is ruled out by the $H_0\to Z a_1$ searches. The gap in sensitivity in the $m_{a_1}$ region between 80 and 130 GeV is likely a combination of a fluctuation in the CMS search and the fact that the ATLAS search does not go below 130 GeV in the mass of the daughter scalar particle. Although the fact that this mass region includes 125 GeV may somewhat diminish the sensitivity due to the appearance of new background processes involving the SM-like Higgs boson, it is likely that future searches in this channel could comprehensively exclude this benchmark point. We also note that $H_0/a_2\to\tau^+\tau^-$ becomes sensitive to this region of larger $m_{a_1}$. Being degenerate in mass in BP1, both $H$ and $a_2$ contribute to the signal and once the cascade decay channels become kinematically inaccessible ($a_2\to h a_1$ at $m_{a_1}= $175 GeV and $H_0\to Za_1$ at $m_{a_1} =$ 209 GeV), this channel yields the dominant constraint, although these masses lie outside of the viable region of parameter space for the specific coupling choices of BP1. Should the common mass scale go beyond $2m_t$,  $H_0/a_2\to t\bar{t}$ would become the best probes of the model. We note that the $a_2\to h a_1$ search only considered parent particle masses down to 400 GeV, meaning that despite its likely strong sensitivity, it is not applicable to BP1.

In BP2, a larger common mass scale of $M=600$ GeV is chosen along with a mild splitting of $m_{a_2} = M+80$ GeV.  A combination of $a_2\to ha_1$ and $H_0\to Za_1$ 
searches rule out the vast majority of the BAU-generating parameter space with only a small island around $m_{a_1}\simeq90$ GeV and $s_{\theta}\simeq0.16$ remaining viable. Again, going below $m_{a_1} = 130$ GeV in the $H_0\to Za_1$ search would significantly improve its constraining power, but the $a_2\to ha_1$ search is able to exclude the majority of the missing parameter space. In fact, the 80 GeV mass splitting was introduced to slightly suppress the $a_2\to ha_1$ partial width, in order to not completely rule out the viable parameter space. Mass splittings greater than $m_W$ snd $m_Z$ would open up the $a_2\to H^\pm W^\pm$ and $a_2\to H_0 Z$ decay channels, respectively, offering new search prospects for this model.

It is now clear how BP3 and BP4 defined  in~\cref{subsec:benchmarks} are derived by taking an experimentally viable $(m_{a_1},s_\theta)$ point from BP1 and BP2, respectively and varying $M$ and $t_\beta$ to explore this complementary parameter plane, in which the impact of flavour constraints in particular can be seen. In both cases, we see that the $b\to s\gamma$ constraint assuming Type-I couplings serves to rule out the low-$t_\beta$ region where the BAU is most enhanced, given our determination that this quantity scales approximately as $t_\beta^{-1}$.

In BP3, where $m_{a_1}$ has been fixed to 100 GeV, the Type-I flavour constraints rule out $M \lesssim 220$ GeV whilst the $H_0\to Z a_1$ significantly constrains the rest of the parameter plane above $M = 200$ GeV (where that search starts) until the potential becomes unbounded around $M \sim 370$ GeV. In Type-II, such a low common mass scale is ruled out by the $t_\beta$-independent lower bound on the charged Higgs mass. One can also clearly see the strength of the $a_2\to h a_1$ search starting at $M = 400$ GeV, suggesting that it would likely provide significant sensitivity to the viable region if lower masses we considered in future analyses. Finally, $H_0/a_2\to \tau^+\tau^-$ searches are also  sensitive to this region of parameter space, although the most sensitive ATLAS search exploiting the full Run 2 dataset also only begins at a resonance mass of 200 GeV.

Finally, BP4 allows for larger $M$, which is only limited by the numerical instability of our code discussed in~\cref{subsec:benchmarks}. We can see that both cascade decay searches die-off at higher masses due to the PDF suppression of the signal and $t_\beta$ values above 1.5 remain viable with respect to the Type-I flavour constraints. In the case of Type-II, the lower bound from flavour constraints of $m_{H^\pm}=M\lesssim540$ GeV would not rule out any further regions of parameter space but is likely to do so in the future as precision improves.

\section{Summary and conclusions}
\label{sec:conclusions}
In this paper, we have performed a comprehensive investigation of baryogenesis within the 2HDM$+a$ model and explored the interplay of theoretical and experimental constraints across parameter space, providing a thorough update of results obtained in \cite{Huber:2022ndk}. In particular, our use of the full one-loop effective potential, our incorporation of the effect of the wall velocity bounds \cite{Ai:2024btx}, and the usage of the transport equations valid for all wall velocities \cite{Cline:2020jre} are pertinent to the resulting BAU, leading to much changed conclusions regarding the preferred parameter space.  Our parameter scans in the $(m_{a_1}, \, s_{\theta})$ plane highlighted regions where the BAU is attainable and scans in the $(M, \, t_{\beta})$ plane emphasised how these said regions can be extended. 
This new region has then been confronted with the latest experimental constraints from CMS and ATLAS, ruling out significant regions of viable parameter space and showing potential for future sensitivity.

The main effect of including the full one-loop effective potential in our analysis was to reduce the expected strength, $\xi_c$, of the EWPT compared to the Hartree approximation, by increasing the critical temperature, with most other relevant quantities remaining relatively unchanged. Given the strong dependence of the expected BAU on $\xi_c$, we found that this led to the requirement of larger mixing angles and somewhat larger singlet pseudoscalar masses than previously expected. We also found that the full one-loop potential led to fewer instances of vacuum trapping, where the barrier between the CPV and EW minima persists to zero temperature and nucleation never occurs. 

An important update to our study of EW baryogenesis in the \THDMa\ was provided by the implementation of the transport equations of \cite{Cline:2020jre}. Not only did this extend the calculation of the BAU to supersonic velocities, it - more relevantly to EW baryogenesis - reduced the BAU for wall velocities $v_w \gtrsim 0.1$ in comparison to the transport equations of \cite{Fromme:2006wx}. We additionally explored how the transport equations vary with wall thickness and transition strength in a simple toy model, and, how these results compared to the analytic BAU estimate of \cite{Huber:2022ndk}. Specifically, we observed that the analytic estimate strongly degrades for transition strengths $\xi \gtrsim 2$, and this is a direct consequence of the non-trivial decrease, including eventual sign-flipping, of the BAU for relatively strong transitions, see \cref{fig:baryo efficiency}. Furthermore, for transitions that are not too strong, i.e. before the BAU flips sign, the coefficient $C_\mathrm{eff}$ is roughly constant in the wall-thickness. The analytic estimate could potentially be re-used by adjusting the constant $C$ in \cref{bau estimate} based on values of $C_\mathrm{eff}$ in \cref{fig:baryo efficiency}. 

Crucial to the results of this work was the implementation of wall velocity bounds using the `ballistic' and `LTE' approximation of \cite{Ai:2024btx}. In many investigations of BSM baryogenesis scenarios, the transport equations input a wall velocity of $v_w = 0.1$, or otherwise, use a scan over the range of velocities $v_w \in \big[ 0.1, 1/\sqrt{3}\big]$ \cite{Cline:2011mm, Basler:2021kgq, Enomoto:2021dkl, Goncalves:2023svb, Liu:2023sey, Aiko:2025tbk}%\KM{citations needed here with some examples}
. Our results showed that the lower bound $v_w^{\mathrm{ball}} = 0.1$ is not realised in the majority of the \THDMa\ parameter space explored. 
Ultimately, the above choice of wall velocity input may be an optimistic one, and suppression to the BAU from faster wall velocities is likely. Even in regions where $v_w^{\mathrm{ball}} < 0.1$, the corresponding wall velocity upper bound $v_w^{\mathrm{LTE}}$ is not strongly constraining, and hence one would have to solve the Boltzmann equation to determine the true wall speed. Indeed, we also showed that $v_w^{\mathrm{LTE}}$ provides a weak lower bound on the BAU, due to the transport equations often predicting a sign change in the BAU at velocities that are similar to $v_w^{\mathrm{LTE}}$ 
(see \cref{fig:baryo curve}). An exact computation of the wall velocity would be beneficial, something which we leave for future investigations.

With the comprehensive analysis used in this work, we found that the expected BAU could easily be suppressed by an order of magnitude or more compared to our previous study. Notably, our previously chosen benchmark points are now deemed not to be viable under our new analysis.
Retaining similar parameter choices to~\cite{Huber:2022ndk}, the main requirement is now relatively large mixing angles of $s_\theta\gtrsim0.15$, and new benchmark points were defined to showcase the new interplay of BAU predictions with experimental bounds. By additionally exploring the $(M,\,t_\beta)$ planes, we showed that it is possible to have a common mass scale for the 2HDM states as high as 700 GeV or more. Whilst we are sure that FOPTs occur for yet higher mass scales, we were limited by numerical instabilities in our bounce-solving algorithm, and further improvements are needed here to determine whether the BAU can be generated in this region.

Ref.~\cite{Huber:2022ndk} already identified that $m_{a_1}<m_h/2$ is essentially ruled out by unobserved Higgs decay channels. For larger masses, the larger mixing angles lead to much better sensitivity to the preferred regions for the BAU. LHC searches for cascade decays such as the previously identified $H_0\to Z a_1\to b\bar{b}\ell^+\ell^-$ as well as the newly identified $a_2\to ha_1\to 4b$ search can rule out much of the viable parameter space, when applicable. Scalar resonance searches in the $\tau^+\tau^-$ final state were also found to have some potential to cover regions of parameter space where the cascade decays were not kinematically accessible. Currently many of these searches are limited by minimum masses for parent and/or daughter particles that would otherwise make them significantly more sensitive than they already are. We therefore encourage the experimental collaborations to continue widening their mass ranges for such exotic scalar searches, particularly towards lower masses. Searches for $b\to s\gamma$ transitions were found to bound the viable parameter space with low $t_\beta$ in Type-I models, whilst completely ruling out much of the Type-II scenarios; future improvements in this direction will further probe interesting regions. 

Although not discussed in the main body, we also verified that our benchmark scenarios do not produce appreciable stochastic gravitational wave (GW) signals during the EWPT, see \cref{grav waves}. The initial spontaneous breaking of CP can also be of first order thanks to one-loop finite-temperature effects on the potential. However, given that the potential barrier would be generated by radiative corrections, the transition would be relatively weak ($\xi \lesssim 1$), and thus unlikely to provide observable signals. Regions in parameter space that feature a one-step transition (see for example \cref{fig:crit key quants}) could also produce an observable signal if the transition is first order, but since this does not produce the required period of transient CP violation, we do not consider their experimental signatures. Moreover, other viable symmetry breaking patterns exist in the \THDMa\ ~\cite{Liu:2023sey, Si:2024vrq}. For example, it is possible to simultaneously break EW and CP symmetry from the origin of field space, potentially yielding a FOPT and the desired transient CP violation. This breaking pattern may be additionally supplemented by a strong FOPT ($\xi \gtrsim 1$) from the mixed to the EW minimum, yielding a promising scenario for viable baryogenesis accompanied by observable GW signals. 

In this work we have limited ourselves to specific benchmark scenarios, highlighting the key parametric dependences of our mechanism. Our benchmarks make several simplifying assumptions, such as a restriction to the alignment limit and the common, heavy mass scale for the 2HDM states. A more detailed exploration of the parameter space is left for future work.
Finally, let us add a comment regarding the well-motivated extension of our \THDMa\ model to include dark 
matter (in the form a dark Dirac fermion $\chi$)~\cite{Ipek:2014gua,No:2015xqa,Goncalves:2016iyg,Bauer:2017ota,Abe:2018bpo,Robens:2021lov}. 
In such a case, and for $m_{a_1} > 2 \, m_{\chi}$ (with $m_{\chi}$ the dark matter mass), the LHC phenomenology of such a model would differ drastically from the one explored in this work, since the state $a_1$ would (dominantly) decay invisibly. We will explore the potential of the LHC to probe the baryogenesis favoured region of such a model in a forthcoming work.  

\section*{Acknowledgements}
This work is supported by the Science and Technology Research Council (STFC) under the Consolidated Grant ST/X000796/1 and the Studentship Grant ST/X508822/1. T. G. is grateful for the hospitality of Universidad Aut{\'o}noma de Madrid during the early stages of this project.
K.M.\ is supported by an Ernest Rutherford Fellowship from the STFC, Grant No.\ ST/X004155/1 and by the STFC Consolidated Grant  No.\ ST/X000583/1.
J.M.N. acknowledges partial financial support by the Spanish Research
Agency (Agencia Estatal de Investigaci\'on) through the grant IFT Centro de Excelencia Severo Ochoa CEX2020-001007-S and the grants PID2021-124704NB-I00 and CNS2023-144536 funded by MCIN/AEI/10.13039/501100011033 and the European Union's Next Generation PRTR. 

\appendix
%\JMN{\section{Appendix: Further details on the \THDMa} \label{appendix a}}
\section{\THDMa \, model details} \label{appendix a}

\subsection{Minimization conditions} \label{minima_App}

The minimisation conditions at zero temperature in the case of generic vevs $\{v_1, \, v_2, \, v_s\}$ are,
\begin{align}
& \mu_{11}\,v_1 - \mu_{12}\,v_2 + \frac{1}{2} \lambda_{1}\,v_1^3 + \frac{1}{2}\lambda_{345}\,v_1\,v_2^2 + \frac{1}{2}\lambda_{a1}\,v_1\,v_s^2 = 0 \,, \\[5pt]
& \mu_{22}\,v_2 - \mu_{12}\,v_1 + \frac{1}{2} \lambda_{2}\,v_2^3 + \frac{1}{2}\lambda_{345}\,v_1^2\,v_2 + \frac{1}{2}\lambda_{a2}\,v_2\,v_s^2 = 0 \,, \\[5pt]
&\mu_{a}^2 \, v_s + \frac{1}{2}\lambda_{a1}\,v_1^2\,v_s + \frac{1}{2}\lambda_{a2}\,v_2^2\,v_s + \lambda_a \, v_s^3 = 0 \,. \label{pseudoscalar minimisation}
\end{align}
Rewriting \cref{pseudoscalar minimisation},
\begin{align}
&\left(\mu_a^2 + \frac{1}{2} \lambda_{\beta}\,v^2 \right) v_s + \lambda_a \, v_s^3 = 0 \,, \\[5pt]
&\left(\mu_a^2 + \frac{1}{2} \lambda_{\beta}\,v^2 \right) + \lambda_a \, v_s^2 = 0 \,\,\, , \,\, v_s \neq 0 \,,
\end{align}
we see that \cref{pseudoscalar minimisation} cannot be satisfied if $m_a^2 \equiv \mu_a^2 + \frac{1}{2} \lambda_{\beta}\,v^2 > 0$ and $\lambda_a > 0$. The first inequality we impose whilst the second inequality is required for the potential to be bounded from below. Therefore, these assumptions are sufficient to forbid a mixed phase ($v_s \neq 0$) in the EW minimum at $T=0$.

%\KM{check general minimisation conditions (https://arxiv.org/abs/2412.07434) to see if our conditions are sufficient to forbid mixed phase at $T=0$}

\subsection{Mixing angles} \label{mixing}

The relations between the fields in \cref{doublet expansion} and the mass eigenstates are given below,
\begin{align}
\begin{bmatrix}
\eta_1 \\[5pt]
\eta_2 
\end{bmatrix}
&= R(\beta)
\begin{bmatrix}
G_0 \\[5pt]
A_0 
\end{bmatrix} \,\, , \quad
\begin{bmatrix}
\phi_1^{+} \\[5pt]
\phi_2^{+}  
\end{bmatrix}
= R(\beta)
\begin{bmatrix}
G_{\pm} \\[5pt]
H_{\pm} 
\end{bmatrix} \,,\\[16pt]
\begin{bmatrix}
h_1 \\[5pt]
h_2
\end{bmatrix}
&= R(\alpha)
\begin{bmatrix}
h_0 \\[5pt]
H_0
\end{bmatrix} \,\, , \quad
\begin{bmatrix}
a \\[5pt]
A_0
\end{bmatrix}
= R(\theta)
\begin{bmatrix}
a_1 \\[5pt]
a_2 
\end{bmatrix} \,\, , \quad
R(\varphi) =
\begin{bmatrix}
c_{\varphi} & -s_{\varphi} \\[5pt]
s_{\varphi} & c_{\varphi}
\end{bmatrix} \,,
\end{align}
where $R(\varphi)$ is the usual $SO(2)$ rotation matrix in the anti-clockwise direction. Notice that the definition of $\alpha$ used in this paper is not the same as what is usually seen in the literature and our choice differs by an additive factor of $\pi/2$. The mixing angles $\beta$, $\alpha$ and $\theta$ are given by,
\begin{align}
 t_{\beta} = \frac{v_2}{v_1} \,\, , \,\, s_{2\alpha} = \frac{(M^2 - \lambda_{345}\,v^2)\, s_{2\beta}}{m_{H_0}^2 - m_{h_0}^2} \,\, , \,\, s_{2\theta} = \frac{2\,\kappa \, v}{m_{a_2}^2 - m_{a_1}^2} \,.
\end{align}

\subsection{Parameters} \label{params}

The relations between the $\mathbb{Z}_2$ basis parameters in the potential and the physical masses and mixings are given by,
\begin{align}
&\mu_{11}^2 = \, M^2 s_{\beta}^2 - \frac{1}{2} m_{h_0}^2 - \frac{1}{2} \left( m_{H_0}^2 - m_{h_0}^2 \right) s_{\beta - \alpha} (s_{\beta - \alpha} - t_{\beta}\,c_{\beta - \alpha}) \,, \\[6pt]
&\mu_{22}^2 = \, M^2 c_{\beta}^2 - \frac{1}{2} m_{h_0}^2 - \frac{1}{2} \left( m_{H_0}^2 - m_{h_0}^2 \right) s_{\beta - \alpha} (s_{\beta - \alpha} + t_{\beta}^{-1}\,c_{\beta - \alpha}) \,, \\[6pt]
&\mu_{12}^2 = \, M^2 s_{\beta} c_{\beta} \,, \\[6pt]
&\lambda_1 v^2 = \, m_{h_0}^2 - t_{\beta}^2 (M^2 - m_{H_0}^2) - (m_{H_0}^2 - m_{h_0}^2) s_{\beta - \alpha} \left( (t_{\beta}^2 - 1) s_{\beta - \alpha} + 2\,t_{\beta} c_{\beta - \alpha} \right) \,, \\[6pt]
&\lambda_2 v^2 = \, m_{h_0}^2 - t_{\beta}^{-2} (M^2 - m_{H_0}^2) - (m_{H_0}^2 - m_{h_0}^2) s_{\beta - \alpha} \left( (t_{\beta}^{-2} - 1) s_{\beta - \alpha} + 2\,t_{\beta}^{-1} c_{\beta - \alpha} \right) \,, \\[6pt]
&\lambda_3 v^2 = \, m_{h_0}^2 + 2M_{H^{\pm}}^2 - M^2 - m_{H_0}^2 + (m_{H_0}^2 - m_{h_0}^2) s_{\beta - \alpha} \left( 2s_{\beta - \alpha} - (t_{\beta} - t_{\beta}^{-1}) c_{\beta - \alpha} \right) \,, \\[6pt]
&\lambda_4 v^2 = \, M^2 + m_{A_0}^2 - 2 m_{H_{\pm}}^2 \,, \\[6pt]
&\lambda_5 v^2 = \, M^2 - m_{A_0}^2 \,.
\end{align}
The expressions greatly simplify in the alignment limit $\alpha = \beta$. Also, the would-be 2HDM pseudoscalar $A_0$ mass is related to the physical mass eigenvalues by $m_{A_0}^2 = s_{\theta}^2 \, m_{a_1}^2 + c_{\theta}^2 \, m_{a_2}^2$. 
Finally, also given is the Higgs basis parameters in terms of the $\mathbb{Z}_2$ ones, and their value in the alignment limit where appropriate,
\begin{align}
&Y_1 = \mu_1^2 c_{\beta}^2 + \mu_2^2 s_{\beta}^2 - \mu_{12}^2 s_{2\beta} \,\, \rightarrow \,\,  -\frac{1}{2} m_{h}^2 \,, \\[6pt]
&Y_2	= \mu_1^2 s_{\beta}^2 + \mu_2^2 c_{\beta}^2 + \mu_{12}^2 s_{2\beta} \,\, \rightarrow \,\,  M^2 -\frac{1}{2} m_{h}^2 \,, \\[6pt]
&Y_3 = \frac{1}{2}(\mu_2^2 - \mu_1^2) s_{2\beta} - \mu_{12}^2 c_{2\beta} \,\, \rightarrow \,\,  0 \,, \\[6pt]
&Z_1 v^2 = \lambda_1 c_{\beta}^4 + \lambda_2 s_{\beta}^4 + \frac{1}{2} \lambda_{345} s_{2\beta}^2 \,\, \rightarrow \,\,  m_{h}^2 \,, \\[6pt]
&Z_2 v^2 = \lambda_1 s_{\beta}^4 + \lambda_2 c_{\beta}^4 + \frac{1}{2} \lambda_{345} s_{2\beta}^2 \,\, \rightarrow \,\,  m_{h}^2 - 4(M^2 - m_{H_0}^2) \, t_{2\beta}^{-2} \,, \\[6pt]
&Z_3 v^2 = \lambda_3 + \frac{1}{4}( \lambda_1 + \lambda_2 - 2\lambda_{345} ) s_{2\beta}^2 \,\, \rightarrow \,\, m_{h}^2 - 2(M^2 - M_{H_{\pm}}^2) \,, \\[6pt]
&Z_4 v^2 = \lambda_4 + \frac{1}{4}( \lambda_1 + \lambda_2 - 2\lambda_{345} ) s_{2\beta}^2 \,\, \rightarrow \,\,  m_{A_0}^2 + m_{H_0}^2 - 2 m_{H_{\pm}}^2 \,, \\[6pt]
&Z_5 v^2 = \lambda_5 + \frac{1}{4}( \lambda_1 + \lambda_2 - 2\lambda_{345} ) s_{2\beta}^2 \,\, \rightarrow \,\,  m_{H_0}^2 - m_{A_0}^2 \,, \\[6pt]
&Z_6 v^2 = -\frac{1}{2}s_{2\beta} ( \lambda_1 c_{\beta}^2 - \lambda_2 s_{\beta}^2 - \lambda_{345} c_{2\beta}) \,\, \rightarrow \,\, 0 \,, \\[6pt]
&Z_7 v^2 = -\frac{1}{2}s_{2\beta} ( \lambda_1 s_{\beta}^2 - \lambda_2 c_{\beta}^2 + \lambda_{345} c_{2\beta}) \,\, \rightarrow \,\,  -2 (M^2 - m_{H_0}^2) \, t_{2\beta}^{-1} \,, \\[6pt]
&\lambda_{a H_1} = \lambda_{a1} c_{\beta}^2 + \lambda_{a2} s_{\beta}^2 \,, \\[6pt]
&\lambda_{a H_2} = \lambda_{a1} s_{\beta}^2 + \lambda_{a2} c_{\beta}^2 \,, \\[6pt]
&\lambda_{a H_3} = \frac{1}{2}(\lambda_{a1}-\lambda_{a2}) \, s_{2\beta} \,.
\end{align} 

\subsection{Yukawa sector} \label{yukawa}

The most general Yukawa Lagrangian for the 2HDM is,
\begin{align}
\mathcal{L}_{\mathrm{Yukawa}} &= -y_{ij}^{u,k} \,\, \overline{Q}_{L,i} \, \tilde{\phi}_k \, u_{R,j} -y_{ij}^{d,k} \,\,\overline{Q}_{L,i} \, \phi_k \, d_{R,j} -y_{ij}^{l,k} \,\, \overline{l}_{L,i} \, \tilde{\phi}_k \, e_{R,j} + \mathrm{h.c.} \\[5pt]
&= -y^{u,k} \,\, \overline{Q}_{L} \, \tilde{\phi}_k \, u_{R} -y^{d,k} \,\,\overline{Q}_{L} \, \phi_k \, d_{R} -y^{l,k} \,\, \overline{l}_{L} \, \tilde{\phi}_k \, e_{R} + \mathrm{h.c.}  \,\,, 
\end{align} 
where $k$ runs over both Higgs doublet and $i$ and $j$ are generation indices. Here, three families are assumed. Assuming the Yukawa matrices have no structure and are generic $3\times3$ complex matrices, one can diagonalise for either doublet for up/down type quarks and leptons. The key point is that one cannot simultaneously diagonalise the Yukawa interactions for both Higgs doublets (assuming no additional symmetries or structure). Once either of the Higgs doublet interactions has been diagonalised, the other will not be diagonal, leading to FCNCs that are severely constrained by experimental measurements. 

Introducing the softly-broken $\mathbb{Z}_2$ symmetry forces the right-handed up/down-type quarks and leptons to couple to one Higgs doublet only. By convention, the right-handed up-type quarks always couple to $\Phi_2$, giving rise to four distinct choices of charge assignments which are summarised below.
\begin{center}
\begin{tabular}{|c|c|c|c|c|c|c|c|}
\hline
  & $Q_{L}$ & $l_{L}$ & $\Phi_1$ & $\Phi_2$ & $u_R$ & $d_R$ & $e_R$ \\ \hline
Type I & $+$ & $+$ & $-$ & $+$ & $+$ & $+$ & $+$ \\ \hline
Type II & $+$ & $+$ & $-$ & $+$ & $+$ & $-$ & $-$ \\ \hline
Type X & $+$ & $+$ & $-$ & $+$ & $+$ & $+$ & $-$ \\ \hline
Type Y & $+$ & $+$ & $-$ & $+$ & $+$ & $-$ & $+$ \\ \hline
\end{tabular}
\captionof{table}{Four different models of 2HDM based on various $\mathbb{Z}_2$ charge assignments to SM matter. \label{tab:2hdm_type_z2}}
\end{center}
Expanding the Yukawa Lagrangian in terms of physical 2HDM mass eigenstates (i.e. $s_{\theta} = 0$) one obtains,
\begin{align}
\mathcal{L}_{\mathrm{Yukawa}} & \supset - \sum_{f=u,d,l} \Bigg[ \left( \frac{m_f}{v} \xi^{f}_{h_0} \,\, \overline{f} \, h_0 \, f + \frac{m_f}{v} \xi^{f}_{H_0} \,\, \overline{f}\, H_0 \, f - i \frac{m_f}{v} \xi^{f}_{A_0} \,\, \overline{f} \, \gamma^5 A_0 \, f \right) \nonumber \\ 
& - \left( \frac{\sqrt{2}V_{ud}}{v} \, \overline{u}\, \left( m_u \, \xi^u_{A_0} P_{L} + m_d \, \xi^d_{A_0} P_{R} \right)\,d\,H^{+} + \frac{\sqrt{2}m_l}{v} \xi_{A_0}^l \, \overline{\nu} \, P_{R}\, e \, H^{+} \right) \Bigg] \,, 
\end{align}
where $V_{ud}$ is the up-down type entry of the CKM-matrix, $P_{L/R} = \frac{1}{2}(1 \mp \gamma_5)$ are the standard projection operators and the couplings $\xi^f_{\phi}$ are summarised in \cref{tab:2hdm_type_couplings}.
\begin{center}
\begin{tabular}{|c|c|c|c|c|c|c|c|c|c|}
\hline
\rule{0pt}{2pt} & $\xi^u_{h_0}$ & $\xi^d_{h_0}$ & $\xi^l_{h_0}$ & $\xi^u_{H_0}$ & $\xi^d_{H_0}$ & $\xi^l_{H_0}$ & $\xi^u_{A_0}$ & $\xi^d_{A_0}$ & $\xi^l_{A_0}$ \\[8pt] \hline
\rule{0pt}{17pt} Type I  & $\frac{s_{\alpha}}{s_{\beta}} \rightarrow 1$ & $\frac{s_{\alpha}}{s_{\beta}} \rightarrow 1$ & $\frac{s_{\alpha}}{s_{\beta}} \rightarrow 1$ & $\frac{c_{\alpha}}{s_{\beta}} \rightarrow t_{\beta}^{-1}$ & $\frac{c_{\alpha}}{s_{\beta}} \rightarrow t_{\beta}^{-1}$ & $\frac{c_{\alpha}}{s_{\beta}} \rightarrow t_{\beta}^{-1}$ & $t_{\beta}^{-1}$ & $-t_{\beta}^{-1}$ & $-t_{\beta}^{-1}$ \\ \hline
\rule{0pt}{17pt} Type II & $\frac{s_{\alpha}}{s_{\beta}} \rightarrow 1$ & $\frac{c_{\alpha}}{c_{\beta}} \rightarrow 1$ & $\frac{c_{\alpha}}{c_{\beta}} \rightarrow 1$ & $\frac{c_{\alpha}}{s_{\beta}} \rightarrow t_{\beta}^{-1}$ & $-\frac{s_{\alpha}}{c_{\beta}} \rightarrow -t_{\beta}$ & $-\frac{s_{\alpha}}{c_{\beta}} \rightarrow -t_{\beta}$ & $t_{\beta}^{-1}$ & $t_{\beta}$ & $t_{\beta}$ \\  \hline
\rule{0pt}{17pt} Type X  & $\frac{s_{\alpha}}{s_{\beta}} \rightarrow 1$ & $\frac{s_{\alpha}}{s_{\beta}} \rightarrow 1$ & $\frac{c_{\alpha}}{c_{\beta}} \rightarrow 1$ & $\frac{c_{\alpha}}{s_{\beta}} \rightarrow t_{\beta}^{-1}$ & $\frac{c_{\alpha}}{s_{\beta}} \rightarrow t_{\beta}^{-1}$ & $-\frac{s_{\alpha}}{c_{\beta}} \rightarrow -t_{\beta}$ & $t_{\beta}^{-1}$ & $-t_{\beta}^{-1}$ & $t_{\beta}$ \\  \hline
\rule{0pt}{17pt} Type Y  & $\frac{s_{\alpha}}{s_{\beta}} \rightarrow 1$ & $\frac{c_{\alpha}}{c_{\beta}} \rightarrow 1$ & $\frac{s_{\alpha}}{s_{\beta}} \rightarrow 1$ & $\frac{c_{\alpha}}{s_{\beta}} \rightarrow t_{\beta}^{-1}$ & $-\frac{s_{\alpha}}{c_{\beta}} \rightarrow -t_{\beta}$ & $\frac{c_{\alpha}}{s_{\beta}} \rightarrow t_{\beta}^{-1}$ & $t_{\beta}^{-1}$ & $t_{\beta}$ & $-t_{\beta}^{-1}$ \\ \hline
\end{tabular}
\captionof{table}{Modified Yukawa couplings for up and down-type quarks and leptons for 2HDM eigenstates. Also shown are their values in the alignment limit ($\alpha=\beta$). \label{tab:2hdm_type_couplings}}
\end{center}
Notice that up-type quarks enjoy the same coupling to the Higgs sector regardless of the model type.

\subsection{Alignment limit} \label{align}
To relate the CP-even mass eigenstates from the \THDMa\ sector to the SM Higgs, it is easiest to do so in the Higgs basis,
\begin{align}
\begin{bmatrix}
\Phi_1 \\[7pt]
\Phi_2
\end{bmatrix}
= R(\beta)
\begin{bmatrix}
H_1 \\[7pt]
H_2
\end{bmatrix} 
\quad \Rightarrow \quad \left\langle H_1^0 \right\rangle = \frac{v}{\sqrt{2}} \,\, , \,\, \left\langle H_2^0 \right\rangle = 0  \,, 
\end{align}
and, expanding $H_1$ and $H_2$, it is not difficult to relate their neutral components to the mass eigenstates $h_0$ and $H_0$,
\begin{align}
h_0 &= c_{\beta - \alpha} \Big(\sqrt{2} \, \mathrm{Re}{(H_1^0)}-v\Big) - s_{\beta - \alpha} \Big(\sqrt{2} \, \mathrm{Re}{(H_2^0)}\Big) \nonumber \,, \\[6pt]
H_0 &= s_{\beta - \alpha} \Big(\sqrt{2} \, \mathrm{Re}{(H_1^0)}-v\Big) + c_{\beta - \alpha} \Big(\sqrt{2} \, \mathrm{Re}{(H_2^0)}\Big) \,.
\end{align} 
Therefore, if you want to align one of the CP-even eigenstates with the SM Higgs, you can either impose $\alpha = \beta$ or $\alpha = \beta - \pi/2$. In this paper, the first choice $\alpha = \beta$ is chosen as this identifies the lighter CP-even eigenstate with the SM Higgs, leaving you with another heavier state $H_0$. Again, it is stressed that our definition of $\alpha$ differs from the majority of the literature but still leads to the same physics.

%\newpage

\section{Effective potential details} \label{appendix b}

\subsection{Debye masses} \label{debye}
To implement the resummation of the leading order IR divergences stemming from the Matsubara zero modes, we need the thermally corrected masses due to screening in the plasma. Only longitudinal components of the gauge bosons receive corrections at one-loop and this includes the previously massless photon \cite{Carrington:1991hz}. Additionally, fermions receive no corrections due to the fact that they do not contain a Matsubara zero mode in their expansion at finite temperature.
\begin{align}
&\overline{m}_t^2 = m_t^2 \,, \\
&\overline{m}_{W_{\pm}^{\mathrm{T}}}^2 = m_{W_{\pm}}^2 \,\, , \,\,\, \overline{m}_{W_{\pm}^{\mathrm{L}}}^2 = m_{W_{\pm}}^2 + 2 g T^2 \,, \\[6pt]
&\overline{m}_{Z_{0}^{\mathrm{T}}}^2 = m_{Z_0}^2 \,\, , \,\,\, \overline{m}_{Z_{0}^{\mathrm{L}}}^2 = \frac{1}{8} \left( f{\left( H_1^0, H_2^0 \right)} + g{\left( H_1^0, H_2^0 \right)} \right) \,, \\[6pt]
&\overline{m}_{\gamma^{\mathrm{T}}}^2 = 0 \,\, , \,\,\, \overline{m}_{\gamma^{\mathrm{L}}}^2 = \frac{1}{8} \left( f{\left( H_1^0, H_2^0 \right)} - g{\left( H_1^0, H_2^0 \right)} \right) \,, \\[6pt]
& f{\left( H_1^0, H_2^0 \right)} = \Big(g^2 + {g'}^2 \Big) \Big( \left| H_1^0 \right|^2 + \left| H_2^0 \right|^2 \Big) \,, \\[6pt]
& g{\left( H_1^0, H_2^0 \right)} = \sqrt{\left(g^2 - {g'}^2 \right)^2 \left(64T^4 + 16T^2 \left( \left| H_1^0 \right|^2 + \left| H_2^0 \right|^2 \right) \right) + \left(g^2 + {g'}^2 \right)^2 \left(\left| H_1^0 \right|^2 + \left| H_2^0 \right|^2 \right)^2} \,, \\[6pt]
& \overline{m}_{\mathrm{2HDM+a}}^2 = \mathrm{Eigenvalues} \left( \frac{\partial^2 V_0}{\partial \phi_i \partial \phi_j} + \Pi \, T^2 \right) \Bigg|_{H_i = H_i^c, \, a = a^c} \,,
\end{align}
with
\begin{align}
&\Pi = \mathrm{Diag}(\Pi_{11}, \Pi_{11}, \Pi_{22}, \Pi_{22}, \Pi_{11}, \Pi_{22}, \Pi_{11}, \Pi_{22}, \Pi_{33}) \,, \\[6pt]
& \Pi_{11} = \frac{1}{24} \left( \frac{9}{2}g_1^2 + \frac{3}{2}g_2^2 + 6y_t^2 + 6Z_1 + 4Z_3 + 2Z_4 + \lambda_{a H_1} \right) \,, \\[6pt]
& \Pi_{22} = \frac{1}{24} \left( \frac{9}{2}g_1^2 + \frac{3}{2}g_2^2 + 6 t_{\beta}^{-2} y_t^2 + 6Z_2 + 4Z_3 + 2Z_4 + \lambda_{a H_2} \right) \,, \\[6pt]
& \Pi_{33} = \frac{1}{24} \left( 6 \lambda_a + 4 \lambda_{a H_1} +4 \lambda_{a H_2} \right)  \,.
\end{align}
\subsection{Counterterms} \label{counterterms}

Preserving the quantities discussed in \cref{effective potential}, and using the shorthand below, the counterterms are determined numerically,
\begin{align}
&N_{\phi_i} = \frac{\partial V_{\mathrm{CW}}}{\partial \phi_i} \Bigg|_{\langle v, \, 0, \, 0, \, 0 \rangle} \,\, , \quad T_{\phi_i \phi_j} = \frac{\partial^2 V_{\mathrm{CW}}}{\partial \phi_i \partial \phi_j} \Bigg|_{\langle v, \, 0, \, 0, \, 0 \rangle} \,\, , \quad \phi = \{ \rho_1, \, \rho_2^{\mathrm{R}}, \, \rho_2^{\mathrm{I}}, \, a \} \,, \\[6pt]
&\delta Y_1 = \frac{1}{2 v} (-3 N_{\rho_1} + v T_{\rho_1 \rho_1}) \,\, ,\quad \delta Y_2 = -\frac{1}{2} (T_{\rho_2^{\mathrm{R}} \rho_2^{\mathrm{R}}} + T_{\rho_2^{\mathrm{I}} \rho_2^{\mathrm{I}}}) \,\, ,\quad \delta Y_3 = \frac{1}{2 v} (3 N_{\rho_2^{\mathrm{R}}} - v T_{\rho_1 \rho_2^{\mathrm{R}}}) \,, \\[6pt]
&\delta Z_1 = \frac{1}{v^3} (N_{\rho_1} - v T_{\rho_1 \rho_1}) \,\, ,\quad \delta Z_5 = \frac{1}{v^2} (T_{\rho_2^{\mathrm{I}}\rho_2^{\mathrm{I}}} - T_{\rho_2^{\mathrm{R}}\rho_2^{\mathrm{R}}}) \,\, ,\quad \delta Z_6 = \frac{1}{v^3} (N_{\rho_2^{\mathrm{R}}} - v T_{\rho_1 \rho_2^{\mathrm{R}}}) \,, \\[6pt]
&\delta \mu_a^2 = - T_{a a} \,\, ,\quad \delta \kappa = - \frac{1}{v} T_{\rho_2^{\mathrm{I}} a} \,, \\[6pt]
&\delta Z_2 = \delta Z_3 = \delta Z_4 = \delta \lambda_a = \delta \lambda_{a H_1} = \delta \lambda_{a H_2} = \delta \lambda_{a H_3} = 0 \,.
\end{align}
IR divergences from the Goldstone bosons appear in the evaluation of the second derivatives $T_{\phi_i \phi_j}$, and, is proportional to $\frac{\partial^2 m^2}{\partial \phi_i \partial \phi_j} \log{m^2}$. The first term is finite but due to the vanishing mass of the Goldstone bosons, the second is divergent. To cure this, we follow the procedure of \cite{Cline:2011mm} and introduce an IR cutoff for the evaluation of $T_{\phi_i \phi_j}$, where vanishing Goldstone masses are replaced by the Higgs mass. The exact value of the cutoff doesn't particularly matter given its logarithmic sensitivity. More sophisticated methods for dealing with these IR divergences can be found in \cite{Martin:2014bca, Elias-Miro:2014pca, Espinosa:2016uaw, Braathen:2016cqe, Espinosa:2017aew}.  

\section{Theoretical constraints} \label{appendix c}

\subsection{Boundedness from below} \label{bfb}
The asymptotic behaviour of the tree-level potential is driven by the quartic couplings, and requiring that the potential is bounded from below translates into inequality constraints on the said quartics. We use the method of copositivity  \cite{Kannike:2012pe, Chakrabortty:2013mha} and in the $\mathbb{Z}_2$ basis, the conditions for being bounded from below are, 
\begin{align}
&\lambda_1 > 0 \quad,\quad \lambda_2 > 0 \quad,\quad \lambda_a > 0 \,, \\[6pt]
&\overline{\lambda}_{12} = \sqrt{\lambda_1 \lambda_2} + \lambda_3 + \mathrm{min}{(0,\lambda_4 - |\lambda_5|)} > 0 \,, \\[6pt]&\overline{\lambda}_{i a} = \sqrt{2\lambda_1 \lambda_a} + \lambda_{ai} > 0 \,\,, \,\,\, i=1,2 \,, \\[6pt]
&\lambda_{a1} \sqrt{\lambda_2} + \lambda_{a2} \sqrt{\lambda_1} + \overline{\lambda}_{12} \sqrt{2\lambda_a} + \sqrt{2\overline{\lambda}_{12} \overline{\lambda}_{1 a} \overline{\lambda}_{2 a}} > 0\,.
\end{align}
The results given here differ from what is seen in \cite{Drozd:2014yla, Muhlleitner:2016mzt, Engeln:2020fld, Biekotter:2021ysx}, who derive their bounds based on the method in \cite{Klimenko:1984qx}. Our results agree with those from \cite{He:2016mls, Arcadi:2022lpp, Dutta:2023cig, Arcadi:2023smv}, and it is unclear how much of a difference using either set of constraints makes. Perhaps it is better to use both sets of inequalities such as was done in \cite{Si:2024vrq}. In any case, as stated in the main body of this paper, we check boundedness from below numerically for the one-loop effective potential.

\subsection{Perturbative unitarity} \label{per uni}

A potential threat to scattering unitarity arises from longitudinal massive gauge bosons, and, considering point interactions of longitudinal gauge bosons, it can be shown that the amplitude grows with the centre of mass energy squared \cite{Cornwall:1974km}. Fortunately, in the SM, this is avoided via the Higgs mechanism \cite{Lee:1977eg, Chanowitz:1985hj, Willenbrock:1987xz, Valencia:1990jp}, but, if the SM is extended we must again worry about unitarity violation. Importantly it will yield upper bounds on the mass of newly predicted particles, for example, this has been done in the 2HDM \cite{Kanemura:1993hm, Arhrib:2000is, Ginzburg:2005dt, Horejsi:2005da, Goodsell:2018fex}. Using the Goldstone equivalence theorem and considering 2-2 scatterings, a bound can be placed on tree-level S-matrix scattering eigenvalues and this translates into bounds on combinations of quartic couplings. In the $\mathbb{Z}_2$ basis, the conditions are,
\begin{align}
&|\lambda_{ai}| < 8\pi \, , \,\, i=1,2 \,, \\
&|\lambda_3 \pm \lambda_4| < 8\pi \,, \\
&|\lambda_3 \pm \lambda_5| < 8\pi \,, \\
&|\lambda_3 + 2\lambda_4 \pm 3\lambda_5| < 8\pi \,, \\
&\Big|\frac{1}{2}\left( \lambda_1+\lambda_2 \pm \sqrt{(\lambda_1-\lambda_2)^2 + 4\lambda_i^2} \right)\Big| < 8\pi \, , \,\, i=4,5 \,, \\
&|x_i| < 8\pi \, , \,\, i=1,2,3 \,,
\end{align} 
where $x_i$ are the three solutions to the following cubic monomial,
\begin{align}
x^3 &-3(\lambda_a + \lambda_1 + \lambda_2)x^2 + \left[ 9((\lambda_1+\lambda_2)\lambda_a + \lambda_1 \lambda_2) - (\lambda_{a1}^2 + \lambda_{a2}^2) - (2\lambda_3 + \lambda_4)^2 \right]x \nonumber \\
&+ \left[ (2\lambda_3 + \lambda_4)(3\lambda_a (2\lambda_3 + \lambda_4) -2\lambda_{a1}\lambda_{a2}) + 3(\lambda_1 \lambda_{a2}^2 + \lambda_2 \lambda_{a1}^2) - 27\lambda_1 \lambda_2 \lambda_a) \right] = 0 \,.
\end{align}
These expressions agree with what were found in \cite{He:2016mls, Arcadi:2022lpp, Arcadi:2023smv}.

\section{Transport, baryogenesis and gravitational waves} \label{appendix d}
\subsection{Transport equations} \label{transport detail}

This appendix is written here for the convenience of the reader and the full computational details can be found in the majority of \cite{Cline:2000nw, Cline:2020jre}. A semiclassical WKB ansatz is used in conjunction with the Dirac equation to derive the dynamics of a fermion in a plasma that passes through a region of varying scalar condensate. The Boltzmann equation for a fermion with complex mass, $m(z) = |m(z)| e^{i \gamma_5 \theta(z)}$, and distribution function, $f$, in the bubble wall frame is given by,
\begin{align}
&(v_g \partial_{z} + F_{z} \partial_{p_z}) f = \mathcal{C}[f] \,,\\[5pt]
&v_g = \frac{p_z}{E} + s_{h}\,s_{k_0} \frac{m^2 \theta'}{2 E^2 \, E_z} \,,\\[5pt]
&F_{z} = -\frac{(m^2)'}{2E} + s_{h}\,s_{k_0} \left( \frac{(m^2 \, \theta')'}{2E E_z} - \frac{m^2 (m^2)' \theta'}{4 E^3 E_z}\right)\,,
\end{align}
where $E^2 = \textbf{p}^2 + m^2$, $E_z^2 = p_z^2 + m^2$, $s_{k_0} = \pm 1$ for particles and anti-particles respectively, and, ignoring the difference between helicity and chirality, $s_h = h \, \mathrm{sign}{(p_z)}$ where $h = \pm 1$ is the helicity. An ansatz for the distribution function is then made that involves perturbing around the equilibrium distribution, and the resulting equations can be linearised in these perturbations,
\begin{align}
f_{\mathrm{ansatz}} = \frac{1}{e^{\beta \left( \gamma_w \left( E_w + v_w p_z \right) - \mu \right)} \mp 1} + \delta f \,,
\end{align} 
where $E_w$ is the conserved energy in the wall frame and can be related to $E$ and $E_z$, $\mu$ is the chemical potential and $\delta f$ is the out-of-equilibrium contribution. Plugging this ansatz into the Boltzmann equation, linearising and splitting the perturbations into odd and even parts, one obtains Boltzmann equations governing the CP-odd and even parts with different source and collision terms,
\begin{align}
&L[\mu_i, \, \delta f_i] = \mathcal{S}_{i} + \mathcal{C}_{i} \,\, , \,\, i = \{ e, o \} \label{boltzmann eqn} \,, \\[5pt]
&L[\mu, \, \delta f] = -\frac{p_z}{E} f_{0, v_w}^{\prime} \, \partial_z \mu + v_w \gamma_w \frac{\partial_z m^2}{2E} \, f_{0, v_w}^{\prime \prime} \, \mu + \frac{p_z}{E} \partial_z \delta f - \frac{\partial_z m^2}{2E} \partial_{p_z} \delta f \,, \\[5pt]
&f_{0, v_w} = \frac{1}{e^{\beta \gamma_w \left( E + v_w p_z \right)} \mp 1} \,, \\[5pt]
&\mathcal{S}_{e} = v_w \gamma_w \frac{\partial_z m^2}{2E} f_{0, v_w}^{\prime} \,, \\[5pt]
&\mathcal{S}_{o} = -v_w \gamma_w s_h \frac{\partial_z (m^2 \partial_z \theta)}{2 E E_z} f_{0, v_w}^{\prime} + v_w \gamma_w s_h \frac{m^2 \partial_z m^2 \, \partial_z \theta}{4 E^2 E_z}\left( \frac{f_{0, v_w}^{\prime}}{E} - \gamma_w f_{0, v_w}^{\prime \prime} \right) \,,
\end{align}
where the prime is the derivative with respect to $\gamma_w E$ and $\mathcal{C}_i$ are undetermined a priori. Considering the CP-odd part of \cref{boltzmann eqn} and taking moments with weights $(p_z / E)^l$ yields a coupled set of ODE's for the chemical potentials and plasma velocities. Focusing on the simplest $l = \{ 0, 1\}$ weights, 
\begin{align}
&\left\langle \left(\frac{p_z}{E}\right)^l L \right\rangle = \left\langle \left(\frac{p_z}{E}\right)^l (\mathcal{S}_i + \mathcal{C}_i) \right\rangle \,, \\[6pt]
&\left\langle L_o \right\rangle = - D_1 \, \partial_z \mu + \partial_z u + v_w \gamma_w \, \partial_z m^2 \, Q_1 \, \mu \,, \\[6pt]
&\left\langle \mathcal{S}_o \right\rangle = -v_w \gamma_w \, h \left[ \partial_z (m^2 \partial_z \theta) \, Q_1^{8o} - m^2 \, \partial_z m^2 \, \partial_z \theta \, \, Q_1^{9o}\right] \,, \\[6pt]
&\left\langle \mathcal{C}_o \right\rangle = K_0 \sum_{i} \Gamma_i \sum_j s_{ij} \frac{\mu_j}{T} \,, \\[6pt]
&\left\langle \left( \frac{p_z}{E} \right) L_o \right\rangle = - D_2 \, \partial_z \mu + R \, \partial_z u + v_w \gamma_w \, \partial_z m^2 \, Q_2 \, \mu + \bar{R} \, u \,, \\[6pt]
&\left\langle \left(\frac{p_z}{E}\right) \mathcal{S}_o \right\rangle = -v_w \gamma_w \, h \left[ \partial_z (m^2 \partial_z \theta) \, Q_2^{8o} - m^2 \, \partial_z m^2 \, \partial_z \theta \, Q_2^{9o}\right] \,, \\[6pt]
&\left\langle \left( \frac{p_z}{E} \right) \mathcal{C}_o \right\rangle = -\Gamma_{\mathrm{tot}}\,u - v_w \left\langle \mathcal{C}_o \right\rangle  \,, 
\end{align}
where $\{ D_i, Q_i, Q_i^{8o}, Q_i^{9o}, R, \bar{R}, K_0 \}$ are simply functions of $m/T$ and $v_w$ whose formulae can be found in \cite{Cline:2020jre}, $\Gamma_i$ is the interaction rate of interest with $s_{ij} = \pm 1$ if the particle is in the initial or final state respectively and $\Gamma_{\mathrm{tot}}$ is the total interaction rate. The forms of the collision moments were justified in \cite{Cline:2000nw} and $\delta f$ and $u$ are related via a factorisation scheme that follows the procedure of \cite{Fromme:2006wx}.
At this point, we raise a small discrepancy in the scaling of equation A2 in \cite{Cline:2020jre} as there should be an additional factor of $T$ on the RHS of said equation. This also leads to the equations in A6 of \cite{Cline:2020jre} being too large on the LHS by a factor of $T$, however, it is important to state that this discrepancy did not affect their results as they worked in units of temperature, and, we find close agreement with their toy model results. Now, checking the scaling of equation A2 with temperature, we calculate,
\begin{align}
\left\langle \frac{p_z^n}{E^m} \, f_{0, v_w}^{(k)} \right\rangle = \frac{1}{N_1} \int d^3p \, \frac{p_z^n}{E^m} \, f_{0, v_w}^{(k)} = \# \, \frac{1}{T^2} T^3 \frac{T^n}{T_m} \frac{1}{T_k} = \# \, T^{n-m-k+1} \, .
\end{align}
Additionally, they claim that the factor $K_0 \approx 1.1$ for a massless fermion, but, there is a missing factor of $T$ and a quick calculations shows, $K_0 = N_0 / N_1 = (9\zeta(3) / \pi^2) \, T \approx 1.1 \, T$. To double check the consistency of our proposed modifications, we take the CP-odd Boltzmann equation for the zeroth moment,
\begin{align}
&\{ D_i, \,Q_i, \,Q_i^{8o}, \,Q_i^{9o}, \,R, \,\bar{R}, \,K_0, \,\partial_z, \,m, \,\mu, \,u, \,\Gamma \} \nonumber \\[5pt]
 \sim \, & \{ T^0, T^{-2}, T^{-2}, T^{-4}, T^0, T^0, T^1, T^1, T^1, T^1, T^1, T^1 \} \,, \\[12pt]
\Rightarrow - & \underbrace{D_1 \, \partial_z \, \mu}_{\sim T^2} + \underbrace{\partial_z u}_{\sim T^2} + \, v_w \gamma_w \underbrace{\partial_z m^2 \, Q_1 \, \mu}_{T^2} \nonumber \\[2pt]
= &-v_w \gamma_w \, h \, \big[ \underbrace{\partial_z(m^2 \, \partial_z \theta) \, Q_{1}^{8o}}_{\sim T^2} + \underbrace{m^2 \, \partial_z m^2 \, \partial_z \theta \, Q_{1}^{9o}}_{\sim T^2} \big] + \underbrace{K_0 \sum_{i} \Gamma_i \sum_{j} s_{ij} \frac{\mu_j}{T}}_{\sim T^2} \,.
\end{align}
The scalings proposed here match with those in \cite{Barni:2025ifb}.

For the transport equations, we track the left- and right-handed top quarks, the left-handed bottom quarks and the Higgs bosons. Following \cite{Fromme:2006cm}, we ignore the bottom quark CP-source due to Yukawa suppression and take the bottom quark and Higgs bosons to be massless. Expanding equation \cref{transport ode} for the considered species yields the system,
\begin{align}
&- D_1^t \, \mu_{t_{\mathrm{L}}}^{\prime} + u_{t_{\mathrm{L}}}^{\prime} + v_w \gamma_w \, (m_{t}^2)^{\prime} \, Q_1^t \, \mu_{t_{\mathrm{L}}} - K_0^t \, \delta \bar{\mathcal{C}}_{t_{\mathrm{L}}}  = S_1^t \nonumber \,, \\[5pt]
&- D_2^t \, \mu_{t_{\mathrm{L}}}^{\prime} + R^t \, u_{t_{\mathrm{L}}}^{\prime} + v_w \gamma_w \, (m_{t}^2)^{\prime} \, Q_2^t \, \mu_{t_{\mathrm{L}}} + (m_t^2)^{\prime} \, \bar{R}^t \, u_{t_{\mathrm{L}}} + \Gamma_{\mathrm{tot}}^{t} \, u_{t_{\mathrm{L}}} + K_0^t \, \delta \bar{\mathcal{C}}_{t_{\mathrm{L}}} = S_2^t  \nonumber \,, \\[5pt] 
&- D_1^t \, \mu_{t_{\mathrm{R}}}^{\prime} + u_{t_{\mathrm{R}}}^{\prime} + v_w \gamma_w \, (m_{t}^2)^{\prime} \, Q_1^t \, \mu_{t_{\mathrm{R}}} - K_0^t \, \delta \bar{\mathcal{C}}_{t_{\mathrm{R}}}  = - S_1^t \nonumber \,, \\[5pt]
&- D_2^t \, \mu_{t_{\mathrm{R}}}^{\prime} + R^t \, u_{t_{\mathrm{R}}}^{\prime} + v_w \gamma_w \, (m_{t}^2)^{\prime} \, Q_2^t \, \mu_{t_{\mathrm{R}}} + (m_t^2)^{\prime} \, \bar{R}^t\, u_{t_{\mathrm{R}}} + \Gamma_{\mathrm{tot}}^{t}\,u_{t_{\mathrm{R}}} + K_0^t \, \delta \bar{\mathcal{C}}_{t_{\mathrm{R}}} = - S_2^t \nonumber \,, \\[5pt] 
&- D_1^b \, \mu_{b_{\mathrm{L}}}^{\prime} + u_{b_{\mathrm{L}}}^{\prime} - K_0^b \, \delta \bar{\mathcal{C}}_{b_{\mathrm{L}}}  = 0 \nonumber \\[5pt]
&- D_2^b \, \mu_{b_{\mathrm{L}}}^{\prime} + R^b \, u_{b_{\mathrm{L}}}^{\prime} + \Gamma_{\mathrm{tot}}^{b}\,u_{b_{\mathrm{L}}} + K_0^b \, \delta \bar{\mathcal{C}}_{b_{\mathrm{L}}} = 0 \nonumber \,, \\[5pt] 
&- D_1^h \, \mu_{h}^{\prime} + u_{h}^{\prime} - K_0^h \, \delta \bar{\mathcal{C}}_{h}  = 0 \nonumber \,, \\[5pt]
&- D_2^h \, \mu_{h}^{\prime} + R^h \, u_{h}^{\prime} + \Gamma_{\mathrm{tot}}^{h}\,u_{h} + K_0^h \, \delta \bar{\mathcal{C}}_{h} = 0 \label{bvp system}\,, 
\end{align}
where the prime now denotes differentiation with respect to $z$. Solving this coupled system of BVP's allows one to construct the left-handed baryon asymmetry, which can then be integrated with suitable weighting functions to yield the final BAU \cite{Cline:2000nw, Cline:2011mm},
\begin{align}
&\mu_{B_{\mathrm{L}}} = \frac{1}{2} \left( 1 + 4 D_0^t \right) \mu_{t_{\mathrm{L}}} + \frac{1}{2} \left( 1 + 4 D_0^b \right) \mu_{b_{\mathrm{L}}} + 2 D_0^t \,\mu_{t_{\mathrm{R}}} \,, \\[5pt]
&\eta_B = \frac{405 \, \Gamma_{ws}}{4 \pi^2 \, v_w \gamma_w \, g_{*} T} \int_{-\infty}^{\infty} dz \, \mu_{B_{\mathrm{L}}} \, f_{\mathrm{sph}} \, e^{-45 \Gamma_{ws} |z| / 4 v_w \gamma_w}\,,  \\[5pt]
&g_{*} = 106.75 \,\, , \,\,\, f_{\mathrm{sph}} = \min{\left[ 1, \,\, 2.4 \, \frac{\Gamma_{ws}}{T} e^{-40\,h(z)/T} \right]} \,, 
\end{align}
where $g_{*}$ is the SM degrees of freedom in the plasma and $f_{\mathrm{spl}}$ takes into account the finite width of the sphaleron when viewed as a quasi-particle. The source and collision terms in the system \cref{bvp system} read,
\begin{align}
& S_i^t = v_w \gamma_w \, \left[ (m_{t}^2 \, \theta_{t}^{\prime})^{\prime} \, Q_i^{8o, t} - m_t^2 \, (m_t^2)^{\prime} \, \theta_t^{\prime} \, \, Q_i^{9o, t}\right]  \, , \,\, i \in \{1, 2\} \,, \\[5pt]
& \delta \bar{\mathcal{C}}_{t_{\mathrm{L}}} = \Gamma_{y} ( \mu_{t_{\mathrm{L}}} - \mu_{t_{\mathrm{R}}} + \mu_{h} ) + 2 \, \Gamma_{m} ( \mu_{t_{\mathrm{L}}} - \mu_{t_{\mathrm{R}}} ) + \Gamma_{W} ( \mu_{t_{\mathrm{L}}} - \mu_{b_{\mathrm{L}}} ) + \overline{\Gamma}_{\mathrm{SS}}[\mu_i] \,, \\[5pt]
& \delta \bar{\mathcal{C}}_{t_{\mathrm{R}}} = \Gamma_{y} ( 2 \mu_{t_{\mathrm{R}}} - \mu_{t_{\mathrm{L}}} - \mu_{b_{\mathrm{L}}} - 2\mu_{h} ) + 2 \, \Gamma_{m} ( \mu_{t_{\mathrm{R}}} - \mu_{t_{\mathrm{L}}} ) - \overline{\Gamma}_{\mathrm{SS}}[\mu_i] \,, \\[5pt]
& \delta \bar{\mathcal{C}}_{b_{\mathrm{L}}} = \Gamma_{y} ( \mu_{b_{\mathrm{L}}} - \mu_{t_{\mathrm{R}}} + \mu_{h} ) + \Gamma_{W} ( \mu_{b_{\mathrm{L}}} - \mu_{t_{\mathrm{L}}} ) + \overline{\Gamma}_{\mathrm{SS}}[\mu_i]\,,  \\[5pt]
& \delta \bar{\mathcal{C}}_{h} = \frac{3}{4} \, \Gamma_{y} ( 2 \mu_{h} + \mu_{t_{\mathrm{L}}} + \mu_{b_{\mathrm{L}}} - 2\mu_{t_{\mathrm{R}}}) + \Gamma_{h} \, \mu_h \,, \\[5pt]
& \overline{\Gamma}_{\mathrm{SS}}[\mu_i] = \Gamma_{\mathrm{SS}} \left( (1+9D_0^t) \mu_{t_{\mathrm{L}}} - (1-9D_0^t) \mu_{t_{\mathrm{R}}} + (1+9D_0^b) \mu_{b_{\mathrm{L}}}  \right) \,, 
\end{align}
where $\{\Gamma_y, \, \Gamma_m, \, \Gamma_W, \, \Gamma_{\mathrm{SS}}, \, \Gamma_h, \, \Gamma_{ws} \}$ are the top Yukawa, top helicity flipping, $W$-boson scattering, strong sphaleron, Higgs number violating and weak sphaleron rates for whose values we use \cite{Huet:1995sh, Fromme:2006cm, Moore:2010jd, Cline:2021dkf},
\begin{align}
\Gamma_{\mathrm{tot}}^i &= \frac{D_2^i}{D_0^i \, D_i} \quad , \quad \{ D_q, \, D_h \} = \{7.4/T, \, 20/T \} \,, \\[4pt]
\Gamma_{ws} &= 1.0 \times 10^{-6} \, T \quad , \quad \Gamma_{\mathrm{SS}} = 1.4 \times 10^{-3} \, T \quad , \quad \Gamma_{y} = 4.2 \times 10^{-3} \, T \,, \\[4pt]
\Gamma_{m} &= \frac{m_t^2}{63 T} \quad , \quad \Gamma_{h} = \frac{m_W^2}{50 T} \quad , \quad \Gamma_W = \Gamma_{\mathrm{tot}}^h \,.
\end{align} 
We note here that some dedicated baryogenesis studies of BSM models use an older estimate of the strong sphaleron rate $\Gamma_{\mathrm{SS}} = 4.9 \times 10^{-4} \, T$, and in comparison to the rate based on \cite{Moore:2010jd}, this increases the BAU by $\mathcal{O}(10\%)$ as left-handed top quarks are converted into other flavours and helicities less efficiently. We note here that whilst finalising this paper, the interaction rates given above were updated \cite{Barni:2025ifb}. 

\subsection{Baryogenesis dependence on $m_{a_1}$ and $s_{\theta}$} \label{BAU dependence}

First, we tackle the $\xi_c$ dependence, and, one way to increase the transition strength is to reduce the critical temperature. Rewriting \cref{crit temp approx} and $\lambda_a$,
\begin{align}
&\lambda_a v_s^2 = \lambda_{\beta} v^2 - m_{a_1}^2 - s_{\theta}^2 \, (m_{a_2}^2 - m_{a_1}^2) \label{lambda relation} \,, \\[5pt]
&T_c^2 = \frac{\sqrt{2 Z_1} \, \lambda_a v_s^2 - \sqrt{\lambda_a} \, Z_1 v^2}{\Pi_{33} \, \sqrt{2 Z_1} - 2 \, \Pi_{11} \, \sqrt{\lambda_a}} = \frac{a_1 \, \lambda_a - a_2 \sqrt{\lambda_a}}{a_3 \, \lambda_a - a_4 \sqrt{\lambda_a} + a_5} \,\, , \,\,\, a_i > 0 \label{crit temp approx 2} \,, 
\end{align}
where all parameters other than $\lambda_a$ are considered to be fixed because we want to find the dependence on the pseudoscalar mass $m_{a_1}$ and mixing angle $s_{\theta}$. 
If this is the case, it is simple to show that ${T_c}{(\lambda_a)}$ (\cref{crit temp approx 2}) is monotonically decreasing (within the domain of the function) if $ - a_5 \, a_1^2 + a_4 \, a_1 \, a_2 - a_3 \, a_2^2  > 0$. This inequality is satisfied for a wide range of model parameter values and requiring the opposite inequality appears to be strongly prevented by theoretical constraints discussed earlier. Thus, we should increase $\lambda_a$ in order to decrease $T_c$, which, through \cref{lambda relation}, is equivalent to decreasing $m_{a_1}$ and $s_{\theta}$ when $\theta \in [0, \pi/2]$. Noting \cref{extrema approx EW} and \cref{extrema approx CP}, decreasing $T_c$ further increases the transition strength by increasing the vevs of the EW and CPV minima, and thus, the strongest transitions are expected for small $m_{a_1}$ and $s_{\theta}$. \newline

Second, we see from the approximation \cref{top phase approx} that there are multiple ways to increase $\Delta \theta_{\mathrm{t}}$ and the first is increasing the CP-odd mixing angle $s_{\theta}$. One has to be careful here as we note that increasing $s_{\theta}$ actually decreases the vev of the CP-minimum $a(T_c) \big|_{\mathrm{CP}}$, potentially opposing the initial increase in $\theta_{\mathrm{t}}$. However, the latter effect is sub-leading in comparison to the former, and hence increasing $s_{\theta}$ is still beneficial for increasing the BAU. Second, decreasing $t_{\beta}$ is clearly beneficial as expected, however, the effect turns out to be less dramatic than \cref{top quark phase} suggests at first glance. Third, decreasing $m_{a_1}$ increases $\Delta \theta_{\mathrm{t}}$, and, it simultaneously increases $a(T_c) \big|_{\mathrm{CP}}$, further increasing $\Delta \theta_{\mathrm{t}}$. Lastly, one can vary the common mass scale $M$ and decreasing this scale should increase the $\Delta \theta_{\mathrm{t}}$, but, the effect is not so strong due to the assumption $M \approx m_{a_2}$. In summary, the largest changes in the top quark phase, $\Delta \theta_{\mathrm{t}}$, are expected for small $m_{a_1}$ and large $s_{\theta}$. \newline

The last important quantity in estimating the BAU is the barrier height as in conjunction with the transition strength, they form a rough estimate (via a thin-wall analysis) of the wall thickness, $L_w T_c \approx \xi_c\, T_c^2 / \sqrt{8 V_{\mathrm{b}}}$. The barrier height \cref{barrier approx} is approximately proportional to two factors. The first is the depth of the minima at the critical temperature which itself is proportional to the vev of the EW minimum ${\rho_1}(T_c) \big|_{\mathrm{EW}}$.  Increasing the barrier height can be achieved by increasing the said vev and this is equivalent to decreasing $T_c$, which, as we have argued, is equivalent to decreasing $m_{a_1}$ and $s_{\theta}$. However, the second factor in \cref{barrier approx} has the opposite behaviour. Therefore, it is difficult to determine a priori how the barrier height, and hence wall-thickness, will change with $m_{a_1}$ and $s_{\theta}$. As discussed previously, there is the difficultly of the the approximation not faithfully reproducing the barrier height along the MEP which is required for the wall-thickness estimate. We observe numerically that first, the analytic estimate does not reproduce the correct patterns, and second, in some regions of $(m_{a_1}, \, s_{\theta})$ space, the approximation strongly breaks down. 

Moving to the one-loop effective potential, ignoring mass-independent contributions, we expand \cref{therm formula},
\begin{align}
\label{asymptotic expansion}
V_T \supset & \sum_{i \in \mathrm{B}} n_i \left( \frac{m_i^2 T^2}{24} - \frac{T}{12\pi} \left( m_i^2 \right)^{3/2} - \frac{m_i^4}{64\pi^2} \log{\left( \frac{m_i^2}{a_{\mathrm{B}} T^2} \right)} \right) \nonumber \\[4pt]
&+ \sum_{i \in \mathrm{F}} n_i \left( \frac{m_i^2 T^2}{48} + \frac{m_i^4}{64\pi^2} \log{\left( \frac{m_i^2}{a_{\mathrm{F}} T^2} \right)} \right) + \mathcal{O} \left( \frac{m^6}{T^2} \right)\,, 
\end{align}
where $a_{\mathrm{B}} \approx 49.76$ and $a_{\mathrm{F}} \approx 3.11$. The first terms proportional to $m^2 T^2$ are nothing but the leading thermal corrections that are included in the Hartree approximation. The first new terms at loop-level are negative cubics, $m^3\,T$, stemming from bosonic particles. Due to degrees of freedom and masses, the major contributions to the thermal function arise from the top quark and gauge bosons whose masses are proportional to $\rho_1$ since they do not couple to the pseudoscalar $a$. Therefore, a much larger negative cubic contribution in $\rho_1$ is generated in comparison to $a$. Thus, in comparison to the CPV minimum, the EW minimum is deepened to a greater extent as well as both minima being stretched. The effect of the former is to increase the critical temperature and thus weaken the transition strength. Again, \cref{extrema approx EW,extrema approx CP} imply that increasing the critical temperature decrease both vevs, however, the generated cubic term for $\rho_1$ and $a$ counteract this and so the vevs do not change substantially. It is seen numerically across the parameter space investigated that using the one-loop potential often slightly increases both vevs. This also leads to the conclusion that $\Delta \theta_{\mathrm{t}}$ does not vary significantly from Hartree to one-loop since the only non-input value in \cref{top phase approx} is $a(T) \big|_{\mathrm{CP}}$. It is difficult to again comment on the wall-thickness but one does not expect large deviations from Hartree to one-loop, and we see from \cref{fig:crit key quants} that the wall-thickness does not change significantly. In conclusion, the main difference between the two approximations will be the transition strength, and, by the BAU estimate \cref{bau estimate}, lower values of BAU are expected.

\subsection{Gravitational waves} 
\label{grav waves}

To assess the GW signals arising from our benchmarks, we use \textit{peak-integrated sensitivity curves} (PISCs) \cite{Schmitz:2020syl} for LISA, DECIGO and BBO \cite{Yagi:2011wg, LISA:2017pwj}, and juxtapose them with peak frequency and signal amplitudes for all relevant benchmark points. Given that our benchmarks do not produce particularly strong transitions (see \cref{fig:bp1_velocity}), the GW spectrum will be dominated by the production of sound waves in the plasma and we ignore the contributions of turbulence and bubble collisions. The peak frequency and amplitude depend on four thermodynamic parameters of the phase transition $\{ \alpha, \, \beta/H_*, \,T_*, \, v_w , \, \kappa_s\}$, where $\alpha$ is the transition strength parameter\footnote{The definition of $\alpha$ used for the study of GWs is typically not the same as the parameter $\alpha_n$ used in this paper. The difference being $e-3p$ is replaced by $e-p/c_-^2$, where $c_-$ is the speed of sound in the broken phase. Since we do not observe large deviations from LTE, $c_-^2 \approx 1/3$ and the values of $\alpha$ and $\alpha_n$ differ by at most $\mathcal{O}(1\%)$.}, $\beta/H_*$ is the inverse duration of the transition in units of the Hubble time $H_*$, $T_*$ is the temperature at which the transition occurs, $v_w$ is the constant velocity of the expanding bubble wall, and $\kappa_s$ is the efficiency factor for the conversion of vacuum energy into sound waves (differing for non-runaway and runaway bubbles) \cite{Schmitz:2020syl},  
\begin{align}
&p = -V_{\mathrm{eff}} \,\, , \,\,\, e=\frac{dp}{dT}-p \,\, , \,\,\, w=e+p \,, \\[5pt]
&\alpha = \frac{1}{3w_{+}}\big[(e-3p)_+ - (e-3p)_{-} \big] \Big|_{T=T_*}\ \,\,,\\[5pt]
&\beta/H_* = T_* \, \frac{d S_3}{dT} \Big|_{T=T_*} \,\,, \\[5pt]
& \kappa_s^{\mathrm{non-runaway}} \simeq \begin{cases}
\frac{\alpha}{0.73 + 0.083\sqrt{\alpha} + \alpha} & ,\,\,v_w \geq v_w^\alpha \\[3pt]
\frac{6.9\,\alpha\,v_w^{6/5}}{1.36-0.037\sqrt{\alpha} + \alpha} & ,\,\,v_w < v_w^\alpha
\end{cases} \,\,\,, \,\,\, v_w^\alpha = \scriptstyle{\left[\frac{1.36-0.037\sqrt{\alpha} + \alpha}{6.9\left( 0.73 + 0.083\sqrt{\alpha} + \alpha\right)}\right]^{5/6}}\,,\\[8pt]
&\kappa_s^{\mathrm{runaway}} \simeq \frac{\alpha_\infty}{\alpha} \frac{\alpha_\infty}{0.73 + 0.083\sqrt{\alpha_\infty} + \alpha_\infty} \,\,\, , \,\,\, \alpha_\infty \simeq 4.9\times 10^{-3} \,\, \xi(T_*)^2 \,\,,
\end{align}
where $\pm$ denotes in-front of (CPV phase) and behind (EW phase) the bubble wall respectively, and $p,\,e$ and $w$ are the pressure, energy and enthalpy density. Since we do not observe strong supercooling, we take $T_* \approx T_n$. For the wall velocity, we take the the average of the upper and lower bound, i.e., $v_w = (v_w^{\scriptscriptstyle{\mathrm{ballistic}}}+v_w^{\scriptscriptstyle{\mathrm{LTE}}})/2$, unless the bubble runs away, in which case $v_w = 1$. The peak frequency $f_s^{\scriptscriptstyle{\mathrm{peak}}}$ and GW signal amplitude $h^2 \, \Omega_s^{\scriptscriptstyle{\mathrm{peak}}}$ produced from sound waves are \cite{Caprini:2015zlo},
\begin{align}
&f_s^{\scriptscriptstyle{\mathrm{peak}}} \simeq 1.9\times10^{-5} \, \mathrm{Hz} \, \left( \frac{g_*(T_*)}{100} \right)^{1/6} \, \left( \frac{T_*}{100\,\mathrm{GeV}} \right) \, \left( \frac{\beta/H_*}{v_w} \right) \,,\\[5pt]
&\Omega_s^{\scriptscriptstyle{\mathrm{peak}}} \simeq 2.65\times10^{-6}  \, \left( \frac{100}{g_*(T_*)} \right)^{1/3}  \, \left( \frac{v_w}{\beta/H_*} \right)  \, \left( \frac{\kappa_s \, \alpha}{1+\alpha} \right)^2  \,,
\end{align}
where $g_*(T_*)$ yields the d.o.f in the plasma for which we use the fit of \cite{Saikawa:2018rcs}.\footnote{Ref. \cite{Saikawa:2018rcs} provides the d.o.f for the SM. This is acceptable for our scenario as the only d.o.f that might contribute is the light pseudoscalar, all other BSM states are Boltzmann suppressed. Or, one can simply use $g_* = 106.75$ and the difference in predictions are minor.} Finally, the PISCs for sound waves and relevant GW observatories are \cite{Schmitz:2020syl},
\begin{align}
&\frac{h^2 \, \Omega_s^{\mathrm{LISA}}}{10^{-14}} \simeq 3.58\times10^{-3}\,x_s^{-4} + 3.26\times10^{-1}\,x_s^{-3} + 1.20\times10^{0}\,x_s^{-2} + 2.48\times10^{0}\,x_s^{-1} \nonumber  \\
&\phantom{h^2 \, \Omega_s^{\mathrm{LISA}} \simeq } + 2.85\times10^{-1}\,x_s^{1} + 1.81\times10^{-2}\,x_s^{2} + 1.50\times10^{-3}\,x_s^{3} \,\,,\\[5pt]
&\frac{h^2 \, \Omega_s^{\mathrm{DECIGO}}}{10^{-14}} \simeq 
3.82\times10^{-1}\,x_s^{-4} + 2.26\times10^{0}\,x_s^{-1.5} + 1.10\times10^{-3}\,x_s^{0} \nonumber \\ 
&\phantom{h^2 \, \Omega_s^{\mathrm{DECIGO}} \simeq } + 2.56\times10^{-6}\,x_s^{1} + 2.91\times10^{-8}\,x_s^{2} + 7.54\times10^{-12}\,x_s^{3}\,\,, \\[5pt]
&\frac{h^2 \, \Omega_s^{\mathrm{BBO}}}{10^{-14}} \simeq 
1.77\times10^{-1}\,x_s^{-4} + 1.06\times10^{0}\,x_s^{-1.5} + 1.35\times10^{-4}\,x_s^{0} \nonumber \\
&\phantom{h^2 \, \Omega_s^{\mathrm{BBO}} \simeq } + 2.23\times10^{-6}\,x_s^{1} + 1.29\times10^{-9}\,x_s^{2} + 2.99\times10^{-12}\,x_s^{3} \,\,,
\end{align}
where $x_s = \tfrac{f_s}{1\,\mathrm{mHz}}$.

\begin{figure}[h] 
\centering
\includegraphics[scale=0.57]{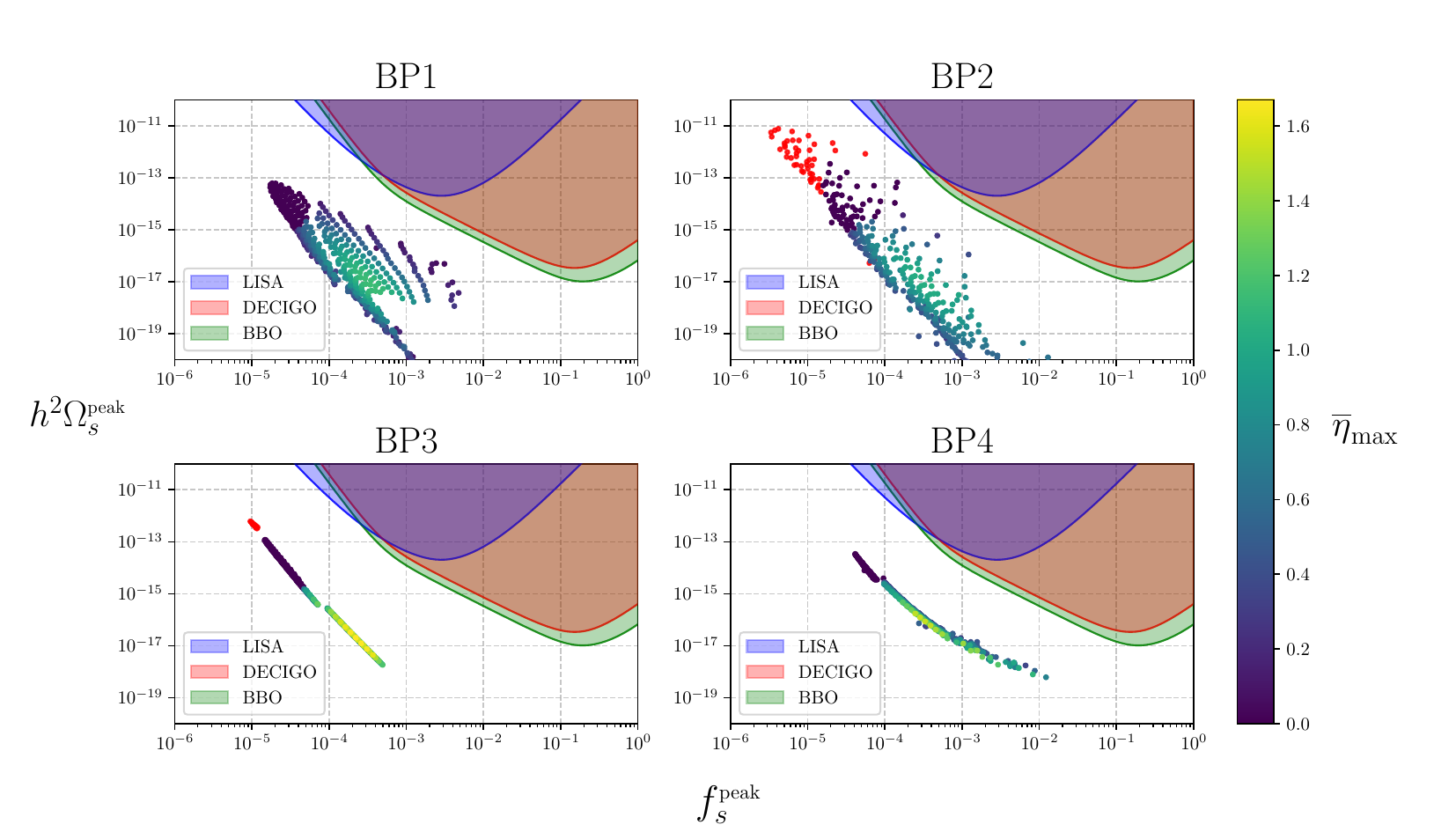}
\caption{\small Peak GW signal amplitude $\Omega_s^{\scriptscriptstyle\mathrm{peak}}$ against peak frequency $f_s^{\scriptscriptstyle\mathrm{peak}}$ produced by sound waves for all points that nucleate in our four benchmarks, contrasted with PISCs for LISA (blue), DECIGO (red) and BBO (green). Also shown is $\overline{\eta}_{\mathrm{max}}$ for each point and transitions that display runaway behaviour are shown in red.} 
\label{fig:grav_wave_signal}
\end{figure}

\bibliographystyle{JHEP}
\bibliography{CPlib}

\providecommand{\href}[2]{#2}\begingroup\raggedright\begin{thebibliography}{100}

\bibitem{Planck:2018vyg}
{\scshape Planck} collaboration, N.~Aghanim et~al., \emph{{Planck 2018 results.
  VI. Cosmological parameters}},
  \href{http://dx.doi.org/10.1051/0004-6361/201833910}{\emph{Astron.
  Astrophys.} {\bf 641} (2020) A6},
  [\href{https://arxiv.org/abs/1807.06209}{{\tt 1807.06209}}].

\bibitem{Sakharov:1967dj}
A.~D. Sakharov, \emph{{Violation of CP Invariance, C asymmetry, and baryon
  asymmetry of the universe}},
  \href{http://dx.doi.org/10.1070/PU1991v034n05ABEH002497}{\emph{Pisma Zh.
  Eksp. Teor. Fiz.} {\bf 5} (1967) 32--35}.

\bibitem{Cohen:1993nk}
A.~G. Cohen, D.~B. Kaplan and A.~E. Nelson, \emph{{Progress in electroweak
  baryogenesis}},
  \href{http://dx.doi.org/10.1146/annurev.ns.43.120193.000331}{\emph{Ann. Rev.
  Nucl. Part. Sci.} {\bf 43} (1993) 27--70},
  [\href{https://arxiv.org/abs/hep-ph/9302210}{{\tt hep-ph/9302210}}].

\bibitem{Trodden:1998ym}
M.~Trodden, \emph{{Electroweak baryogenesis}},
  \href{http://dx.doi.org/10.1103/RevModPhys.71.1463}{\emph{Rev. Mod. Phys.}
  {\bf 71} (1999) 1463--1500},
  [\href{https://arxiv.org/abs/hep-ph/9803479}{{\tt hep-ph/9803479}}].

\bibitem{Morrissey:2012db}
D.~E. Morrissey and M.~J. Ramsey-Musolf, \emph{{Electroweak baryogenesis}},
  \href{http://dx.doi.org/10.1088/1367-2630/14/12/125003}{\emph{New J. Phys.}
  {\bf 14} (2012) 125003}, [\href{https://arxiv.org/abs/1206.2942}{{\tt
  1206.2942}}].

\bibitem{Konstandin:2013caa}
T.~Konstandin, \emph{{Quantum Transport and Electroweak Baryogenesis}},
  \href{http://dx.doi.org/10.3367/UFNe.0183.201308a.0785}{\emph{Phys. Usp.}
  {\bf 56} (2013) 747--771}, [\href{https://arxiv.org/abs/1302.6713}{{\tt
  1302.6713}}].

\bibitem{Kuzmin:1985mm}
V.~A. Kuzmin, V.~A. Rubakov and M.~E. Shaposhnikov, \emph{{On the Anomalous
  Electroweak Baryon Number Nonconservation in the Early Universe}},
  \href{http://dx.doi.org/10.1016/0370-2693(85)91028-7}{\emph{Phys. Lett. B}
  {\bf 155} (1985) 36}.

\bibitem{DOnofrio:2015gop}
M.~D'Onofrio and K.~Rummukainen, \emph{{Standard model cross-over on the
  lattice}}, \href{http://dx.doi.org/10.1103/PhysRevD.93.025003}{\emph{Phys.
  Rev. D} {\bf 93} (2016) 025003},
  [\href{https://arxiv.org/abs/1508.07161}{{\tt 1508.07161}}].

\bibitem{Gavela:1993ts}
M.~B. Gavela, P.~Hernandez, J.~Orloff and O.~Pene, \emph{{Standard model CP
  violation and baryon asymmetry}},
  \href{http://dx.doi.org/10.1142/S0217732394000629}{\emph{Mod. Phys. Lett. A}
  {\bf 9} (1994) 795--810}, [\href{https://arxiv.org/abs/hep-ph/9312215}{{\tt
  hep-ph/9312215}}].

\bibitem{Gavela:1994ds}
M.~B. Gavela, M.~Lozano, J.~Orloff and O.~Pene, \emph{{Standard model CP
  violation and baryon asymmetry. Part 1: Zero temperature}},
  \href{http://dx.doi.org/10.1016/0550-3213(94)00409-9}{\emph{Nucl. Phys. B}
  {\bf 430} (1994) 345--381}, [\href{https://arxiv.org/abs/hep-ph/9406288}{{\tt
  hep-ph/9406288}}].

\bibitem{Gavela:1994dt}
M.~B. Gavela, P.~Hernandez, J.~Orloff, O.~Pene and C.~Quimbay, \emph{{Standard
  model CP violation and baryon asymmetry. Part 2: Finite temperature}},
  \href{http://dx.doi.org/10.1016/0550-3213(94)00410-2}{\emph{Nucl. Phys. B}
  {\bf 430} (1994) 382--426}, [\href{https://arxiv.org/abs/hep-ph/9406289}{{\tt
  hep-ph/9406289}}].

\bibitem{ACME:2018yjb}
{\scshape ACME} collaboration, V.~Andreev et~al., \emph{{Improved limit on the
  electric dipole moment of the electron}},
  \href{http://dx.doi.org/10.1038/s41586-018-0599-8}{\emph{Nature} {\bf 562}
  (2018) 355--360}.

\bibitem{nEDM:2020crw}
{\scshape nEDM} collaboration, C.~Abel et~al., \emph{{Measurement of the
  permanent electric dipole moment of the neutron}},
  \href{http://dx.doi.org/10.1103/PhysRevLett.124.081803}{\emph{Phys. Rev.
  Lett.} {\bf 124} (2020) 081803},
  [\href{https://arxiv.org/abs/2001.11966}{{\tt 2001.11966}}].

\bibitem{Roussy:2022cmp}
T.~S. Roussy et~al., \emph{{An improved bound on the electron{\textquoteright}s
  electric dipole moment}},
  \href{http://dx.doi.org/10.1126/science.adg4084}{\emph{Science} {\bf 381}
  (2023) adg4084}, [\href{https://arxiv.org/abs/2212.11841}{{\tt 2212.11841}}].

\bibitem{Griffith:2009zz}
W.~C. Griffith, M.~D. Swallows, T.~H. Loftus, M.~V. Romalis, B.~R. Heckel and
  E.~N. Fortson, \emph{{Improved Limit on the Permanent Electric Dipole Moment
  of Hg-199}},
  \href{http://dx.doi.org/10.1103/PhysRevLett.102.101601}{\emph{Phys. Rev.
  Lett.} {\bf 102} (2009) 101601}, [\href{https://arxiv.org/abs/0901.2328}{{\tt
  0901.2328}}].

\bibitem{Fromme:2006cm}
L.~Fromme, S.~J. Huber and M.~Seniuch, \emph{{Baryogenesis in the two-Higgs
  doublet model}},
  \href{http://dx.doi.org/10.1088/1126-6708/2006/11/038}{\emph{JHEP} {\bf 11}
  (2006) 038}, [\href{https://arxiv.org/abs/hep-ph/0605242}{{\tt
  hep-ph/0605242}}].

\bibitem{Dorsch:2016nrg}
G.~Dorsch, S.~Huber, T.~Konstandin and J.~No, \emph{{A Second Higgs Doublet in
  the Early Universe: Baryogenesis and Gravitational Waves}},
  \href{http://dx.doi.org/10.1088/1475-7516/2017/05/052}{\emph{JCAP} {\bf 05}
  (2017) 052}, [\href{https://arxiv.org/abs/1611.05874}{{\tt 1611.05874}}].

\bibitem{Basler:2021kgq}
P.~Basler, L.~Biermann, M.~M{\"u}hlleitner and J.~M{\"u}ller,
  \emph{{Electroweak baryogenesis in the CP-violating two-Higgs doublet
  model}}, \href{http://dx.doi.org/10.1140/epjc/s10052-023-11192-9}{\emph{Eur.
  Phys. J. C} {\bf 83} (2023) 57},
  [\href{https://arxiv.org/abs/2108.03580}{{\tt 2108.03580}}].

\bibitem{Huber:2022ndk}
S.~J. Huber, K.~Mimasu and J.~M. No, \emph{{Baryogenesis from transitional CP
  violation in the early Universe}},
  \href{http://dx.doi.org/10.1103/PhysRevD.107.075042}{\emph{Phys. Rev. D} {\bf
  107} (2023) 075042}, [\href{https://arxiv.org/abs/2208.10512}{{\tt
  2208.10512}}].

\bibitem{Cline:2020jre}
J.~M. Cline and K.~Kainulainen, \emph{{Electroweak baryogenesis at high bubble
  wall velocities}},
  \href{http://dx.doi.org/10.1103/PhysRevD.101.063525}{\emph{Phys. Rev. D} {\bf
  101} (2020) 063525}, [\href{https://arxiv.org/abs/2001.00568}{{\tt
  2001.00568}}].

\bibitem{Paschos:1976ay}
E.~A. Paschos, \emph{{Diagonal Neutral Currents}},
  \href{http://dx.doi.org/10.1103/PhysRevD.15.1966}{\emph{Phys. Rev. D} {\bf
  15} (1977) 1966}.

\bibitem{Glashow:1976nt}
S.~L. Glashow and S.~Weinberg, \emph{{Natural Conservation Laws for Neutral
  Currents}}, \href{http://dx.doi.org/10.1103/PhysRevD.15.1958}{\emph{Phys.
  Rev. D} {\bf 15} (1977) 1958}.

\bibitem{Battye:2020jeu}
R.~A. Battye, A.~Pilaftsis and D.~G. Viatic, \emph{{Domain wall constraints on
  two-Higgs-doublet models with $Z_2$ symmetry}},
  \href{http://dx.doi.org/10.1103/PhysRevD.102.123536}{\emph{Phys. Rev. D} {\bf
  102} (2020) 123536}, [\href{https://arxiv.org/abs/2010.09840}{{\tt
  2010.09840}}].

\bibitem{Gunion:2002zf}
J.~F. Gunion and H.~E. Haber, \emph{{The CP conserving two Higgs doublet model:
  The Approach to the decoupling limit}},
  \href{http://dx.doi.org/10.1103/PhysRevD.67.075019}{\emph{Phys. Rev. D} {\bf
  67} (2003) 075019}, [\href{https://arxiv.org/abs/hep-ph/0207010}{{\tt
  hep-ph/0207010}}].

\bibitem{ATLAS:2022vkf}
{\scshape ATLAS} collaboration, G.~Aad et~al., \emph{{A detailed map of Higgs
  boson interactions by the ATLAS experiment ten years after the discovery}},
  \href{http://dx.doi.org/10.1038/s41586-022-04893-w}{\emph{Nature} {\bf 607}
  (2022) 52--59}, [\href{https://arxiv.org/abs/2207.00092}{{\tt 2207.00092}}].

\bibitem{Branco:2011iw}
G.~C. Branco, P.~M. Ferreira, L.~Lavoura, M.~N. Rebelo, M.~Sher and J.~P.
  Silva, \emph{{Theory and phenomenology of two-Higgs-doublet models}},
  \href{http://dx.doi.org/10.1016/j.physrep.2012.02.002}{\emph{Phys. Rept.}
  {\bf 516} (2012) 1--102}, [\href{https://arxiv.org/abs/1106.0034}{{\tt
  1106.0034}}].

\bibitem{Weinberg:1973am}
E.~J. Weinberg, \emph{{Radiative corrections as the origin of spontaneous
  symmetry breaking}}.
\newblock PhD thesis, Harvard U., 1973.
\newblock \href{https://arxiv.org/abs/hep-th/0507214}{{\tt hep-th/0507214}}.

\bibitem{Bernon:2017jgv}
J.~Bernon, L.~Bian and Y.~Jiang, \emph{{A new insight into the phase transition
  in the early Universe with two Higgs doublets}},
  \href{http://dx.doi.org/10.1007/JHEP05(2018)151}{\emph{JHEP} {\bf 05} (2018)
  151}, [\href{https://arxiv.org/abs/1712.08430}{{\tt 1712.08430}}].

\bibitem{Dolan:1973qd}
L.~Dolan and R.~Jackiw, \emph{{Symmetry Behavior at Finite Temperature}},
  \href{http://dx.doi.org/10.1103/PhysRevD.9.3320}{\emph{Phys. Rev. D} {\bf 9}
  (1974) 3320--3341}.

\bibitem{Linde:1978px}
A.~D. Linde, \emph{{Phase Transitions in Gauge Theories and Cosmology}},
  \href{http://dx.doi.org/10.1088/0034-4885/42/3/001}{\emph{Rept. Prog. Phys.}
  {\bf 42} (1979) 389}.

\bibitem{Gross:1980br}
D.~J. Gross, R.~D. Pisarski and L.~G. Yaffe, \emph{{QCD and Instantons at
  Finite Temperature}},
  \href{http://dx.doi.org/10.1103/RevModPhys.53.43}{\emph{Rev. Mod. Phys.} {\bf
  53} (1981) 43}.

\bibitem{Arnold:1992rz}
P.~B. Arnold and O.~Espinosa, \emph{{The Effective potential and first order
  phase transitions: Beyond leading-order}},
  \href{http://dx.doi.org/10.1103/PhysRevD.47.3546}{\emph{Phys. Rev. D} {\bf
  47} (1993) 3546}, [\href{https://arxiv.org/abs/hep-ph/9212235}{{\tt
  hep-ph/9212235}}].

\bibitem{Espinosa:1992gq}
J.~R. Espinosa, M.~Quiros and F.~Zwirner, \emph{{On the phase transition in the
  scalar theory}},
  \href{http://dx.doi.org/10.1016/0370-2693(92)90129-R}{\emph{Phys. Lett. B}
  {\bf 291} (1992) 115--124}, [\href{https://arxiv.org/abs/hep-ph/9206227}{{\tt
  hep-ph/9206227}}].

\bibitem{Parwani:1991gq}
R.~R. Parwani, \emph{{Resummation in a hot scalar field theory}},
  \href{http://dx.doi.org/10.1103/PhysRevD.45.4695}{\emph{Phys. Rev. D} {\bf
  45} (1992) 4695}, [\href{https://arxiv.org/abs/hep-ph/9204216}{{\tt
  hep-ph/9204216}}].

\bibitem{Dine:1992wr}
M.~Dine, R.~G. Leigh, P.~Y. Huet, A.~D. Linde and D.~A. Linde, \emph{{Towards
  the theory of the electroweak phase transition}},
  \href{http://dx.doi.org/10.1103/PhysRevD.46.550}{\emph{Phys. Rev. D} {\bf 46}
  (1992) 550--571}, [\href{https://arxiv.org/abs/hep-ph/9203203}{{\tt
  hep-ph/9203203}}].

\bibitem{Boyd:1993tz}
C.~G. Boyd, D.~E. Brahm and S.~D.~H. Hsu, \emph{{Resummation methods at finite
  temperature: The Tadpole way}},
  \href{http://dx.doi.org/10.1103/PhysRevD.48.4963}{\emph{Phys. Rev. D} {\bf
  48} (1993) 4963--4973}, [\href{https://arxiv.org/abs/hep-ph/9304254}{{\tt
  hep-ph/9304254}}].

\bibitem{Andersen:2004fp}
J.~O. Andersen and M.~Strickland, \emph{{Resummation in hot field theories}},
  \href{http://dx.doi.org/10.1016/j.aop.2004.09.017}{\emph{Annals Phys.} {\bf
  317} (2005) 281--353}, [\href{https://arxiv.org/abs/hep-ph/0404164}{{\tt
  hep-ph/0404164}}].

\bibitem{Curtin:2016urg}
D.~Curtin, P.~Meade and H.~Ramani, \emph{{Thermal Resummation and Phase
  Transitions}},
  \href{http://dx.doi.org/10.1140/epjc/s10052-018-6268-0}{\emph{Eur. Phys. J.
  C} {\bf 78} (2018) 787}, [\href{https://arxiv.org/abs/1612.00466}{{\tt
  1612.00466}}].

\bibitem{Curtin:2022ovx}
D.~Curtin, J.~Roy and G.~White, \emph{{Gravitational waves and tadpole
  resummation: Efficient and easy convergence of finite temperature QFT}},
  \href{http://dx.doi.org/10.1103/PhysRevD.109.116001}{\emph{Phys. Rev. D} {\bf
  109} (2024) 116001}, [\href{https://arxiv.org/abs/2211.08218}{{\tt
  2211.08218}}].

\bibitem{Bahl:2024ykv}
H.~Bahl, M.~Carena, A.~Ireland and C.~E.~M. Wagner, \emph{{Improved thermal
  resummation for multi-field potentials}},
  \href{http://dx.doi.org/10.1007/JHEP09(2024)153}{\emph{JHEP} {\bf 09} (2024)
  153}, [\href{https://arxiv.org/abs/2404.12439}{{\tt 2404.12439}}].

\bibitem{Bittar:2025lcr}
P.~Bittar, S.~Roy and C.~E.~M. Wagner, \emph{{Self Consistent Thermal
  Resummation: A Case Study of the Phase Transition in 2HDM}},
  \href{https://arxiv.org/abs/2504.02024}{{\tt 2504.02024}}.

\bibitem{Weinberg:1987vp}
E.~J. Weinberg and A.-q. Wu, \emph{{UNDERSTANDING COMPLEX PERTURBATIVE
  EFFECTIVE POTENTIALS}},
  \href{http://dx.doi.org/10.1103/PhysRevD.36.2474}{\emph{Phys. Rev. D} {\bf
  36} (1987) 2474}.

\bibitem{Gould:2021oba}
O.~Gould and T.~V.~I. Tenkanen, \emph{{On the perturbative expansion at high
  temperature and implications for cosmological phase transitions}},
  \href{http://dx.doi.org/10.1007/JHEP06(2021)069}{\emph{JHEP} {\bf 06} (2021)
  069}, [\href{https://arxiv.org/abs/2104.04399}{{\tt 2104.04399}}].

\bibitem{Laine:2000rm}
M.~Laine and K.~Rummukainen, \emph{{Two Higgs doublet dynamics at the
  electroweak phase transition: A Nonperturbative study}},
  \href{http://dx.doi.org/10.1016/S0550-3213(00)00736-7}{\emph{Nucl. Phys. B}
  {\bf 597} (2001) 23--69}, [\href{https://arxiv.org/abs/hep-lat/0009025}{{\tt
  hep-lat/0009025}}].

\bibitem{Laine:2017hdk}
M.~Laine, M.~Meyer and G.~Nardini, \emph{{Thermal phase transition with full
  2-loop effective potential}},
  \href{http://dx.doi.org/10.1016/j.nuclphysb.2017.04.023}{\emph{Nucl. Phys. B}
  {\bf 920} (2017) 565--600}, [\href{https://arxiv.org/abs/1702.07479}{{\tt
  1702.07479}}].

\bibitem{Kainulainen:2019kyp}
K.~Kainulainen, V.~Keus, L.~Niemi, K.~Rummukainen, T.~V.~I. Tenkanen and
  V.~Vaskonen, \emph{{On the validity of perturbative studies of the
  electroweak phase transition in the Two Higgs Doublet model}},
  \href{http://dx.doi.org/10.1007/JHEP06(2019)075}{\emph{JHEP} {\bf 06} (2019)
  075}, [\href{https://arxiv.org/abs/1904.01329}{{\tt 1904.01329}}].

\bibitem{Niemi:2020hto}
L.~Niemi, M.~J. Ramsey-Musolf, T.~V.~I. Tenkanen and D.~J. Weir,
  \emph{{Thermodynamics of a Two-Step Electroweak Phase Transition}},
  \href{http://dx.doi.org/10.1103/PhysRevLett.126.171802}{\emph{Phys. Rev.
  Lett.} {\bf 126} (2021) 171802},
  [\href{https://arxiv.org/abs/2005.11332}{{\tt 2005.11332}}].

\bibitem{Niemi:2021qvp}
L.~Niemi, P.~Schicho and T.~V.~I. Tenkanen, \emph{{Singlet-assisted electroweak
  phase transition at two loops}},
  \href{http://dx.doi.org/10.1103/PhysRevD.103.115035}{\emph{Phys. Rev. D} {\bf
  103} (2021) 115035}, [\href{https://arxiv.org/abs/2103.07467}{{\tt
  2103.07467}}].

\bibitem{Ramsey-Musolf:2024ykk}
M.~J. Ramsey-Musolf, T.~V.~I. Tenkanen and V.~Q. Tran, \emph{{Refining
  Gravitational Wave and Collider Physics Dialogue via Singlet Scalar
  Extension}},  \href{https://arxiv.org/abs/2409.17554}{{\tt 2409.17554}}.

\bibitem{Niemi:2024axp}
L.~Niemi, M.~J. Ramsey-Musolf and G.~Xia, \emph{{Nonperturbative study of the
  electroweak phase transition in the real scalar singlet extended standard
  model}}, \href{http://dx.doi.org/10.1103/PhysRevD.110.115016}{\emph{Phys.
  Rev. D} {\bf 110} (2024) 115016},
  [\href{https://arxiv.org/abs/2405.01191}{{\tt 2405.01191}}].

\bibitem{Niemi:2024vzw}
L.~Niemi and T.~V.~I. Tenkanen, \emph{{Investigating two-loop effects for
  first-order electroweak phase transitions}},
  \href{http://dx.doi.org/10.1103/PhysRevD.111.075034}{\emph{Phys. Rev. D} {\bf
  111} (2025) 075034}, [\href{https://arxiv.org/abs/2408.15912}{{\tt
  2408.15912}}].

\bibitem{Coleman:1977py}
S.~R. Coleman, \emph{{The Fate of the False Vacuum. 1. Semiclassical Theory}},
  \href{http://dx.doi.org/10.1103/PhysRevD.16.1248}{\emph{Phys. Rev. D} {\bf
  15} (1977) 2929--2936}.

\bibitem{Callan:1977pt}
C.~G. Callan, Jr. and S.~R. Coleman, \emph{{The Fate of the False Vacuum. 2.
  First Quantum Corrections}},
  \href{http://dx.doi.org/10.1103/PhysRevD.16.1762}{\emph{Phys. Rev. D} {\bf
  16} (1977) 1762--1768}.

\bibitem{Linde:1980tt}
A.~D. Linde, \emph{{Fate of the False Vacuum at Finite Temperature: Theory and
  Applications}},
  \href{http://dx.doi.org/10.1016/0370-2693(81)90281-1}{\emph{Phys. Lett. B}
  {\bf 100} (1981) 37--40}.

\bibitem{Anderson:1991zb}
G.~W. Anderson and L.~J. Hall, \emph{{The Electroweak phase transition and
  baryogenesis}}, \href{http://dx.doi.org/10.1103/PhysRevD.45.2685}{\emph{Phys.
  Rev. D} {\bf 45} (1992) 2685--2698}.

\bibitem{Cline:1999wi}
J.~M. Cline, G.~D. Moore and G.~Servant, \emph{{Was the electroweak phase
  transition preceded by a color broken phase?}},
  \href{http://dx.doi.org/10.1103/PhysRevD.60.105035}{\emph{Phys. Rev. D} {\bf
  60} (1999) 105035}, [\href{https://arxiv.org/abs/hep-ph/9902220}{{\tt
  hep-ph/9902220}}].

\bibitem{Baum:2020vfl}
S.~Baum, M.~Carena, N.~R. Shah, C.~E.~M. Wagner and Y.~Wang, \emph{{Nucleation
  is more than critical}: {A case study of the electroweak phase transition in
  the NMSSM}}, \href{http://dx.doi.org/10.1007/JHEP03(2021)055}{\emph{JHEP}
  {\bf 03} (2021) 055}, [\href{https://arxiv.org/abs/2009.10743}{{\tt
  2009.10743}}].

\bibitem{Biekotter:2021ysx}
T.~Biek\"otter, S.~Heinemeyer, J.~M. No, M.~O. Olea and G.~Weiglein,
  \emph{{Fate of electroweak symmetry in the early Universe: Non-restoration
  and trapped vacua in the N2HDM}},
  \href{http://dx.doi.org/10.1088/1475-7516/2021/06/018}{\emph{JCAP} {\bf 06}
  (2021) 018}, [\href{https://arxiv.org/abs/2103.12707}{{\tt 2103.12707}}].

\bibitem{Goncalves:2021egx}
D.~Gon{\c{c}}alves, A.~Kaladharan and Y.~Wu, \emph{{Electroweak phase
  transition in the 2HDM: Collider and gravitational wave complementarity}},
  \href{http://dx.doi.org/10.1103/PhysRevD.105.095041}{\emph{Phys. Rev. D} {\bf
  105} (2022) 095041}, [\href{https://arxiv.org/abs/2108.05356}{{\tt
  2108.05356}}].

\bibitem{Biekotter:2022kgf}
T.~Biek{\"o}tter, S.~Heinemeyer, J.~M. No, M.~O. Olea-Romacho and G.~Weiglein,
  \emph{{The trap in the early Universe: impact on the interplay between
  gravitational waves and LHC physics in the 2HDM}},
  \href{http://dx.doi.org/10.1088/1475-7516/2023/03/031}{\emph{JCAP} {\bf 03}
  (2023) 031}, [\href{https://arxiv.org/abs/2208.14466}{{\tt 2208.14466}}].

\bibitem{Athron:2022mmm}
P.~Athron, C.~Bal{\'a}zs and L.~Morris, \emph{{Supercool subtleties of
  cosmological phase transitions}},
  \href{http://dx.doi.org/10.1088/1475-7516/2023/03/006}{\emph{JCAP} {\bf 03}
  (2023) 006}, [\href{https://arxiv.org/abs/2212.07559}{{\tt 2212.07559}}].

\bibitem{Wainwright:2011kj}
C.~L. Wainwright, \emph{{CosmoTransitions: Computing Cosmological Phase
  Transition Temperatures and Bubble Profiles with Multiple Fields}},
  \href{http://dx.doi.org/10.1016/j.cpc.2012.04.004}{\emph{Comput. Phys.
  Commun.} {\bf 183} (2012) 2006--2013},
  [\href{https://arxiv.org/abs/1109.4189}{{\tt 1109.4189}}].

\bibitem{Athron:2019nbd}
P.~Athron, C.~Bal\'azs, M.~Bardsley, A.~Fowlie, D.~Harries and G.~White,
  \emph{{BubbleProfiler: finding the field profile and action for cosmological
  phase transitions}},
  \href{http://dx.doi.org/10.1016/j.cpc.2019.05.017}{\emph{Comput. Phys.
  Commun.} {\bf 244} (2019) 448--468},
  [\href{https://arxiv.org/abs/1901.03714}{{\tt 1901.03714}}].

\bibitem{Guada:2020xnz}
V.~Guada, M.~Nemev{\v{s}}ek and M.~Pintar, \emph{{FindBounce: Package for
  multi-field bounce actions}},
  \href{http://dx.doi.org/10.1016/j.cpc.2020.107480}{\emph{Comput. Phys.
  Commun.} {\bf 256} (2020) 107480},
  [\href{https://arxiv.org/abs/2002.00881}{{\tt 2002.00881}}].

\bibitem{Basler:2024aaf}
P.~Basler, L.~Biermann, M.~M{\"u}hlleitner, J.~M{\"u}ller, R.~Santos and
  J.~Viana, \emph{{BSMPT v3 A Tool for Phase Transitions and Primordial
  Gravitational Waves in Extended Higgs Sectors}},
  \href{https://arxiv.org/abs/2404.19037}{{\tt 2404.19037}}.

\bibitem{Athron:2024xrh}
P.~Athron, C.~Balazs, A.~Fowlie, L.~Morris, W.~Searle, Y.~Xiao et~al.,
  \emph{{PhaseTracer2: from the effective potential to gravitational waves}},
  \href{http://dx.doi.org/10.1140/epjc/s10052-025-14258-y}{\emph{Eur. Phys. J.
  C} {\bf 85} (2025) 559}, [\href{https://arxiv.org/abs/2412.04881}{{\tt
  2412.04881}}].

\bibitem{Hua:2025fap}
B.~Hua and J.~Zhu, \emph{{VacuumTunneling: A package to solve bounce equation
  with renormalization factor}},  \href{https://arxiv.org/abs/2501.15236}{{\tt
  2501.15236}}.

\bibitem{Quiros:1999jp}
M.~Quiros, \emph{{Finite temperature field theory and phase transitions}},  in
  \emph{{ICTP Summer School in High-Energy Physics and Cosmology}}, 1, 1999.
\newblock \href{https://arxiv.org/abs/hep-ph/9901312}{{\tt hep-ph/9901312}}.

\bibitem{Khlebnikov:1988sr}
S.~Y. Khlebnikov and M.~E. Shaposhnikov, \emph{{The Statistical Theory of
  Anomalous Fermion Number Nonconservation}},
  \href{http://dx.doi.org/10.1016/0550-3213(88)90133-2}{\emph{Nucl. Phys. B}
  {\bf 308} (1988) 885--912}.

\bibitem{Rubakov:1996vz}
V.~A. Rubakov and M.~E. Shaposhnikov, \emph{{Electroweak baryon number
  nonconservation in the early universe and in high-energy collisions}},
  \href{http://dx.doi.org/10.1070/PU1996v039n05ABEH000145}{\emph{Usp. Fiz.
  Nauk} {\bf 166} (1996) 493--537},
  [\href{https://arxiv.org/abs/hep-ph/9603208}{{\tt hep-ph/9603208}}].

\bibitem{Patel:2011th}
H.~H. Patel and M.~J. Ramsey-Musolf, \emph{{Baryon Washout, Electroweak Phase
  Transition, and Perturbation Theory}},
  \href{http://dx.doi.org/10.1007/JHEP07(2011)029}{\emph{JHEP} {\bf 07} (2011)
  029}, [\href{https://arxiv.org/abs/1101.4665}{{\tt 1101.4665}}].

\bibitem{Joyce:1994zt}
M.~Joyce, T.~Prokopec and N.~Turok, \emph{{Nonlocal electroweak baryogenesis.
  Part 2: The Classical regime}},
  \href{http://dx.doi.org/10.1103/PhysRevD.53.2958}{\emph{Phys. Rev. D} {\bf
  53} (1996) 2958--2980}, [\href{https://arxiv.org/abs/hep-ph/9410282}{{\tt
  hep-ph/9410282}}].

\bibitem{Cline:2000nw}
J.~M. Cline, M.~Joyce and K.~Kainulainen, \emph{{Supersymmetric electroweak
  baryogenesis}},
  \href{http://dx.doi.org/10.1088/1126-6708/2000/07/018}{\emph{JHEP} {\bf 07}
  (2000) 018}, [\href{https://arxiv.org/abs/hep-ph/0006119}{{\tt
  hep-ph/0006119}}].

\bibitem{Kainulainen:2002th}
K.~Kainulainen, T.~Prokopec, M.~G. Schmidt and S.~Weinstock,
  \emph{{Semiclassical force for electroweak baryogenesis: Three-dimensional
  derivation}}, \href{http://dx.doi.org/10.1103/PhysRevD.66.043502}{\emph{Phys.
  Rev. D} {\bf 66} (2002) 043502},
  [\href{https://arxiv.org/abs/hep-ph/0202177}{{\tt hep-ph/0202177}}].

\bibitem{Prokopec:2004ic}
T.~Prokopec, M.~G. Schmidt and S.~Weinstock, \emph{{Transport equations for
  chiral fermions to order h-bar and electroweak baryogenesis. Part II}},
  \href{http://dx.doi.org/10.1016/j.aop.2004.06.001}{\emph{Annals Phys.} {\bf
  314} (2004) 267--320}, [\href{https://arxiv.org/abs/hep-ph/0406140}{{\tt
  hep-ph/0406140}}].

\bibitem{Fromme:2006wx}
L.~Fromme and S.~J. Huber, \emph{{Top transport in electroweak baryogenesis}},
  \href{http://dx.doi.org/10.1088/1126-6708/2007/03/049}{\emph{JHEP} {\bf 03}
  (2007) 049}, [\href{https://arxiv.org/abs/hep-ph/0604159}{{\tt
  hep-ph/0604159}}].

\bibitem{Cline:2011mm}
J.~M. Cline, K.~Kainulainen and M.~Trott, \emph{{Electroweak Baryogenesis in
  Two Higgs Doublet Models and B meson anomalies}},
  \href{http://dx.doi.org/10.1007/JHEP11(2011)089}{\emph{JHEP} {\bf 11} (2011)
  089}, [\href{https://arxiv.org/abs/1107.3559}{{\tt 1107.3559}}].

\bibitem{Espinosa:2010hh}
J.~R. Espinosa, T.~Konstandin, J.~M. No and G.~Servant, \emph{{Energy Budget of
  Cosmological First-order Phase Transitions}},
  \href{http://dx.doi.org/10.1088/1475-7516/2010/06/028}{\emph{JCAP} {\bf 06}
  (2010) 028}, [\href{https://arxiv.org/abs/1004.4187}{{\tt 1004.4187}}].

\bibitem{Konstandin:2014zta}
T.~Konstandin, G.~Nardini and I.~Rues, \emph{{From Boltzmann equations to
  steady wall velocities}},
  \href{http://dx.doi.org/10.1088/1475-7516/2014/09/028}{\emph{JCAP} {\bf 09}
  (2014) 028}, [\href{https://arxiv.org/abs/1407.3132}{{\tt 1407.3132}}].

\bibitem{Dorsch:2018pat}
G.~C. Dorsch, S.~J. Huber and T.~Konstandin, \emph{{Bubble wall velocities in
  the Standard Model and beyond}},
  \href{http://dx.doi.org/10.1088/1475-7516/2018/12/034}{\emph{JCAP} {\bf 12}
  (2018) 034}, [\href{https://arxiv.org/abs/1809.04907}{{\tt 1809.04907}}].

\bibitem{Laurent:2020gpg}
B.~Laurent and J.~M. Cline, \emph{{Fluid equations for fast-moving electroweak
  bubble walls}},
  \href{http://dx.doi.org/10.1103/PhysRevD.102.063516}{\emph{Phys. Rev. D} {\bf
  102} (2020) 063516}, [\href{https://arxiv.org/abs/2007.10935}{{\tt
  2007.10935}}].

\bibitem{Friedlander:2020tnq}
A.~Friedlander, I.~Banta, J.~M. Cline and D.~Tucker-Smith, \emph{{Wall speed
  and shape in singlet-assisted strong electroweak phase transitions}},
  \href{http://dx.doi.org/10.1103/PhysRevD.103.055020}{\emph{Phys. Rev. D} {\bf
  103} (2021) 055020}, [\href{https://arxiv.org/abs/2009.14295}{{\tt
  2009.14295}}].

\bibitem{Dorsch:2021ubz}
G.~C. Dorsch, S.~J. Huber and T.~Konstandin, \emph{{On the wall velocity
  dependence of electroweak baryogenesis}},
  \href{http://dx.doi.org/10.1088/1475-7516/2021/08/020}{\emph{JCAP} {\bf 08}
  (2021) 020}, [\href{https://arxiv.org/abs/2106.06547}{{\tt 2106.06547}}].

\bibitem{DeCurtis:2022hlx}
S.~De~Curtis, L.~D. Rose, A.~Guiggiani, {\'A}.~G. Muyor and G.~Panico,
  \emph{{Bubble wall dynamics at the electroweak phase transition}},
  \href{http://dx.doi.org/10.1007/JHEP03(2022)163}{\emph{JHEP} {\bf 03} (2022)
  163}, [\href{https://arxiv.org/abs/2201.08220}{{\tt 2201.08220}}].

\bibitem{Dorsch:2021nje}
G.~C. Dorsch, S.~J. Huber and T.~Konstandin, \emph{{A sonic boom in bubble wall
  friction}},
  \href{http://dx.doi.org/10.1088/1475-7516/2022/04/010}{\emph{JCAP} {\bf 04}
  (2022) 010}, [\href{https://arxiv.org/abs/2112.12548}{{\tt 2112.12548}}].

\bibitem{Laurent:2022jrs}
B.~Laurent and J.~M. Cline, \emph{{First principles determination of bubble
  wall velocity}},
  \href{http://dx.doi.org/10.1103/PhysRevD.106.023501}{\emph{Phys. Rev. D} {\bf
  106} (2022) 023501}, [\href{https://arxiv.org/abs/2204.13120}{{\tt
  2204.13120}}].

\bibitem{DeCurtis:2023hil}
S.~De~Curtis, L.~Delle~Rose, A.~Guiggiani, {\'A}.~Gil~Muyor and G.~Panico,
  \emph{{Collision integrals for cosmological phase transitions}},
  \href{http://dx.doi.org/10.1007/JHEP05(2023)194}{\emph{JHEP} {\bf 05} (2023)
  194}, [\href{https://arxiv.org/abs/2303.05846}{{\tt 2303.05846}}].

\bibitem{Dorsch:2023tss}
G.~C. Dorsch and D.~A. Pinto, \emph{{Bubble wall velocities with an extended
  fluid Ansatz}},
  \href{http://dx.doi.org/10.1088/1475-7516/2024/04/027}{\emph{JCAP} {\bf 04}
  (2024) 027}, [\href{https://arxiv.org/abs/2312.02354}{{\tt 2312.02354}}].

\bibitem{DeCurtis:2024hvh}
S.~De~Curtis, L.~Delle~Rose, A.~Guiggiani, {\'A}.~Gil~Muyor and G.~Panico,
  \emph{{Non-linearities in cosmological bubble wall dynamics}},
  \href{http://dx.doi.org/10.1007/JHEP05(2024)009}{\emph{JHEP} {\bf 05} (2024)
  009}, [\href{https://arxiv.org/abs/2401.13522}{{\tt 2401.13522}}].

\bibitem{Ai:2025bjw}
W.-Y. Ai, M.~Carosi, B.~Garbrecht, C.~Tamarit and M.~Vanvlasselaer,
  \emph{{Bubble wall dynamics from nonequilibrium quantum field theory}},
  \href{https://arxiv.org/abs/2504.13725}{{\tt 2504.13725}}.

\bibitem{Branchina:2025jou}
C.~Branchina, A.~Conaci, S.~De~Curtis and L.~Delle~Rose, \emph{{Electroweak
  Phase Transition and Bubble Wall Velocity in Local Thermal Equilibrium}},
  \href{https://arxiv.org/abs/2504.21213}{{\tt 2504.21213}}.

\bibitem{Ramsey-Musolf:2025jyk}
M.~J. Ramsey-Musolf and J.~Zhu, \emph{{Bubble wall velocity from Kadanoff-Baym
  equations: fluid dynamics and microscopic interactions}},
  \href{https://arxiv.org/abs/2504.13724}{{\tt 2504.13724}}.

\bibitem{Ai:2021kak}
W.-Y. Ai, B.~Garbrecht and C.~Tamarit, \emph{{Bubble wall velocities in local
  equilibrium}},
  \href{http://dx.doi.org/10.1088/1475-7516/2022/03/015}{\emph{JCAP} {\bf 03}
  (2022) 015}, [\href{https://arxiv.org/abs/2109.13710}{{\tt 2109.13710}}].

\bibitem{Ai:2023see}
W.-Y. Ai, B.~Laurent and J.~van~de Vis, \emph{{Model-independent bubble wall
  velocities in local thermal equilibrium}},
  \href{http://dx.doi.org/10.1088/1475-7516/2023/07/002}{\emph{JCAP} {\bf 07}
  (2023) 002}, [\href{https://arxiv.org/abs/2303.10171}{{\tt 2303.10171}}].

\bibitem{Ai:2024btx}
W.-Y. Ai, B.~Laurent and J.~van~de Vis, \emph{{Bounds on the bubble wall
  velocity}}, \href{http://dx.doi.org/10.1007/JHEP02(2025)119}{\emph{JHEP} {\bf
  02} (2025) 119}, [\href{https://arxiv.org/abs/2411.13641}{{\tt 2411.13641}}].

\bibitem{Steinhardt:1981ct}
P.~J. Steinhardt, \emph{{Relativistic Detonation Waves and Bubble Growth in
  False Vacuum Decay}},
  \href{http://dx.doi.org/10.1103/PhysRevD.25.2074}{\emph{Phys. Rev. D} {\bf
  25} (1982) 2074}.

\bibitem{Krajewski:2024gma}
T.~Krajewski, M.~Lewicki and M.~Zych, \emph{{Bubble-wall velocity in local
  thermal equilibrium: hydrodynamical simulations vs analytical treatment}},
  \href{http://dx.doi.org/10.1007/JHEP05(2024)011}{\emph{JHEP} {\bf 05} (2024)
  011}, [\href{https://arxiv.org/abs/2402.15408}{{\tt 2402.15408}}].

\bibitem{Konstandin:2010dm}
T.~Konstandin and J.~M. No, \emph{{Hydrodynamic obstruction to bubble
  expansion}},
  \href{http://dx.doi.org/10.1088/1475-7516/2011/02/008}{\emph{JCAP} {\bf 02}
  (2011) 008}, [\href{https://arxiv.org/abs/1011.3735}{{\tt 1011.3735}}].

\bibitem{Eriksson:2025owh}
M.~Eriksson and M.~Laine, \emph{{Entropy production at electroweak bubble walls
  from scalar field fluctuations}},
  \href{http://dx.doi.org/10.1088/1475-7516/2025/09/027}{\emph{JCAP} {\bf 09}
  (2025) 027}, [\href{https://arxiv.org/abs/2507.07755}{{\tt 2507.07755}}].

\bibitem{Krajewski:2024xuz}
T.~Krajewski, M.~Lewicki, M.~Vasar, V.~Vaskonen, H.~Veerm{\"a}e and M.~Zych,
  \emph{{Thermalization effects on the dynamics of growing vacuum bubbles}},
  \href{http://dx.doi.org/10.1007/JHEP03(2025)178}{\emph{JHEP} {\bf 03} (2025)
  178}, [\href{https://arxiv.org/abs/2411.15094}{{\tt 2411.15094}}].

\bibitem{Bodeker:2009qy}
D.~Bodeker and G.~D. Moore, \emph{{Can electroweak bubble walls run away?}},
  \href{http://dx.doi.org/10.1088/1475-7516/2009/05/009}{\emph{JCAP} {\bf 05}
  (2009) 009}, [\href{https://arxiv.org/abs/0903.4099}{{\tt 0903.4099}}].

\bibitem{Bodeker:2017cim}
D.~Bodeker and G.~D. Moore, \emph{{Electroweak Bubble Wall Speed Limit}},
  \href{http://dx.doi.org/10.1088/1475-7516/2017/05/025}{\emph{JCAP} {\bf 05}
  (2017) 025}, [\href{https://arxiv.org/abs/1703.08215}{{\tt 1703.08215}}].

\bibitem{Kainulainen:2024qpm}
K.~Kainulainen and N.~Venkatesan, \emph{{Systematic moment expansion for
  electroweak baryogenesis}},
  \href{http://dx.doi.org/10.1088/1475-7516/2024/08/058}{\emph{JCAP} {\bf 08}
  (2024) 058}, [\href{https://arxiv.org/abs/2407.13639}{{\tt 2407.13639}}].

\bibitem{Barni:2025ifb}
G.~Barni, \emph{{Electroweak Baryogenesis with BARYONET: a self-contained
  review of the WKB approach}},  \href{https://arxiv.org/abs/2510.21915}{{\tt
  2510.21915}}.

\bibitem{vandeVis:2025plm}
J.~van~de Vis, P.~Schicho, L.~Niemi, B.~Laurent, J.~Hirvonen and O.~Gould,
  \emph{{WallGo investigates: Theoretical uncertainties in the bubble wall
  velocity}},  \href{https://arxiv.org/abs/2510.27691}{{\tt 2510.27691}}.

\bibitem{Moore:1997im}
G.~D. Moore, \emph{{Computing the strong sphaleron rate}},
  \href{http://dx.doi.org/10.1016/S0370-2693(97)01046-0}{\emph{Phys. Lett. B}
  {\bf 412} (1997) 359--370}, [\href{https://arxiv.org/abs/hep-ph/9705248}{{\tt
  hep-ph/9705248}}].

\bibitem{McDonald:1995hp}
J.~McDonald, \emph{{Cosmological domain wall evolution and spontaneous CP
  violation from a gauge singlet scalar sector}},
  \href{http://dx.doi.org/10.1016/0370-2693(95)00716-X}{\emph{Phys. Lett. B}
  {\bf 357} (1995) 19--28}.

\bibitem{Espinosa:2011eu}
J.~R. Espinosa, B.~Gripaios, T.~Konstandin and F.~Riva, \emph{{Electroweak
  Baryogenesis in Non-minimal Composite Higgs Models}},
  \href{http://dx.doi.org/10.1088/1475-7516/2012/01/012}{\emph{JCAP} {\bf 01}
  (2012) 012}, [\href{https://arxiv.org/abs/1110.2876}{{\tt 1110.2876}}].

\bibitem{Chao:2017oux}
W.~Chao, \emph{{CP Violation at the Finite Temperature}},
  \href{http://dx.doi.org/10.1016/j.physletb.2019.07.025}{\emph{Phys. Lett. B}
  {\bf 796} (2019) 102--106}, [\href{https://arxiv.org/abs/1706.01041}{{\tt
  1706.01041}}].

\bibitem{Altmannshofer:2020shb}
W.~Altmannshofer, S.~Gori, N.~Hamer and H.~H. Patel, \emph{{Electron EDM in the
  complex two-Higgs doublet model}},
  \href{http://dx.doi.org/10.1103/PhysRevD.102.115042}{\emph{Phys. Rev. D} {\bf
  102} (2020) 115042}, [\href{https://arxiv.org/abs/2009.01258}{{\tt
  2009.01258}}].

\bibitem{CMS:2016xnc}
{\scshape CMS} collaboration, V.~Khachatryan et~al., \emph{{Search for neutral
  resonances decaying into a Z boson and a pair of b jets or $\tau$ leptons}},
  \href{http://dx.doi.org/10.1016/j.physletb.2016.05.087}{\emph{Phys. Lett. B}
  {\bf 759} (2016) 369--394}, [\href{https://arxiv.org/abs/1603.02991}{{\tt
  1603.02991}}].

\bibitem{CMS:2019wml}
{\scshape CMS} collaboration, \emph{{Search for 2HDM neutral Higgs bosons
  through the $\mathrm{H} \to \mathrm{Z}\mathrm{A} \to
  \ell^{+}\ell^{-}\mathrm{b}\overline{\mathrm{b}}$ process in proton-proton
  collisions at $\sqrt{s} = 13~\mathrm{TeV}$}}, {\emph{CMS-PAS-HIG-18-012}
  (2019) }.

\bibitem{ATLAS:2018oht}
{\scshape ATLAS} collaboration, M.~Aaboud et~al., \emph{{Search for a heavy
  Higgs boson decaying into a $Z$ boson and another heavy Higgs boson in the
  $\ell\ell bb$ final state in $pp$ collisions at $\sqrt{s}=13$ TeV with the
  ATLAS detector}},
  \href{http://dx.doi.org/10.1016/j.physletb.2018.07.006}{\emph{Phys. Lett. B}
  {\bf 783} (2018) 392--414}, [\href{https://arxiv.org/abs/1804.01126}{{\tt
  1804.01126}}].

\bibitem{CMS:2018rmh}
{\scshape CMS} collaboration, A.~M. Sirunyan et~al., \emph{{Search for
  additional neutral MSSM Higgs bosons in the $\tau\tau$ final state in
  proton-proton collisions at $\sqrt{s}=$ 13 TeV}},
  \href{http://dx.doi.org/10.1007/JHEP09(2018)007}{\emph{JHEP} {\bf 09} (2018)
  007}, [\href{https://arxiv.org/abs/1803.06553}{{\tt 1803.06553}}].

\bibitem{ATLAS:2020zms}
{\scshape ATLAS} collaboration, G.~Aad et~al., \emph{{Search for heavy Higgs
  bosons decaying into two tau leptons with the ATLAS detector using $pp$
  collisions at $\sqrt{s}=13$ TeV}},
  \href{http://dx.doi.org/10.1103/PhysRevLett.125.051801}{\emph{Phys. Rev.
  Lett.} {\bf 125} (2020) 051801},
  [\href{https://arxiv.org/abs/2002.12223}{{\tt 2002.12223}}].

\bibitem{Barducci:2019xkq}
D.~Barducci, K.~Mimasu, J.~M. No, C.~Vernieri and J.~Zurita, \emph{{Enlarging
  the scope of resonant di-Higgs searches: Hunting for Higgs-to-Higgs cascades
  in $4b$ final states at the LHC and future colliders}},
  \href{http://dx.doi.org/10.1007/JHEP02(2020)002}{\emph{JHEP} {\bf 02} (2020)
  002}, [\href{https://arxiv.org/abs/1910.08574}{{\tt 1910.08574}}].

\bibitem{CMS:2018qmt}
{\scshape CMS} collaboration, A.~M. Sirunyan et~al., \emph{{Search for resonant
  pair production of Higgs bosons decaying to bottom quark-antiquark pairs in
  proton-proton collisions at 13 TeV}},
  \href{http://dx.doi.org/10.1007/JHEP08(2018)152}{\emph{JHEP} {\bf 08} (2018)
  152}, [\href{https://arxiv.org/abs/1806.03548}{{\tt 1806.03548}}].

\bibitem{CMS:2022dwd}
{\scshape CMS} collaboration, A.~Tumasyan et~al., \emph{{A portrait of the
  Higgs boson by the CMS experiment ten years after the discovery.}},
  \href{http://dx.doi.org/10.1038/s41586-022-04892-x}{\emph{Nature} {\bf 607}
  (2022) 60--68}, [\href{https://arxiv.org/abs/2207.00043}{{\tt 2207.00043}}].

\bibitem{CMS:2024bds}
{\scshape CMS} collaboration, \emph{{Search for a new heavy scalar boson
  decaying into a Higgs boson and a new scalar particle in the four b-quarks
  final state using proton-proton collisions at $sqrt{s}=$ 13 TeV}}, .

\bibitem{HFLAV:2022esi}
Y.~S. Amhis et~al., \emph{{Averages of b-hadron, c-hadron, and
  \ensuremath{\tau}-lepton properties as of 2021}},
  \href{http://dx.doi.org/10.1103/PhysRevD.107.052008}{\emph{Phys. Rev. D} {\bf
  107} (2023) 052008}.

\bibitem{Hermann:2012fc}
T.~Hermann, M.~Misiak and M.~Steinhauser, \emph{{$\bar{B}\to X_s \gamma$ in the
  Two Higgs Doublet Model up to Next-to-Next-to-Leading Order in QCD}},
  \href{http://dx.doi.org/10.1007/JHEP11(2012)036}{\emph{JHEP} {\bf 11} (2012)
  036}, [\href{https://arxiv.org/abs/1208.2788}{{\tt 1208.2788}}].

\bibitem{HFLAV:2019otj}
{\scshape HFLAV} collaboration, Y.~S. Amhis et~al., \emph{{Averages of
  b-hadron, c-hadron, and $\tau $-lepton properties as of 2018}},
  \href{http://dx.doi.org/10.1140/epjc/s10052-020-8156-7}{\emph{Eur. Phys. J.
  C} {\bf 81} (2021) 226}, [\href{https://arxiv.org/abs/1909.12524}{{\tt
  1909.12524}}].

\bibitem{Atkinson:2021eox}
O.~Atkinson, M.~Black, A.~Lenz, A.~Rusov and J.~Wynne, \emph{{Cornering the Two
  Higgs Doublet Model Type II}},
  \href{http://dx.doi.org/10.1007/JHEP04(2022)172}{\emph{JHEP} {\bf 04} (2022)
  172}, [\href{https://arxiv.org/abs/2107.05650}{{\tt 2107.05650}}].

\bibitem{Skiba:1992mg}
W.~Skiba and J.~Kalinowski, \emph{{$B_s \to \tau^{+} \tau^{-}$ decay in a two
  Higgs doublet model}},
  \href{http://dx.doi.org/10.1016/0550-3213(93)90470-A}{\emph{Nucl. Phys. B}
  {\bf 404} (1993) 3--19}.

\bibitem{Logan:2000iv}
H.~E. Logan and U.~Nierste, \emph{{$B_{s,d} \to \ell^+ \ell^-$ in a two Higgs
  doublet model}},
  \href{http://dx.doi.org/10.1016/S0550-3213(00)00417-X}{\emph{Nucl. Phys. B}
  {\bf 586} (2000) 39--55}, [\href{https://arxiv.org/abs/hep-ph/0004139}{{\tt
  hep-ph/0004139}}].

\bibitem{Dolan:2014ska}
M.~J. Dolan, F.~Kahlhoefer, C.~McCabe and K.~Schmidt-Hoberg, \emph{{A taste of
  dark matter: Flavour constraints on pseudoscalar mediators}},
  \href{http://dx.doi.org/10.1007/JHEP03(2015)171}{\emph{JHEP} {\bf 03} (2015)
  171}, [\href{https://arxiv.org/abs/1412.5174}{{\tt 1412.5174}}].

\bibitem{Enomoto:2021dkl}
K.~Enomoto, S.~Kanemura and Y.~Mura, \emph{{Electroweak baryogenesis in aligned
  two Higgs doublet models}},
  \href{http://dx.doi.org/10.1007/JHEP01(2022)104}{\emph{JHEP} {\bf 01} (2022)
  104}, [\href{https://arxiv.org/abs/2111.13079}{{\tt 2111.13079}}].

\bibitem{Goncalves:2023svb}
D.~Gon\c{c}alves, A.~Kaladharan and Y.~Wu, \emph{{Gravitational waves, bubble
  profile, and baryon asymmetry in the complex 2HDM}},
  \href{http://dx.doi.org/10.1103/PhysRevD.108.075010}{\emph{Phys. Rev. D} {\bf
  108} (2023) 075010}, [\href{https://arxiv.org/abs/2307.03224}{{\tt
  2307.03224}}].

\bibitem{Liu:2023sey}
S.~Liu and L.~Wang, \emph{{Spontaneous CP violation electroweak baryogenesis
  and gravitational wave through multistep phase transitions}},
  \href{http://dx.doi.org/10.1103/PhysRevD.107.115008}{\emph{Phys. Rev. D} {\bf
  107} (2023) 115008}, [\href{https://arxiv.org/abs/2302.04639}{{\tt
  2302.04639}}].

\bibitem{Aiko:2025tbk}
M.~Aiko, M.~Endo, S.~Kanemura and Y.~Mura, \emph{{Electroweak baryogenesis in
  2HDM without EDM cancellation}},
  \href{http://dx.doi.org/10.1007/JHEP07(2025)236}{\emph{JHEP} {\bf 07} (2025)
  236}, [\href{https://arxiv.org/abs/2504.07705}{{\tt 2504.07705}}].

\bibitem{Si:2024vrq}
Z.-g. Si, H.-x. Wang, L.~Wang and Y.~Zhang, \emph{{Exploring multi-step
  electroweak phase transitions in the 2HDM+$a$}},
  \href{http://dx.doi.org/10.1140/epjc/s10052-025-13936-1}{\emph{Eur. Phys. J.
  C} {\bf 85} (2025) 273}, [\href{https://arxiv.org/abs/2410.15975}{{\tt
  2410.15975}}].

\bibitem{Ipek:2014gua}
S.~Ipek, D.~McKeen and A.~E. Nelson, \emph{{A Renormalizable Model for the
  Galactic Center Gamma Ray Excess from Dark Matter Annihilation}},
  \href{http://dx.doi.org/10.1103/PhysRevD.90.055021}{\emph{Phys. Rev. D} {\bf
  90} (2014) 055021}, [\href{https://arxiv.org/abs/1404.3716}{{\tt
  1404.3716}}].

\bibitem{No:2015xqa}
J.~M. No, \emph{{Looking through the pseudoscalar portal into dark matter:
  Novel mono-Higgs and mono-Z signatures at the LHC}},
  \href{http://dx.doi.org/10.1103/PhysRevD.93.031701}{\emph{Phys. Rev. D} {\bf
  93} (2016) 031701}, [\href{https://arxiv.org/abs/1509.01110}{{\tt
  1509.01110}}].

\bibitem{Goncalves:2016iyg}
D.~Goncalves, P.~A.~N. Machado and J.~M. No, \emph{{Simplified Models for Dark
  Matter Face their Consistent Completions}},
  \href{http://dx.doi.org/10.1103/PhysRevD.95.055027}{\emph{Phys. Rev. D} {\bf
  95} (2017) 055027}, [\href{https://arxiv.org/abs/1611.04593}{{\tt
  1611.04593}}].

\bibitem{Bauer:2017ota}
M.~Bauer, U.~Haisch and F.~Kahlhoefer, \emph{{Simplified dark matter models
  with two Higgs doublets: I. Pseudoscalar mediators}},
  \href{http://dx.doi.org/10.1007/JHEP05(2017)138}{\emph{JHEP} {\bf 05} (2017)
  138}, [\href{https://arxiv.org/abs/1701.07427}{{\tt 1701.07427}}].

\bibitem{Abe:2018bpo}
{\scshape LHC Dark Matter Working Group} collaboration, T.~Abe et~al.,
  \emph{{LHC Dark Matter Working Group: Next-generation spin-0 dark matter
  models}}, \href{http://dx.doi.org/10.1016/j.dark.2019.100351}{\emph{Phys.
  Dark Univ.} {\bf 27} (2020) 100351},
  [\href{https://arxiv.org/abs/1810.09420}{{\tt 1810.09420}}].

\bibitem{Robens:2021lov}
T.~Robens, \emph{{The THDMa revisited}},
  \href{https://arxiv.org/abs/2106.02962}{{\tt 2106.02962}}.

\bibitem{Carrington:1991hz}
M.~E. Carrington, \emph{{The Effective potential at finite temperature in the
  Standard Model}},
  \href{http://dx.doi.org/10.1103/PhysRevD.45.2933}{\emph{Phys. Rev. D} {\bf
  45} (1992) 2933--2944}.

\bibitem{Martin:2014bca}
S.~P. Martin, \emph{{Taming the Goldstone contributions to the effective
  potential}}, \href{http://dx.doi.org/10.1103/PhysRevD.90.016013}{\emph{Phys.
  Rev. D} {\bf 90} (2014) 016013}, [\href{https://arxiv.org/abs/1406.2355}{{\tt
  1406.2355}}].

\bibitem{Elias-Miro:2014pca}
J.~Elias-Miro, J.~R. Espinosa and T.~Konstandin, \emph{{Taming Infrared
  Divergences in the Effective Potential}},
  \href{http://dx.doi.org/10.1007/JHEP08(2014)034}{\emph{JHEP} {\bf 08} (2014)
  034}, [\href{https://arxiv.org/abs/1406.2652}{{\tt 1406.2652}}].

\bibitem{Espinosa:2016uaw}
J.~R. Espinosa, M.~Garny and T.~Konstandin, \emph{{Interplay of Infrared
  Divergences and Gauge-Dependence of the Effective Potential}},
  \href{http://dx.doi.org/10.1103/PhysRevD.94.055026}{\emph{Phys. Rev. D} {\bf
  94} (2016) 055026}, [\href{https://arxiv.org/abs/1607.08432}{{\tt
  1607.08432}}].

\bibitem{Braathen:2016cqe}
J.~Braathen and M.~D. Goodsell, \emph{{Avoiding the Goldstone Boson Catastrophe
  in general renormalisable field theories at two loops}},
  \href{http://dx.doi.org/10.1007/JHEP12(2016)056}{\emph{JHEP} {\bf 12} (2016)
  056}, [\href{https://arxiv.org/abs/1609.06977}{{\tt 1609.06977}}].

\bibitem{Espinosa:2017aew}
J.~R. Espinosa and T.~Konstandin, \emph{{Resummation of Goldstone Infrared
  Divergences: A Proof to All Orders}},
  \href{http://dx.doi.org/10.1103/PhysRevD.97.056020}{\emph{Phys. Rev. D} {\bf
  97} (2018) 056020}, [\href{https://arxiv.org/abs/1712.08068}{{\tt
  1712.08068}}].

\bibitem{Kannike:2012pe}
K.~Kannike, \emph{{Vacuum Stability Conditions From Copositivity Criteria}},
  \href{http://dx.doi.org/10.1140/epjc/s10052-012-2093-z}{\emph{Eur. Phys. J.
  C} {\bf 72} (2012) 2093}, [\href{https://arxiv.org/abs/1205.3781}{{\tt
  1205.3781}}].

\bibitem{Chakrabortty:2013mha}
J.~Chakrabortty, P.~Konar and T.~Mondal, \emph{{Copositive Criteria and
  Boundedness of the Scalar Potential}},
  \href{http://dx.doi.org/10.1103/PhysRevD.89.095008}{\emph{Phys. Rev. D} {\bf
  89} (2014) 095008}, [\href{https://arxiv.org/abs/1311.5666}{{\tt
  1311.5666}}].

\bibitem{Drozd:2014yla}
A.~Drozd, B.~Grzadkowski, J.~F. Gunion and Y.~Jiang, \emph{{Extending
  two-Higgs-doublet models by a singlet scalar field - the Case for Dark
  Matter}}, \href{http://dx.doi.org/10.1007/JHEP11(2014)105}{\emph{JHEP} {\bf
  11} (2014) 105}, [\href{https://arxiv.org/abs/1408.2106}{{\tt 1408.2106}}].

\bibitem{Muhlleitner:2016mzt}
M.~Muhlleitner, M.~O.~P. Sampaio, R.~Santos and J.~Wittbrodt, \emph{{The N2HDM
  under Theoretical and Experimental Scrutiny}},
  \href{http://dx.doi.org/10.1007/JHEP03(2017)094}{\emph{JHEP} {\bf 03} (2017)
  094}, [\href{https://arxiv.org/abs/1612.01309}{{\tt 1612.01309}}].

\bibitem{Engeln:2020fld}
I.~Engeln, P.~Ferreira, M.~M. M{\"u}hlleitner, R.~Santos and J.~Wittbrodt,
  \emph{{The Dark Phases of the N2HDM}},
  \href{http://dx.doi.org/10.1007/JHEP08(2020)085}{\emph{JHEP} {\bf 08} (2020)
  085}, [\href{https://arxiv.org/abs/2004.05382}{{\tt 2004.05382}}].

\bibitem{Klimenko:1984qx}
K.~G. Klimenko, \emph{{On Necessary and Sufficient Conditions for Some Higgs
  Potentials to Be Bounded From Below}},
  \href{http://dx.doi.org/10.1007/BF01034825}{\emph{Theor. Math. Phys.} {\bf
  62} (1985) 58--65}.

\bibitem{He:2016mls}
X.-G. He and J.~Tandean, \emph{{New LUX and PandaX-II Results Illuminating the
  Simplest Higgs-Portal Dark Matter Models}},
  \href{http://dx.doi.org/10.1007/JHEP12(2016)074}{\emph{JHEP} {\bf 12} (2016)
  074}, [\href{https://arxiv.org/abs/1609.03551}{{\tt 1609.03551}}].

\bibitem{Arcadi:2022lpp}
G.~Arcadi, N.~Benincasa, A.~Djouadi and K.~Kannike,
  \emph{{Two-Higgs-doublet-plus-pseudoscalar model: Collider, dark matter, and
  gravitational wave signals}},
  \href{http://dx.doi.org/10.1103/PhysRevD.108.055010}{\emph{Phys. Rev. D} {\bf
  108} (2023) 055010}, [\href{https://arxiv.org/abs/2212.14788}{{\tt
  2212.14788}}].

\bibitem{Dutta:2023cig}
J.~Dutta, J.~Lahiri, C.~Li, G.~Moortgat-Pick, S.~F. Tabira and J.~A. Ziegler,
  \emph{{Dark matter phenomenology in 2HDMS in light of the 95 GeV excess}},
  \href{http://dx.doi.org/10.1140/epjc/s10052-024-13176-9}{\emph{Eur. Phys. J.
  C} {\bf 84} (2024) 926}, [\href{https://arxiv.org/abs/2308.05653}{{\tt
  2308.05653}}].

\bibitem{Arcadi:2023smv}
G.~Arcadi, G.~Busoni, D.~Cabo-Almeida and N.~Krishnan, \emph{{Is there a scalar
  or pseudoscalar at 95~GeV?}},
  \href{http://dx.doi.org/10.1103/PhysRevD.110.115028}{\emph{Phys. Rev. D} {\bf
  110} (2024) 115028}, [\href{https://arxiv.org/abs/2311.14486}{{\tt
  2311.14486}}].

\bibitem{Cornwall:1974km}
J.~M. Cornwall, D.~N. Levin and G.~Tiktopoulos, \emph{{Derivation of Gauge
  Invariance from High-Energy Unitarity Bounds on the s Matrix}},
  \href{http://dx.doi.org/10.1103/PhysRevD.10.1145}{\emph{Phys. Rev. D} {\bf
  10} (1974) 1145}.

\bibitem{Lee:1977eg}
B.~W. Lee, C.~Quigg and H.~B. Thacker, \emph{{Weak Interactions at Very
  High-Energies: The Role of the Higgs Boson Mass}},
  \href{http://dx.doi.org/10.1103/PhysRevD.16.1519}{\emph{Phys. Rev. D} {\bf
  16} (1977) 1519}.

\bibitem{Chanowitz:1985hj}
M.~S. Chanowitz and M.~K. Gaillard, \emph{{The TeV Physics of Strongly
  Interacting W's and Z's}},
  \href{http://dx.doi.org/10.1016/0550-3213(85)90580-2}{\emph{Nucl. Phys. B}
  {\bf 261} (1985) 379--431}.

\bibitem{Willenbrock:1987xz}
S.~S.~D. Willenbrock, \emph{{Pair Production of $W$ and $Z$ Bosons and the
  Goldstone Boson Equivalence Theorem}},
  \href{http://dx.doi.org/10.1016/S0003-4916(88)80016-2}{\emph{Annals Phys.}
  {\bf 186} (1988) 15}.

\bibitem{Valencia:1990jp}
G.~Valencia and S.~Willenbrock, \emph{{The Goldstone Boson Equivalence Theorem
  and the Higgs Resonance}},
  \href{http://dx.doi.org/10.1103/PhysRevD.42.853}{\emph{Phys. Rev. D} {\bf 42}
  (1990) 853--859}.

\bibitem{Kanemura:1993hm}
S.~Kanemura, T.~Kubota and E.~Takasugi, \emph{{Lee-Quigg-Thacker bounds for
  Higgs boson masses in a two doublet model}},
  \href{http://dx.doi.org/10.1016/0370-2693(93)91205-2}{\emph{Phys. Lett. B}
  {\bf 313} (1993) 155--160}, [\href{https://arxiv.org/abs/hep-ph/9303263}{{\tt
  hep-ph/9303263}}].

\bibitem{Arhrib:2000is}
A.~Arhrib, \emph{{Unitarity constraints on scalar parameters of the standard
  and two Higgs doublets model}},  in \emph{{Workshop on Noncommutative
  Geometry, Superstrings and Particle Physics}}, 12, 2000.
\newblock \href{https://arxiv.org/abs/hep-ph/0012353}{{\tt hep-ph/0012353}}.

\bibitem{Ginzburg:2005dt}
I.~F. Ginzburg and I.~P. Ivanov, \emph{{Tree-level unitarity constraints in the
  most general 2HDM}},
  \href{http://dx.doi.org/10.1103/PhysRevD.72.115010}{\emph{Phys. Rev. D} {\bf
  72} (2005) 115010}, [\href{https://arxiv.org/abs/hep-ph/0508020}{{\tt
  hep-ph/0508020}}].

\bibitem{Horejsi:2005da}
J.~Horejsi and M.~Kladiva, \emph{{Tree-unitarity bounds for THDM Higgs masses
  revisited}}, \href{http://dx.doi.org/10.1140/epjc/s2006-02472-3}{\emph{Eur.
  Phys. J. C} {\bf 46} (2006) 81--91},
  [\href{https://arxiv.org/abs/hep-ph/0510154}{{\tt hep-ph/0510154}}].

\bibitem{Goodsell:2018fex}
M.~D. Goodsell and F.~Staub, \emph{{Improved unitarity constraints in
  Two-Higgs-Doublet-Models}},
  \href{http://dx.doi.org/10.1016/j.physletb.2018.11.030}{\emph{Phys. Lett. B}
  {\bf 788} (2019) 206--212}, [\href{https://arxiv.org/abs/1805.07310}{{\tt
  1805.07310}}].

\bibitem{Huet:1995sh}
P.~Huet and A.~E. Nelson, \emph{{Electroweak baryogenesis in supersymmetric
  models}}, \href{http://dx.doi.org/10.1103/PhysRevD.53.4578}{\emph{Phys. Rev.
  D} {\bf 53} (1996) 4578--4597},
  [\href{https://arxiv.org/abs/hep-ph/9506477}{{\tt hep-ph/9506477}}].

\bibitem{Moore:2010jd}
G.~D. Moore and M.~Tassler, \emph{{The Sphaleron Rate in SU(N) Gauge Theory}},
  \href{http://dx.doi.org/10.1007/JHEP02(2011)105}{\emph{JHEP} {\bf 02} (2011)
  105}, [\href{https://arxiv.org/abs/1011.1167}{{\tt 1011.1167}}].

\bibitem{Cline:2021dkf}
J.~M. Cline and B.~Laurent, \emph{{Electroweak baryogenesis from light fermion
  sources: A critical study}},
  \href{http://dx.doi.org/10.1103/PhysRevD.104.083507}{\emph{Phys. Rev. D} {\bf
  104} (2021) 083507}, [\href{https://arxiv.org/abs/2108.04249}{{\tt
  2108.04249}}].

\bibitem{Schmitz:2020syl}
K.~Schmitz, \emph{{New Sensitivity Curves for Gravitational-Wave Signals from
  Cosmological Phase Transitions}},
  \href{http://dx.doi.org/10.1007/JHEP01(2021)097}{\emph{JHEP} {\bf 01} (2021)
  097}, [\href{https://arxiv.org/abs/2002.04615}{{\tt 2002.04615}}].

\bibitem{Yagi:2011wg}
K.~Yagi and N.~Seto, \emph{{Detector configuration of DECIGO/BBO and
  identification of cosmological neutron-star binaries}},
  \href{http://dx.doi.org/10.1103/PhysRevD.83.044011}{\emph{Phys. Rev. D} {\bf
  83} (2011) 044011}, [\href{https://arxiv.org/abs/1101.3940}{{\tt
  1101.3940}}].

\bibitem{LISA:2017pwj}
{\scshape LISA} collaboration, P.~Amaro-Seoane et~al., \emph{{Laser
  Interferometer Space Antenna}},  \href{https://arxiv.org/abs/1702.00786}{{\tt
  1702.00786}}.

\bibitem{Caprini:2015zlo}
C.~Caprini et~al., \emph{{Science with the space-based interferometer eLISA.
  II: Gravitational waves from cosmological phase transitions}},
  \href{http://dx.doi.org/10.1088/1475-7516/2016/04/001}{\emph{JCAP} {\bf 04}
  (2016) 001}, [\href{https://arxiv.org/abs/1512.06239}{{\tt 1512.06239}}].

\bibitem{Saikawa:2018rcs}
K.~Saikawa and S.~Shirai, \emph{{Primordial gravitational waves, precisely: The
  role of thermodynamics in the Standard Model}},
  \href{http://dx.doi.org/10.1088/1475-7516/2018/05/035}{\emph{JCAP} {\bf 05}
  (2018) 035}, [\href{https://arxiv.org/abs/1803.01038}{{\tt 1803.01038}}].

\end{thebibliography}\endgroup

\end{document}